\documentclass[12pt]{article}

\usepackage[english]{babel}
\usepackage[utf8]{inputenc}
\usepackage[round,semicolon]{natbib}
\usepackage{geometry}
\usepackage{setspace}
\usepackage{amsmath}
\usepackage{amsfonts}
\usepackage{amsthm}
\usepackage{amssymb}
\usepackage{mathrsfs}
\usepackage{bbm}
\usepackage{amstext}
\usepackage{bm}

\usepackage{ulem}
\usepackage{soul}
\usepackage{mdframed}
\usepackage{cancel}
\usepackage{footmisc}

\usepackage{multirow}
\usepackage{babel}
\usepackage{array}
\usepackage{rotating}
\usepackage{longtable}
\usepackage{float}
\usepackage{booktabs}
\usepackage{lscape}

\usepackage{caption}
\usepackage{subcaption}

\usepackage[ruled]{algorithm2e}

\usepackage{xcolor}
\usepackage{listings}

\definecolor{codegreen}{rgb}{0,0.6,0}
\definecolor{codegray}{rgb}{0.5,0.5,0.5}
\definecolor{codepurple}{rgb}{0.58,0,0.82}
\definecolor{backcolour}{rgb}{0.95,0.95,0.92}

\lstdefinestyle{mystyle}{
    backgroundcolor=\color{backcolour},   
    commentstyle=\color{codegreen},
    keywordstyle=\color{magenta},
    numberstyle=\tiny\color{codegray},
    stringstyle=\color{codepurple},
    basicstyle=\ttfamily\footnotesize,
    breakatwhitespace=false,         
    breaklines=true,    captionpos=b,                    
    keepspaces=true, numbers=left,                    
    numbersep=5pt, showspaces=false,                
    showstringspaces=false, showtabs=false, tabsize=2}
\usepackage{color}
\usepackage{graphicx}
\usepackage{xcolor}

\makeatletter
\def\maxwidth{ %
  \ifdim\Gin@nat@width>\linewidth
    \linewidth
  \else
    \Gin@nat@width
  \fi
}
\makeatother

\definecolor{fgcolor}{rgb}{0.345, 0.345, 0.345}

\usepackage{framed}
\makeatletter
 {\par\unskip\endMakeFramed%
 \at@end@of@kframe}
\makeatother

\definecolor{shadecolor}{rgb}{.97, .97, .97}
\definecolor{messagecolor}{rgb}{0, 0, 0}
\definecolor{warningcolor}{rgb}{1, 0, 1}
\definecolor{errorcolor}{rgb}{1, 0, 0}
\usepackage{listings}
\usepackage{alltt}

\usepackage{url}
\usepackage[hidelinks]{hyperref}
\hypersetup{
    colorlinks=true,
    linkcolor=[rgb]{0.6, 0.1058824, 0.1176471}, 
    filecolor=[rgb]{0.6, 0.1058824, 0.1176471}, 
    urlcolor=[rgb]{0.6, 0.1058824, 0.1176471}, 
    citecolor=[rgb]{0.6, 0.1058824, 0.1176471}, 
    }
\usepackage{chngcntr}
\counterwithin*{section}{part}

\usepackage{todonotes}

\usepackage{rotfloat}
\usepackage{pdfpages}
\usepackage{indentfirst}
\usepackage{enumitem}

\newtheorem{theorem}{Theorem}
\newtheorem{remark}{Remark}

\newtheorem{lemma}{Lemma}

\newtheorem{example}{Example}
\newtheorem{definition}{Definition}

\newtheorem{corollary}{Corollary}

\newcommand{\E}{\mathbb{E}}
\newcommand{\indep}{\perp\!\!\!\perp}
\newcommand{\var}{\text{Var}}
\newcommand{\cov}{\text{Cov}}

\title{On the Asymptotics of the Minimax Linear Estimator}

\begin{document}

\author{Jing Kong \\
Department of Economics, University of Southern California}
\maketitle

\begin{abstract}
Many causal estimands, such as average treatment effects under unconfoundedness, can be written as continuous linear functionals of an unknown regression function. We study a weighting estimator that sets weights by a minimax procedure: solving a convex optimization problem that trades off worst‑case conditional bias against variance. Despite its growing use, general root‑$n$ theory for this method has been limited. This paper fills that gap.
Under regularity conditions, we show that the minimax linear estimator is root‑$n$ consistent and asymptotically normal, and we derive its asymptotic variance. These results justify ignoring worst‑case bias when forming large‑sample confidence intervals and make inference less sensitive to the scaling of the function class. With a mild variance condition, the estimator attains the semiparametric efficiency bound, so an augmentation step commonly used in the literature is not needed to achieve first‑order optimality.
Evidence from simulations and three empirical applications, including job‑training and minimum‑wage policies, points to a simple rule: in designs satisfying our regularity conditions, standard‑error confidence intervals suffice; otherwise, bias‑aware intervals remain important.


\end{abstract}

\newpage

\section{Introduction}
Many empirical questions reduce to estimating a continuous linear functional of an unknown regression function. Familiar examples include the estimands in causal inference such as the average treatment effect (ATE) under unconfoundedness (\citealp{rosenbaum1983central}) or selection on observables. Minimax (or equivalently balancing) weighting methods (e.g., \citealp{zubizarreta2015stable}, \citealp{kallus2020generalized}, \citealp{ben2021balancing}) have become increasingly popular because they target the estimand directly and are straightforward to implement. Yet large-sample theory for minimax weighting schemes that choose weights by trading off worst-case bias against variance has not been fully characterized. This paper is the first to show that the minimax linear estimator is root-$n$ consistent, asymptotically normal, and semiparametrically efficient for broad classes of continuous linear functionals under standard regularity (Donsker‑type) conditions. 

The minimax principle is straightforward. Given a candidate function class $\mathcal{F}$ for the outcome regression, minimax weights are chosen by solving a single convex optimization problem: they minimize the worst‑case conditional bias over $\mathcal{F}$ subject to a variance penalty. One approach to conduct the inference with such weighting estimators uses bias‑aware confidence intervals that explicitly add a worst‑case bias bound to the standard error to guarantee coverage (\citealp{armstrong2021finite}). This is valuable in complex or irregular settings, but can be conservative in benign cases. 
Motivated by the practical importance of classical root-$n$ normal inference with standard-error-only intervals ($\pm 1.96 \hat{\sigma}$ where $\hat{\sigma}$ is an estimated standard error), we develop asymptotic theory for the minimax linear estimator. We then translate these results into clear guidance on when standard-error confidence intervals (CI) suffice and when bias-aware CIs remain important.

This paper leverages a tight link between minimax weights and the ideal weighting function. By the Riesz representation theorem, any continuous linear functional admits an inner product representation with a unique ``Riesz representer'' (the ideal weight). Because the optimal minimax weights directly estimate this Riesz representer, the resulting estimator inherits good asymptotic properties. Our first result shows a double‐robustness‐type property of the minimax weighting estimator. Specifically, whenever the weighting function converges to the Riesz representer, if either the outcome regression or the Riesz representer function lies in the function class $\mathcal{F}$, the weighting estimator remains consistent (though not necessarily root-$n$ consistent). 

Second, we establish the convergence rate and the limiting distribution of the minimax linear estimator under certain regularity conditions. Suppose both the outcome regression function and the Riesz representer function (strictly speaking, up to a multiplicative constant) belong to $\mathcal{F}$. Then, under standard empirical-process constraints on the function class $\mathcal{F}$ (e.g., Donsker-type conditions) and other mild regularity conditions, 
the minimax weights converge to the Riesz representer and, more importantly, the worst-case conditional bias vanishes faster than $1/\sqrt{n}$. Therefore, 
the minimax linear estimator is root-$n$ consistent and asymptotically normally distributed. Because bias is second order under these conditions, we can safely ignore the bias term: standard‑error–only CIs are valid and insensitive to the scaling of the function class. For example, if $\mathcal{F}$ is a Lipschitz class, any reasonable Lipschitz constant yields the same CI asymptotically.
In addition, under mild conditions on the variance function, the estimator achieves the semiparametric efficiency bound. This implies that, an augmentation step proposed in the literature (e.g., \citealp{hirshberg2021augmented}), adding a regression adjustment to the minimax weights, is not needed for first-order optimality under our conditions.

We also characterize the asymptotic behavior of the minimax linear estimator when the conditional variance function is known and used in the procedure, which clarifies the role of heteroskedasticity in efficiency. Although conditional variance is rarely known in practice, it yields valuable theoretical implications and suggests how feasible procedures using an estimated variance function may be developed in future work. Beyond linear functionals, we show (Appendices D and E) that smooth functions of several minimax linear estimators inherit root‑$n$ normality, enabling valid inference for a broad class of non-linear functionals that depend on multiple regression functions.

In the simulation study, we consider causal parameter estimation under both Donsker and non-Donsker function classes. In Donsker settings, the minimax weights converge toward the Riesz representer as $n$ grows, and the worst-case conditional bias is small relative to the standard error. Augmentation adds little in these cases because the pure weighting estimator is already efficient. In non-Donsker settings, the weights do not closely track the Riesz representer even in larger samples, and the worst-case bias is sizable and sensitive to the scaling of the function class.

Finally, we apply the method in three empirical settings for causal inference and, in each, report point estimates with two CIs: one using only the standard error and one that adds the worst‑case bias. First, for the minimum‑wage change and teenage employment (a county‑level difference-in-differences), a univariate specification makes the Lipschitz class Donsker: the worst‑case bias is negligible and essentially invariant to the scaling constant $C$; the bias-aware and traditional intervals are nearly identical. Second, in the National Supported Work evaluation, multiple continuous covariates render the Lipschitz class non‑Donsker: the worst‑case bias is non-negligible and highly $C$-sensitive, so credible intervals should explicitly account for bias. Third, in the JTPA randomized experiment, we consider both regimes. Across these applications, the patterns align with what the theory implies.

Taken together, our results deliver both a theoretical and a practical contribution. We show that the pure minimax linear estimator is root‑$n$, asymptotically normal, and efficient for continuous linear functionals under regularity conditions. This yields a simple rule of thumb for practice: in regular (Donsker) specifications with large samples, standard-error-only CIs suffice; otherwise, keep inference bias‑aware by adding a worst‑case bias term.

\textbf{Related Literature.} This paper builds upon two strands of literature. First, minimax linear estimation of linear functionals. The idea of constructing linear estimators for a (general) linear functional of a regression function by solving a worst-case bias-variance tradeoff dates back to \citet{donoho1994statistical} and \citet{low1995bias}. An earlier paper is
\citet{heckman1988minimax}, which derives minimax linear estimators for the slope coefficient in a partially linear model when the nuisance function is Lipschitz and shows that the worst-case mean squared error is of order $1/n$ under mild design conditions. More recently, this viewpoint has been applied to treatment-effect problems.  \citet{armstrong2018optimal} provide (near-)optimal inference for a linear functional under convex classes (e.g., smoothness or shape constraints) and take the treatment effect under a regression discontinuity design as an illustrative example. In a follow-up, \citet{armstrong2021finite} extend this to treatment effects under unconfoundedness. They give valid inference procedures that do not rely on overlap assumptions or sufficient smoothness conditions, by explicitly accounting for the non-negligible bias bound. However, these papers do not provide root-$n$ asymptotic distributions or characterize how the worst-case bias behaves as $n$ grows. This paper fill this gap by showing that, under regularity conditions, the worst-case bias is second order and the minimax linear estimator is root-$n$ asymptotically normal.

In parallel, the causal inference literature has developed balancing-weighting methods, e.g., \citet{zubizarreta2015stable}, \citet{athey2018approximate}, \citet{ben2021balancing}, \citet{hirshberg2021augmented}, \citet{chernozhukov2022automatic}, to enforce covariate balance for the ATE and extended it to estimate more general targets.\footnote{Balancing-weighting methods not only mitigate the extreme‐weight instability of traditional inverse propensity score weighting estimator (e.g. \citealp{hirano2003efficient}) when estimated propensity scores approach zero or one (e.g. \citealp{athey2018approximate}, \citealp{chernozhukov2022automatic}), but they also extend naturally to continuous linear functionals whose Riesz representers are not available in closed form.}
Crucially, whenever both impose the same convex function class constraint and criterion (e.g.\ worst‐case mean-square-error), balancing‐weighting and minimax‐linear estimators are identical: both choose weights by solving a single convex program trading maximum imbalance (worst‐case bias) against a penalty on weight magnitude (variance). Among these balancing approaches (e.g., \citealp{athey2018approximate}, \citealp{bruns2025augmented}) this program is typically written in terms of a set of basis expansions, such as a linear sieve, with a norm constraint on the coefficient vector, so that the maximal imbalance reduces to a familiar vector‐norm of the basis‐moment discrepancy. By contrast, the minimax‐linear literature typically considers a general formulation of convex function class.

Although augmented balancing methods, which combine linear weighting with a (nonlinear) outcome regression adjustment (e.g. \citealp{hirshberg2021augmented}, \citealp{chernozhukov2022automatic}), are known to achieve $\sqrt{n}$-consistent and asymptotic normality under certain conditions, the pure balancing/minimax linear estimators for general continuous linear functionals have been less studied under classical $\sqrt{n}$-asymptotics. Recent work has begun to close this gap, but only in specialized treatment effect or population mean settings, such as \citet{chan2016globally}, \citet{wong2018kernel}, \citet{hirshberg2019minimax}, \citet{wang2020minimal}.
For example, 
\citet{hirshberg2019minimax} establishes asymptotic properties for the treatment‐specific mean on a target population, and they confine their analysis to regression functions in the unit ball of a reproducing‐kernel Hilbert space.
To the best of our knowledge, our paper is the first to bridge the pure minimax linear approach with standard root-$n$ asymptotic theory for continuous linear functionals under Donsker-type conditions.
In particular, we show that the pure minimax weighting estimator is itself efficient under these regularity conditions. This aligns with the optimality or near-optimality of the minimax linear/affine estimator for convex function class $\mathcal{F}$ (e.g. \citealp{donoho1994statistical} and \citealp{armstrong2018optimal}
 etc.). In other words, a commonly used augmentation step is unnecessary for efficiency in this scenario. 
 
 However, if the condition of the convexity on functional class $\mathcal{F}$ fails, the augmented or nonlinear estimator may dominate the pure linear estimator. For example, under high-dimensional settings with sparsity (a nonconvex functional class), nonlinear estimators, such as the debiased lasso estimator (e.g. \citealp{javanmard2014confidence} and \citealp{zhang2014confidence}), the double machine learning estimator (e.g. \citealp{chernozhukov2018double}), show better performance. A recent strand, known as automatic double machine learning estimator (Auto-DML) (\citealp{chernozhukov2022automatic}, \citealp{chernozhukov2022dynamic}), shares the same objective as the minimax linear approach: directly recovering the Riesz representer function to get stable weights. These papers show that the Riesz representer is the unique minimizer of a quadratic ``Riesz-regression'' loss (\citealp{chernozhukov2024automatic}), which only depends on the parameter-defining functional and requires no closed functional form for the Riesz representer itself. Their algorithm therefore learns the Riesz representer with any machine-learning method (Lasso, random forests, neural nets) and plugs the result into a Neyman-orthogonal score. \citet{bruns2025augmented} show that under a linearity assumption, the Auto-DML Lasso minimum distance problem (\citealp{chernozhukov2022automatic}) is equivalent to the balancing weights optimization problem.

The remainder of this paper is organized as follows. Section 2 introduces the model setup and the minimax linear estimator. Section 3 shows the asymptotic properties of the minimax linear estimator. Section 4 presents a limited Monte Carlo study. Section 5 discusses three empirical applications. Additional details and proofs are collected in the Appendices. 

\bigskip
Notation: For a given function class $\mathcal{F}$, the gauge of a function $f$ is defined as $||f||_\mathcal{F} = \inf\{\alpha>0: f\in \alpha \mathcal{F}\}$.
Write $\mathcal{F}_r$ to denote the localized class $\{f\in \mathcal{F}: ||f||_{L^2(P)} \leq r\}$, $g\mathcal{F}$ to denote the class of products $\{gf: f\in \mathcal{F}\}$,
and $h(\cdot, \mathcal{F})$ to denote the image class $\{h(\cdot, f): f\in \mathcal{F}\}$. Write $\overline{\text{span}}\mathcal{F}$ to denote the closure of $\text{span}\mathcal{F}$ in the space of square integrable functions and 
$(\overline{\text{span}}\mathcal{F})^\bot $ to denote the orthogonal complement of $\overline{\text{span}}\mathcal{F}$.
The function class $\mathcal{F}$ is pointwise bounded if $\sup_{f\in \mathcal{F}}|f(z)| < \infty$, 
uniformly bounded if $\sup_{f\in \mathcal{F}}||f||_\infty < \infty$ where $||f||_\infty = \sup_{z\in \mathcal{Z}}|f(z)|$,
and pointwise closed if $f\in \mathcal{F}$ whenever for a sequence $f_j \in \mathcal{F}$ it holds that $\lim_{j\to \infty} f_j(z) = f(z)$ for all $z \in \mathcal{Z}$. Denote a standard normal variable by $\bar{z}$ and the corresponding cumulative distribution function by $\Phi(\cdot)$.

\section{Setup and minimax linear estimator}

\subsection{Setup}
We observe a random sample $\{(Z_i, Y_i)\}_{i=1}^n \overset{\text{i.i.d.}}{\sim} P$ where $Z_i \in \mathcal{Z}$ collects all observed covariates (possibly including treatment indicators) and $Y_i \in \mathbb{R}$ is the scalar outcome. The data follow the regression model
\begin{equation}
    Y_i = f(Z_i) + \varepsilon_i,
    \label{eq:yequation}
\end{equation}
where $f:\mathcal{Z} \to \mathbb{R}$ is the (unknown) conditional mean $f(z) = \E[Y|Z=z]$ and $\varepsilon_i$ is a mean-zero error satisfying $\E[\varepsilon_i|Z_i] = 0$ and $\var[\varepsilon_i|Z_i] = \sigma^2(Z_i) \leq \bar{\sigma}^2 <\infty$. Let $h(z, f)$ denote a functional of $f$, which depends on a data observation $z$ and for each fixed $z \in \mathcal{Z}$ the map $f \mapsto h(z,f)$ is linear.
We are interested in parameters of the form
\begin{equation}
    \psi(f) = \E[h(Z, f)],
\end{equation}
which is an expectation of some known formula $h(Z,f)$. Since expectation itself is linear, it follows immediately that $\psi(f)$ is a linear functional of $f$.
We further assume that $\psi(f)$ is a continuous linear functional on $L^2(P_Z)$, i.e., there exists a constant $C_\psi >0$ such that $|\psi(f)| \leq C_\psi ||f||_{L^2(P_Z)}$ for all $f \in L^2(P_Z)$.
A few examples of causal or semiparametric targets fall into this framework:
\begin{example}
    (ATE) Let $Z = (D,X)$ where $D \in \{0,1\}$ is a binary treatment indicator and $X$ collects observed controls. In the potential outcomes framework, each unit has a pair of potential responses $\{Y(1), Y(0)\}$ and we observe only $Y = D Y(1) + (1-D)Y(0)$. Define the conditional mean $f(d,x) = \E[Y|D=d,X=x]$ and $h(z,f) = f(1,x)-f(0,x)$. Then our target parameter is
    \begin{equation}
        \psi(f) = \E[f(1,X) - f(0,X)].
    \end{equation}
    If potential outcomes $(Y(1), Y(0))$ are mean independent of the treatment indicator $D$ conditional on control covariates $X$ (i.e., unconfoundedness assumption), then this target coincides with the ATE, $\psi(f) = \E[Y(1)-Y(0)]$.
\end{example}

\begin{example}
    (Partially Linear Model Coefficient) Let $Z = (X_1, X_2)$ where the conditional expectation function takes the partially linear form 
    \begin{equation}
        f(X_1, X_2) = X_1\beta + g(X_2)
    \end{equation}
    where $\beta \in \mathbb{R}$ is an unknown scalar and $g(\cdot)$ is an arbitrary ``nuisance'' square-integrable function. The target parameter is 
    \begin{equation}
        \psi(f) = \beta.
    \end{equation}
    In particular, if $X_1$ is a binary treatment indicator and one imposes $g(X_2)$ as the control-only component, this reduces to the constant effects ATE setting (under unconfoundedness), $\E[Y(1)-Y(0)|X_2] = \beta$.
\end{example}

\subsection{The Riesz representation of a linear functional}
To construct estimators or confidence intervals for $\psi(f) = E[h(Z,f)]$, it is crucial to recognize that $\psi(f)$ is a continuous linear functional on the Hilbert space $L^2(P_Z)$ of square-integrable functions of $Z$. Define the inner product in $L^2(P_Z)$ as 
\begin{equation}
    \langle f_1, f_2 \rangle = \int f_1(z) f_2(z) dP_Z(z)  = \E[f_1(Z) f_2(Z)]
\end{equation}
for any $f_1, f_2 \in L^2(P_Z)$.

Let $\mathcal{F} \subset L^2(P_Z)$ be a convex class of candidate regression functions and write $\overline{\text{span}}\mathcal{F}$ for its closed linear span in $L^2(P_Z)$. By the Riesz representation theorem, for any continuous linear functional $\psi(\cdot)$ on $\overline{\text{span}}\mathcal{F}$,
there exists a unique function $\gamma^* \in \overline{\text{span}}\mathcal{F}$ with $\E[(\gamma^*(Z))^2]< \infty$ such that 
\begin{equation}
    \psi(f) = \langle \gamma^*, f \rangle  = \int \gamma^*(z)f(z) dP_Z(z)  \text{ for all } f \in \overline{\text{span}}\mathcal{F}.
    \label{eq:Rproperty}
\end{equation}
We call $\gamma^*$ the \textit{Riesz representer} of $\psi(\cdot)$ on $\overline{\text{span}}\mathcal{F}$. 

\begin{remark}
    If $\overline{\text{span}}\mathcal{F} = L^2(P_Z)$, then $\gamma^*$ is uniquely determined in the full ambient space. However, if instead $\overline{\text{span}}\mathcal{F}$ is a strict subspace of $L^2(P_Z)$, then a function $\gamma^\#$ satisfying the equality $\int \gamma^\#(z)f(z) dP(z) = \psi(f)$ for all $f \in \overline{\text{span}}\mathcal{F}$ is not unique in $L^2(P_Z)$ space. We return to this phenomenon and its implications in Section \ref{section:hetero}.
\end{remark}



\setcounter{example}{0}
\begin{example}
    (ATE, cont'd) In the ATE setting we take $\mathcal{F}$ to be such that $\overline{\text{span}}\mathcal{F} = L^2(P_Z)$, so the Riesz representer for the ATE is unique in the full Hilbert space of square-integrable functions of $(d,x)$. It is the well-known inverse propensity score function (e.g. \citealp{chernozhukov2018double}):
    \begin{equation}
        \gamma^*_{\text{ATE, FULL}}(d,x) = \frac{d}{e(x)} - \frac{1-d}{1-e(x)},
        \label{eq:RR_ATE_L2}
    \end{equation}
    where $e(x) = \E[D|X=x]$ is the propensity score function.
\end{example}

\begin{example}
    (Partially Linear Model Coefficient, cont'd) In this example, we restrict attention to a strict subspace 
    \begin{equation}
        \mathcal{F} = \{f(x_1,x_2) = x_1\beta + g(x_2)|\beta \in \mathbb{R}, g \in L^2(P_{X_2})\} \subsetneq L^2(P_Z).
    \end{equation}
    Its closed linear span is itself so $\overline{\text{span}}\mathcal{F}  \subsetneq L^2(P_Z)$. 
    The unique Riesz representer $\gamma^* \in \overline{\text{span}}\mathcal{F}$ is 
    \begin{equation}
        \gamma^*(x_1,x_2) = \frac{x_1 - \tilde{g}(x_2)}{\E[(X_1 - \tilde{g}(X_2))^2]},
    \end{equation}
    where $\tilde{g}(x_2) = \E[X_1|X_2=x_2]$. 
    In particular, if $X_1 = D$ and $X_2 = X$, which is the ATE problem under constant treatment effects constraint, the Riesz representer
    $\gamma^*_{\text{ATE, PL}}(d,x) = \frac{d - e(x)}{\E[(D - e(X))^2]}$ can be viewed as a projection of \eqref{eq:RR_ATE_L2} onto the partially linear space $\mathcal{F}$:
    \begin{equation}
        \gamma^*_{\text{ATE, PL}} = \min_{\gamma \in \mathcal{F}} \E\left[
        \left(\frac{D}{e(X)} - \frac{1-D}{1-e(X)} - \gamma\right)^2\right].
        \label{eq:RR_ATE_PL_proj}
    \end{equation}
\end{example}

If the Riesz representer $\gamma^*$ were known, a natural ``oracle'' estimator of the linear functional $\psi(f) = \E[h(Z,f)]$ is simply the sample analogue of \eqref{eq:Rproperty}:
\begin{equation}
    \hat{\psi}_{\text{oracle}} = \frac{1}{n}\sum_{i=1}^n \gamma^*(Z_i) Y_i.
    \label{eq:oracle}
\end{equation}
The estimation error of $\hat{\psi}_{\text{oracle}}$ can be decomposed as 
\begin{equation}
    \hat{\psi}_{\text{oracle}} - \psi = \frac{1}{n}\sum_{i=1}^n \gamma^*(Z_i) f(Z_i) - \E[\gamma^*(Z_i) f(Z_i)] + \frac{1}{n}\sum_{i=1}^n \gamma^*(Z_i) \varepsilon_i.
\end{equation}
Under mild regularity - e.g. a Lindeberg's condition on the given weights $\gamma^*(Z_i)$ - the oracle estimator is $\sqrt{n}$-consistent and asymptotically normally distributed.  While the ``oracle'' estimator is infeasible in practice (since $\gamma^*$ is unknown), the analysis above shows that a purely linear estimator for $\psi(f)$ can achieve desirable asymptotic properties - provided one can estimate the Riesz representer well.

\begin{remark}
    In the ATE problem, \citet{hirano2003efficient} show that the inverse propensity weighting estimator
    \begin{equation}
        \hat{\psi}_{\text{IPW}} = \frac{1}{n}\sum_{i=1}^n \left(
        \frac{D_i}{\hat{e}(X_i)}  -\frac{1-D_i}{1-\hat{e}(X_i)}\right)Y_i,
    \end{equation}
    which has a pure linear form in outcomes, achieves $\sqrt{n}$-consistency and asymptotic normality under regularity conditions. While the inverse propensity score provides a closed‐form Riesz representer for the average treatment effect, it may be hard to derive an explicit Riesz representer in general cases.
    Moreover, in finite samples the inversion step of estimated $\hat{e}(x)$ can be highly unstable whenever $e(x)$ is close to 0 or 1, as discussed in \citet{ben2021balancing}, \citet{chernozhukov2022automatic} etc. By contrast, the linear estimator we consider later in Section 2.3 never requires one to posit or invert a particular formula for the Riesz representer. Instead, it directly estimates the entire weighting function $\gamma^*(z)$ by solving a convex program.
\end{remark}

\subsection{Minimax linear estimator}
We begin in a design-based, finite-sample framework: conditional on the fixed design of $Z_1,...,Z_n$,  we analyze estimators that are linear in the outcomes, targeting the parameter $\frac{1}{n}\sum_{i=1}^n h(Z_i,f)$.
Later sections show that, under standard conditions, the same weights yield valid estimation and inference for the population functional \(\E[h(Z,f)]\) asymptotically.
Specifically, we consider estimators of the form
\begin{equation}
    \hat{\psi}_{\text{linear}} = \sum_{i=1}^n k(Z_i)Y_i,
    \label{eq:linear}
\end{equation}
where the weights $k(Z_i)$ depend only on $Z_1,...,Z_n$. (Any overall $1/n$-scaling could be absorbed into $k(\cdot)$.)
For example, if one were given the the Riesz representer $\gamma^*$, one could set $k(Z_i) = \frac{1}{n}\gamma^*(Z_i)$, in which case $\hat{\psi}_{\text{linear}}$ coincides with the infeasible oracle estimator defined by \eqref{eq:oracle} in Section 2.2.
We posit that $f$ lies in a known function class $\mathcal{F}$ which, throughout this paper, is assumed to be convex and centrosymmetric (i.e., $f\in \mathcal{F} \Rightarrow -f \in \mathcal{F}$).

Conditional on $Z_1,...,Z_n$, the maximum bias of the linear estimator $\hat{\psi}_{\text{linear}}$ over $f \in \mathcal{F}$ is given by:
\begin{equation}
    \overline{\text{bias}}_{\mathcal{F},h} := \sup_{f\in \mathcal{F}} \E_f \left(\hat{\psi}_{\text{linear}} - \frac{1}{n}\sum_{i=1}^n h(Z_i,f) \,\middle|\, Z_1,...,Z_n\right) = \sup_{f \in \mathcal{F}} \left[\sum_{i=1}^n k(Z_i)f(Z_i) - \frac{1}{n}\sum_{i=1}^n h(Z_i,f)\right]
\label{eq:maxbias_sample}
\end{equation}
and variance $\var(\hat{\psi}_{\text{linear}}) = \sum_{i=1}^n k(Z_i)^2\sigma^2(Z_i)$. By centrosymmetry of $\mathcal{F}$, the minimum bias is given by $-\overline{\text{bias}}_{\mathcal{F},h}$.
If the error terms $\varepsilon_i$ are normally distributed, a valid one-sided $100 \cdot (1-\alpha) \%$ confidence interval (CI) based on $\hat{\psi}_{\text{linear}}$ in finite samples:
\begin{equation}
    [\hat{c}, \infty), \quad \hat{c} = \hat{\psi}_{\text{linear}} - \overline{\text{bias}}_{\mathcal{F},h} - \bar{z}_{1-\alpha} \text{sd}(\hat{\psi}_{\text{linear}}),
\end{equation}
where $\bar{z}_{1-\alpha}$ is the $1-\alpha$ quantile of a standard normal distribution and $\text{sd}(\hat{\psi}_{\text{linear}}) = \sqrt{\var(\hat{\psi}_{\text{linear}})}$. Subtracting the worst-case bias ensures uniform coverage of this CI over $f \in \mathcal{F}$.

The optimal selection of the weighting function $k$ depends on the chosen evaluation criterion for estimators. A natural criterion is the maximum root-mean-squared-error (RMSE), defined as:
\begin{equation}
    R_{\mathcal{F},h} = \left[\overline{\text{bias}}_{\mathcal{F},h}^2 + \var(\hat{\psi}_{\text{linear}})\right]^{1/2}.
\label{eq:maxRMSE_sample}
\end{equation}
Under homoskedasticity, i.e., $Var(\varepsilon_i|Z_i) = \bar{\sigma}^2$ for all $i$, the worst-case squared RMSE can be written as
\begin{equation}
    \left\{\sup_{f \in \mathcal{F}} \left[\sum_{i=1}^n k(Z_i)f(Z_i) - \frac{1}{n}\sum_{i=1}^n h(Z_i,f)\right]\right\}^2 + \bar{\sigma}^2\sum_{i=1}^n k(Z_i)^2.
    \label{eq:maxRMSE_hirshberg}
\end{equation}
Moreover, under heteroskedasticity, the objective function \eqref{eq:maxRMSE_hirshberg} can also be interpreted as the sum of the squared maximum bias and an upper bound for the variance when $Var(\varepsilon_i|Z_i) \leq \bar{\sigma}^2$.

More generally, one may minimize the worst-case bias subject to a bound on variance or equivalently standard deviation:
\begin{equation}
    \min \overline{\text{bias}}_{\mathcal{F},h} \ s.t.\  \text{sd}(\hat{\psi}_{\text{linear}}) \leq c
    \label{eq:general_criterion_cons}
\end{equation}
for some standard deviation budget $c$.
For example, to obtain the weights that minimize the RMSE criterion in \eqref{eq:maxRMSE_hirshberg}, we can solve \eqref{eq:general_criterion_cons} for a fixed $c$ and then vary $c$ until the resulting bias-variance pair minimizes the RMSE.
This constrained problem \eqref{eq:general_criterion_cons}
can be written in unconstrained form as 
\begin{equation}
    \mathcal{L}_{\mathcal{F},h} = \overline{\text{bias}}_{\mathcal{F},h} + \lambda \cdot \text{sd}(\hat{\psi}_{\text{linear}}),
    \label{eq:general_criterion_uncons}
\end{equation}
where $\lambda \geq 0$ is a Lagrangian multiplier. For a given sample size, there exists a value $\lambda_n^\star$ such that the minimizer of $\mathcal{L}_{\mathcal{F},h}(k, \lambda_n^\star)$ coincides with the optimal weights that minimize
\eqref{eq:maxRMSE_hirshberg}.
Another example is one-sided CIs: if we set $\lambda = 
\frac{1}{2}(\bar{z}_{1-\alpha} + \bar{z}_\beta)$, then minimizing $\mathcal{L}_{\mathcal{F},h}$ is equivalent to minimizing the $\beta$-quantile of the worst-case excess length for one-sided CIs: 
\begin{equation}
    q_\beta (\hat{c}, \mathcal{F}) = \sup_{f \in \mathcal{F}} q_{f,\beta} \left(\frac{1}{n}\sum_{i=1}^n h(Z_i,f) - \hat{c}\right)
    \label{eq:onesidedCI_excess}
\end{equation}
where $\frac{1}{n}\sum_{i=1}^n h(Z_i,f) - \hat{c}$ is the excess length of the CI $[\hat{c}, \infty)$ and $q_{f,\beta}(\cdot)$ denotes the $\beta$-th quantile under the function $f$. As shown in \citet{armstrong2021finite}, for a one-sided CI based on $\hat{\psi}_{\text{linear}}$ under normal errors, it holds that 
\begin{equation}
    q_\beta (\hat{c}, \mathcal{F}) = 2\overline{\text{bias}}_{\mathcal{F},h} + \text{sd}(\hat{\psi}_{\text{linear}}) (\bar{z}_{1-\alpha} + \bar{z}_\beta).
\end{equation}
Therefore, both RMSE and one-sided CI objectives can be viewed as special cases of the general criterion \eqref{eq:general_criterion_uncons}.
The weights $k(\cdot)$ that minimize \eqref{eq:general_criterion_uncons} give us the minimax linear estimator.

\section{Asymptotic Results}
\subsection{Optimal weights and the Riesz representer}
In light of the bias-variance trade-off discussed in Section 2.3, one can choose the optimal weights in the linear estimator $\hat{\psi}_{\text{linear}}$ by minimizing the general criterion \eqref{eq:general_criterion_uncons}. Equivalently (and by a slight abuse of notation) we set 
$$\hat{\gamma}_i = nk(Z_i),$$ 
so that the weight vector $\{\hat{\gamma}_i\}_{i=1}^n$ minimizes
\begin{equation}
    \left\{\sup_{f \in \mathcal{F}} \left[\frac{1}{n}\sum_{i=1}^n \gamma_i f(Z_i) - \frac{1}{n}\sum_{i=1}^n h(Z_i,f)\right]\right\} + \lambda\sqrt{\frac{\bar\sigma^2}{n^2}\sum_{i=1}^n \gamma_i^2}.
    \label{eq:maxRMSE_hirshberg_gamma}
\end{equation}
Writing the weights as $\hat{\gamma}_i$ makes the connection to the Riesz representer $\gamma^*(Z_i)$ transparent; later, when we invoke standard minimax linear results (e.g. \citealp{donoho1994statistical}, \citealp{armstrong2018optimal}), it will be convenient to revert to the notation $k(Z_i) = \hat{\gamma}_i/n$.

The first term in braces in \eqref{eq:maxRMSE_hirshberg_gamma} is the sample analogue of the Riesz representer identity,
\begin{equation}
    \E[h(Z, f)] - \E[\gamma^*(Z) f(Z)] = 0, \quad \forall f\in \mathcal{F}.
\end{equation}
Hence, if the sample analogue $\frac{1}{n}\sum_{i=1}^n \hat\gamma_i f(Z_i)$ is close to $\frac{1}{n}\sum_{i=1}^n h(Z_i,f)$ for every $f \in \mathcal{F}$, then the vector $\hat{\gamma}_i$ is expected to lie near the Riesz representer values $\{\gamma^*(Z_i)\}_{i=1}^n$. The second term in \eqref{eq:maxRMSE_hirshberg_gamma}, which penalizes the $\ell_2$-norm of $\hat{\gamma}_i$, controls the variance/standard derivation of the estimator. 

To build intuition for how the optimal (minimax linear) weights approximate the Riesz representer, Figure \ref{fig:Illustration_Case1_onesim}-\ref{fig:Illustration_Case2} below show their discrepancy for estimating the ATT under two data‐generating scenarios (see Case 1 ($\mathcal{F}$ is a Donsker class) and Case 2 (non-Donsker class) in Section \ref{section:simulation} for details). Specifically,
Figure \ref{fig:Illustration_Case1_onesim} shows, for a single simulated dataset under Case 1 with sample sizes $n=100, 400,$ and $800$, the estimated minimax weights $\hat{\gamma}$ (blue line) against the true Riesz representer $\gamma^*$ (red line).\footnote{
For simplicity, Figure \ref{fig:Illustration_Case1_onesim} only plots the estimated weights and Riesz weights for observations in the control group.} under Case 1 with $n=100, 400, 800$. As $n$ grows, the optimal weights concentrate increasingly tightly around the true Riesz representer function.
\begin{figure}[H]
    \centering
    \includegraphics[width=1\linewidth]{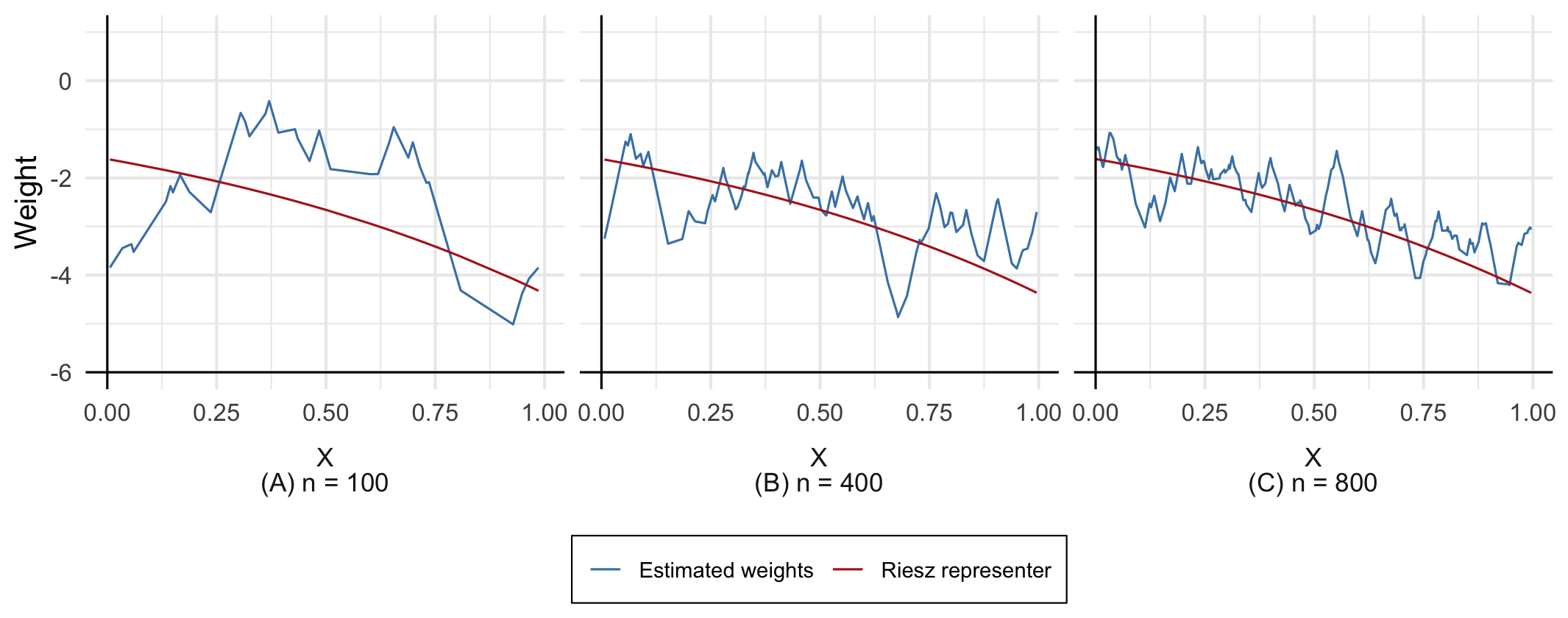}
    \caption{Estimated versus true Riesz weights under Donsker class}
    \label{fig:Illustration_Case1_onesim}
\end{figure}
Figure \ref{fig:Illustration_Case1}-\ref{fig:Illustration_Case2} plots the mean-squared discrepancy $\frac{1}{n}\sum_{i=1}^n (\hat{\gamma}_i - \gamma^*(Z_i))^2$
between estimated weights $\hat{\gamma}$ and the true Riesz representer $\gamma^*$ over 500 replications under Case 1 and Case 2.
\begin{figure}[H]
    \centering
    \includegraphics[width=0.9\linewidth]{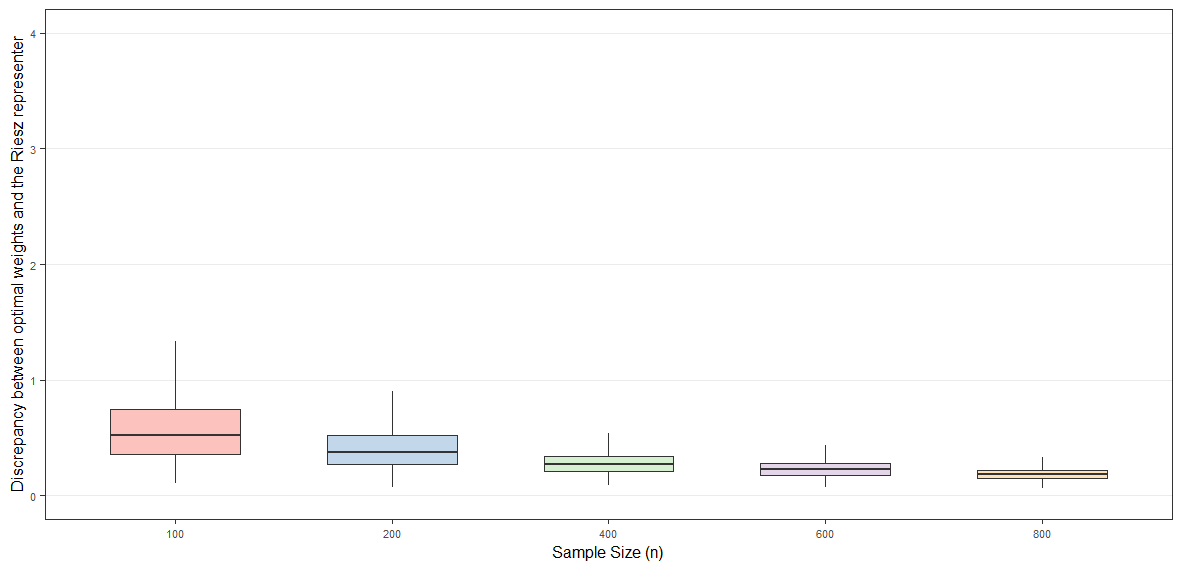}
    \caption{Discrepancy between optimal weights and Riesz representer under Donsker class}
    \label{fig:Illustration_Case1}
\end{figure}
\begin{figure}[H]
    \centering
    \includegraphics[width=0.9\linewidth]{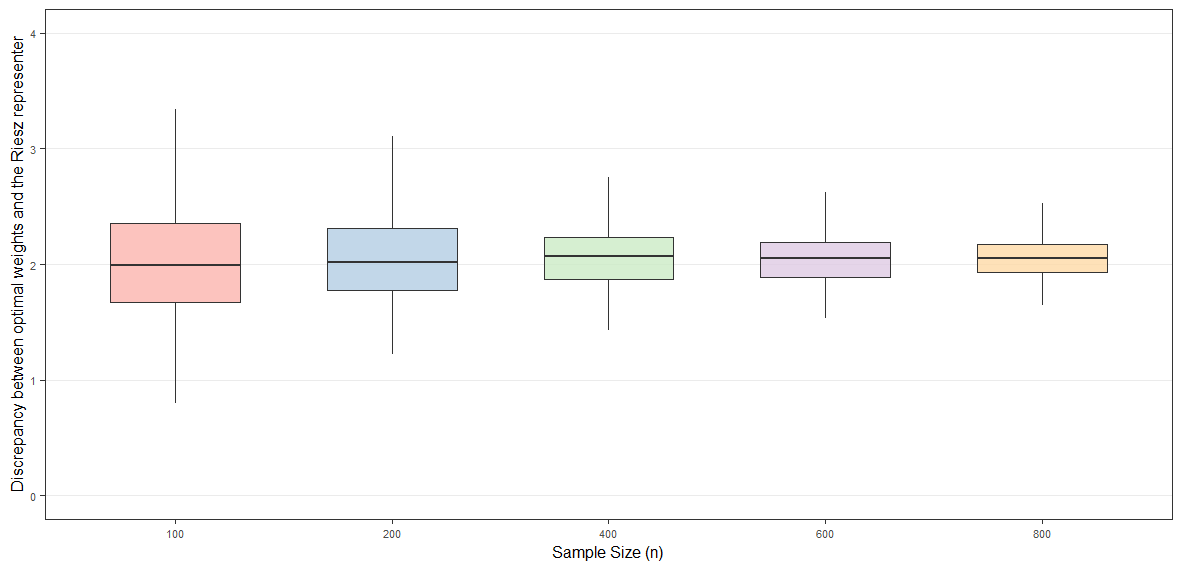}
    \caption{Discrepancy between optimal weights and Riesz representer under non-Donsker class}
    \label{fig:Illustration_Case2}
\end{figure}

Figure \ref{fig:Illustration_Case1} shows results when the candidate function class $\mathcal{F}$ is Lipschitz with a single covariate (Donsker).
At each sample size $n\in \{100,200, 400, 600, 800\}$, we plot the boxplots of the discrepancy over 500 simulations. We see that both the median and the upper-tail of the discrepancy shrink toward zero as $n$ grows. 
Figure \ref{fig:Illustration_Case2} repeats the same experiment but with a Lipschitz class on three covariates (non-Donsker). In this case, the boxplots remain bounded away from zero even as $n$ increases. Together, these plots straightforwardly illustrate that under the standard empirical-process constraints, the minimax‐linear weights closely approximate the Riesz representer as the sample size increases,
whereas relaxations beyond Donsker could break this convergence.

Motivated by graphical evidence, it is reasonable to believe that, under certain Donsker-type conditions, the following equation holds:
\begin{equation}
    \frac{1}{n}\sum_{i=1}^n (\hat{\gamma}_i - \gamma^*(Z_i))^2 \to_p 0.
    \label{eq:HWwights_RP}
\end{equation}
Formally, Theorem 1 in \citet{hirshberg2021augmented} shows that the weights by minimizing the RMSE criterion in \eqref{eq:maxRMSE_hirshberg_gamma} satisfy \eqref{eq:HWwights_RP}. In the subsequent section, I immediately show the optimal weights which minimize the general criterion \eqref{eq:general_criterion_uncons} with a fixed $\lambda$ also satisfy \eqref{eq:HWwights_RP} in Theorem \ref{thm:root-n}. Crucially, this ``closeness" yields a double-robustness property and the formal statement is given next.

\begin{theorem}
    Assume that $\gamma^*(Z)$ is the Riesz representer for a continuous linear functional $\psi(f)$ on the space $\overline{\text{span}}\mathcal{F}$ and there exists known weights $\hat{\gamma}_i$ such that $\frac{1}{n}\sum_{i=1}^n (\hat{\gamma}_i - \gamma^*(Z_i))^2 \to_p 0$, then as long as $f$ satisfies $\psi(f) = \E[\gamma^*(Z)f(Z)]$, then a linear estimator $\hat{\psi}_{\text{linear}} = \frac{1}{n}\sum_{i=1}^n \hat{\gamma}_i Y_i$ satisfies 
    \begin{equation}
        \hat{\psi}_{\text{linear}} - \psi(f) \to_p 0.
    \end{equation}
    \label{thm:consistency}
\end{theorem}

Theorem \ref{thm:consistency} highlights the robustness of the minimax linear estimator to the model misspecification: consistency obtains whenever the Riesz representer $\gamma^*$ of the functional $\E[h(Z, g)]$ on $\overline{\text{span}}\mathcal{F}$ satisfies $\E[h(Z,f)] = \E[\gamma^*(Z)f(Z)]$, even if the regression function $f$ itself does not lie inside $\overline{\text{span}}\mathcal{F}$.
In the ATE setting, the Riesz representer of $\E[f(1,X) - f(0,X)]$ on $L^2(P_Z)$ space is the usual IPW function, $\gamma^*(d,x) = \frac{d}{e(x)} - \frac{1-d}{1-e(x)}$. Theorem \ref{thm:consistency} shows the \textit{double robustness} of the minimax linear estimator: if either the outcome regression function $f$ lies in $\overline{\text{span}}\mathcal{F}$ or the IPW function belongs to $\overline{\text{span}}\mathcal{F}$, the minimax linear estimator remains consistent for $\psi(f)$.

Theorem \ref{thm:consistency} assures us that the linear estimator with weights converging to the Riesz representer function is consistent under the double-robustness regime described above. For empirical work, however, consistency alone is not enough—we also need a tractable sampling distribution in order to build confidence intervals and conduct hypothesis tests. The next subsection therefore strengthens the result.

\subsection{Root-$n$ consistency for homoskedasticity-based weights}
\label{section:homo}
Though the true variance structure, $\sigma^2(Z_i) \leq  \bar{\sigma}^2$ for all $Z_i \in \mathcal{Z}$, may vary with $Z_i$, for weight construction, we treat it as homoskedastic with variance bound $\bar{\sigma}^2$. Under this simplifying assumption, we have 
$$\var\left(\sum_{i=1}^n k(Z_i)Y_i|Z_i\right) = \bar{\sigma}^2\sum_{i=1}^n k(Z_i)^2.$$
We then choose weights $k(Z_i)$ that minimize the $\beta$-quantile of excess of an one-sided CI in \eqref{eq:onesidedCI_excess}, which is equivalent to solving the minimization problem given by \eqref{eq:general_criterion_uncons} with $\lambda = \frac{1}{2}(\bar{z}_{\beta}+\bar{z}_{1-\alpha})$.
We will first show the $\sqrt{n}$-consistency for the minimax linear estimator based on the criterion of interval length and then discuss about the weights based on minimizing the worst-case RMSE.

For notational abuse, we define $\delta = \bar\sigma \lambda  = \bar\sigma (\bar{z}_{\beta}+\bar{z}_{1-\alpha})/2$ and let 
\begin{equation}
    \hat{\psi}_{\text{linear}, \delta} = \sum_{i=1}^n k_\delta(Z_i)Y_i
    \label{eq:minimaxlinearestimator_fixeddelta}
\end{equation}
where $k_\delta(Z_i)$ is the solution to the following optimization problem 
\begin{equation}
    \min_{\{k(Z_i)\}_{i=1}^n} \sup_{f\in \mathcal{F}} \left\{\left[\sum_{i=1}^n k(Z_i)f(Z_i) - \frac{1}{n}\sum_{i=1}^n h(Z_i,f)\right]\right\} + \delta \sqrt{\sum_{i=1}^n k(Z_i)^2}
    \label{eq:minopt_delta_homo}
\end{equation}
\cite{armstrong2018optimal} and \cite{armstrong2021finite} showed that a one-sided confidence interval (CI) based on $\hat{\psi}_{\text{linear}, \delta}$ is minimax optimal for $\beta$-quantile of excess length among all $1-\alpha$ CIs
for $\psi$ where $\beta = \Phi(\delta/\bar\sigma -\bar{z}_{1-\alpha})$ in finite samples with normal errors. 
We now drop the Gaussianity assumption and study the estimator $\hat{\psi}_{\text{linear},\delta}$ as $n \to \infty$.

\begin{theorem}
    Let $\mathcal{F}$ be a pre-specified convex and centrosymmetric function class and $\gamma^*$ is the Riesz representer for a continuous linear functional $\psi(f) = \E[h(Z,f)]$ on $\overline{\text{span}}\mathcal{F}$.
    Assume that $\mathcal{F}$ is uniformly bounded, $\mathcal{F}$, $\gamma^*\mathcal{F}$, and $h(\cdot,\mathcal{F})$ are Donsker, and $h(Z,\cdot)$ is pointwise bounded and mean-square equicontinuous on $\mathcal{F}$ in the sense that 
    $\sup_{f \in \mathcal{F}} |h(z,f)| < \infty$ for each $z \in \mathcal{Z}$ and $\lim_{r \to 0} \sup_{f\in \mathcal{F}_r} ||h(\cdot,f)||_{L^2(P)} = 0$.
    If the Riesz representer $\gamma^* \in a\mathcal{F}$ for some constant $a$, for any given constant $\delta$, then the weights that solve 
    \eqref{eq:minopt_delta_homo} satisfy
        \begin{equation}
        \frac{1}{n}\sum_{i=1}^n [nk_\delta(Z_i) - \gamma^*(Z_i)]^2 \to_p 0.
            \label{eq:fixeddetla_weightapprox}
        \end{equation}
    and the worst-case bias defined as
        \begin{equation}
            \overline{\text{bias}}_{\mathcal{F},h} = \sup_{f\in \mathcal{F}} \left\{\left[\sum_{i=1}^n k_\delta(Z_i)f(Z_i) - \frac{1}{n}\sum_{i=1}^n h(Z_i,f)\right]\right\}
            \label{eq:worstbias_a=1}
        \end{equation}
        is of order $o_p(1/\sqrt{n})$.
        \label{thm:root-n}
\end{theorem}
The Donsker condition on $\mathcal{F}$ is satisfied by familiar smoothness classes. A canonical example is the H\"{o}lder class $H^\alpha(Z)$ on a bounded, convex domain $Z\subset\mathbb{R}^p$. Informally, $H^\alpha$ collects functions whose partial derivatives up to order $\lfloor\alpha\rfloor$ (the greatest integer strictly smaller than $\alpha$) exist and whose $\lfloor\alpha\rfloor$th derivatives obey a Lipschitz‑type bound; see Example F.1 in \ref{app:donsker} for the formal definition. The unit ball $H^\alpha(Z)_1$ is $P$‑Donsker whenever $\alpha>p/2$. 
As a special case, the Lipschitz class is simply $H^{1}(Z)$; hence in one dimension ($p=1$) this class is Donsker (since $1>1/2$). Other examples covered by our condition include unit balls of bounded‑kernel reproducing‑kernel Hilbert spaces (e.g., \citealp{hirshberg2019minimax}); see Example F.3 in \ref{app:donsker} for details.

The Riesz representation theorem ensures the uniqueness and existence of the Riesz representer function $\gamma^* \in \overline{\text{span}}\mathcal{F}$ for the linear functional $\psi(f)$ defined on the space $\overline{\text{span}}\mathcal{F}$.
But in Theorem \ref{thm:root-n} we assume a stronger condition that there exists a constant $a$ such that $\gamma^* \in a\mathcal{F}$. Equivalently, $\gamma^*$ does not lie on the ``boundary'' of $\overline{\text{span}}\mathcal{F}$. Boundary elements of $\overline{\text{span}}\mathcal{F}$ need not share the same properties as functions in $\mathcal{F}$ or $a\mathcal{F}$, and so may be considered ``irregular'' from the perspective of our imposed smoothness/shape constraints on the convex class $\mathcal{F}$.
In other words, the condition that $\gamma^* \in a\mathcal{F}$ rules out those irregular functions to ensure the good performance of the weights and estimators.
Specifically, Theorem \ref{thm:root-n} shows the worst-case bias converges to 0 faster than root-$n$ and the weights are close to the Riesz representer $\gamma^*$ in the sense of \eqref{eq:fixeddetla_weightapprox}. These results help us to establish $\sqrt{n}$-consistency and asymptotically normality in the following corollary.
\begin{corollary}
    Under the conditions in Theorem \ref{thm:root-n} and $\sigma^2(z) \leq \bar\sigma^2$ for all $z \in \mathcal{Z}$, if the regression function $f \in a\mathcal{F}$ for some constant $a$, then the linear estimator $\hat{\psi}_{\text{linear},\delta}$ defined in \eqref{eq:minimaxlinearestimator_fixeddelta}
		satisfies 
		\begin{equation}
		\sqrt{n}\left(\hat{\psi}_{\text{linear},\delta} - \psi(f)\right) \to_d \mathcal{N}(0, V)
		\label{eq:asydist_homo}
		\end{equation}
		where 
        \begin{equation}
            V=\E(\gamma^*(Z_i)^2\sigma^2(Z_i)) + \var(h(Z_i,f)).
            \label{eq:asyvar}
        \end{equation}
        In addition, an estimator satisfying \eqref{eq:asydist_homo} with \eqref{eq:asyvar} is asymptotically efficient on the model space $S$ if $S = \overline{\text{span}}\mathcal{F}$ and $\gamma^*(\cdot) \sigma^2(\cdot)  \in \overline{\text{span}}\mathcal{F}$.
        \label{coro:asynormal}
\end{corollary}

\begin{remark}
    Theorem \ref{thm:root-n} and Corollary \ref{coro:asynormal} ensures $\sqrt{n}$-consistency of $\hat{\psi}_{\text{linear},\delta}$ based on the criterion of $\beta$-quantile of excess length of CIs. Instead, we can choose the criterion of worst-case RMSE and denote the linear estimator by $\hat\psi_{\text{linear}, \text{rmse}} = \sum_{i=1}^n k_{\text{rmse}}(Z_i) Y_i$ with the weights $k_{\text{rmse}}(Z_i)$
minimaxes the RMSE. When $n$ goes to infinity, we conjecture, though without strict proof, that the minimax linear estimator based on worst-case RMSE criterion should enjoy the same properties as $\hat\psi_{\text{linear}, \delta}$.
\end{remark}

Theorem \ref{thm:root-n} and Corollary \ref{coro:asynormal} imply that, for any fixed $\delta>0$, the worst-case bias of $\hat\psi_{\text{linear}, \delta}$ is $o_p(n^{-1/2})$, which is negligible compared to the standard deviation. In particular, for finite-sample minimax CIs, we usually need to compute a nontrivial worst-case bias correction (e.g. via a Lipschitz constant $C$ if we let $\mathcal{F}$ be a Lipschitz function class) to guarantee uniform coverage. But under the conditions of Theorem \ref{thm:root-n}, as long as $n$ is sufficiently large, the bias-aware term becomes vanishingly small compared to the variance term. As a result, for asymptotic intervals one can safely ignore the bias bound adjustment and a standard two-sided ``$\pm \bar{z}_{1-\alpha/2}\hat{se}$'' interval will attain correct coverage asymptotically, where $\hat{se}$ is a consistent estimator for the asymptotic standard deviation. More importantly, because the bias vanishes so quickly, a conservative choice of the Lipschitz constant $C$ drops out of the leading term. In finite sample, such over-bounding will inflate the interval, but Theorem \ref{thm:root-n} ensures that any such conservatism becomes harmless to first order if the sample is large enough.

The conclusions of Theorem \ref{thm:root-n} and Corollary \ref{coro:asynormal} hold for any fixed $\delta>0$. In other words, the particular choice of $\delta$ (equivalently, the target quantile $\beta$ or minimum power level $\beta$ of your one-sided minimax test) does not affect $\sqrt{n}$-rate or asymptotic variance. In practice, one may pick $\delta$ to target a desired minimum power level for a one‑sided minimax test, e.g. $$\beta = \Phi(\delta/\bar\sigma-\bar{z}_{1-\alpha}).$$ A rule of thumb is to choose $\delta/\bar\sigma$ around 3 to 5, which corresponds to the minimum power level exceeding $90\%$.

Finally, the limiting variance \eqref{eq:asyvar} in Corollary \ref{coro:asynormal}
coincides with the semiparametric efficiency bound in literature (e.g., \citet{hirshberg2021augmented}) if the variance function is relatively simple compared to the Riesz representer so that $\gamma^*(Z)\sigma^2(Z) \in \overline{\text{span}}\mathcal{F}$. The augmented minimax linear estimator 
\begin{equation}
    \hat{\psi}_{\text{AML}} = \frac{1}{n}\sum_{i=1}^n h(Z_i, \hat{f}) + \frac{1}{n}\sum_{i=1}^n \hat{\gamma}_i (Y_i -f(Z_i))
\end{equation}
which uses a regression augmentation $\hat{f}$ in addition to weighting, is shown by \citet{hirshberg2021augmented} to be efficient under conditions similar to those in Corollary \ref{coro:asynormal}. Our corollary shows that, under our regularity conditions, augmentation is not first‑order necessary because the pure minimax linear estimator is already efficient.
\begin{remark}
    The $\sqrt{n}$-consistency and asymptotic efficiency of the pure minimax linear estimator rest on the convexity and Donsker-type conditions imposed on $\mathcal F$.  If either convexity or the Donsker property fails, those guarantees no longer hold. In such settings, one might instead consider augmented weighting estimators such as in \citet{hirshberg2021augmented} and \citet{chernozhukov2022automatic}. 
    \citet{chernozhukov2022automatic} establish desirable properties of Auto-DML estimators in high-dimensional settings under sparsity structure.
\end{remark}

\setcounter{example}{0}
\begin{example}
    (ATE, cont'd) The asymptotic variance of the minimax linear estimator with the weights \eqref{eq:minimaxlinearestimator_fixeddelta} for estimating ATE can be expressed as:
    \begin{equation*}
        \begin{aligned}
            &\E[\gamma^*(Z_i)^2\sigma^2(Z_i)] + \text{Var}(h(Z_i,f)) \\
            &= \E\left\{\E\left[\frac{D_i^2}{e^2(X_i)}+\frac{(1-D_i)^2}{(1-e(X_i))^2} - \frac{2D_i(1-D_i)}{e(X_i)(1-e(X_i))}|X_i,D_i\right]\sigma^2(D_i,X_i)\right\} + \E\left[ (\beta(X_i)- \beta)^2\right]\\    
            &= \E\left\{\E\left[\frac{D_i}{e^2(X_i)}+\frac{(1-D_i)}{(1-e(X_i))^2}|X_i,D_i\right]\sigma^2(D_i,X_i)\right\} + \E\left[ (\beta(X_i)- \beta)^2\right] \\
            &= \E\left[\frac{\sigma^2(1,X_i)}{e(X_i)}+\frac{\sigma^2(0,X_i)}{1-e(X_i)}+(\beta(X_i)- \beta)^2\right].
        \end{aligned}
    \end{equation*}
    where $\beta(X_i):= \E[Y_i(1)-Y_i(0)|X_i]$ and $\beta: = \E[Y_i(1)-Y_i(0)] = \E[\beta(X_i)]$.
    This coincides with the semi-parametric efficiency bound in \cite{hahn1998role}, and thus aligns with our Corollary \ref{coro:asynormal} that the variance $V$ indeed attains the efficiency bound as $\gamma^*(Z)\sigma^2(Z) \in L^2(P_Z)$.
\end{example}

\begin{example}
    (Partially Linear Model Coefficient, cont'd) If the true variance function is constant, i.e., $\sigma^2(z) = \bar{\sigma}^2$ for all $z \in \mathcal{Z}$, then $\gamma^*(z)\sigma^2(z) = \bar{\sigma}^2 \gamma^*(z)$ belongs to the partially linear function class $\mathcal{F}$. In this case, the minimax linear estimator therefore achieves the efficiency bound. However, once \(\sigma^2(d,x)\) depends nontrivially on \(d\) and \(x\), the product \(\gamma^*(d,x)\sigma^2(d,x)\) will generally fail to be a partially linear model, i.e., \(\gamma^*(d,x)\sigma^2(d,x) \not \in \overline{\text{span}}\mathcal{F}\). 
    So the minimax linear estimator under this heteroskedasticity case is not efficient. 
    Indeed, the semiparametric efficiency bound for partially linear model under heteroskedasticity (e.g. \citealp{chamberlain1992efficiency}, \citealp{ma2006efficient}) can be written as
    \begin{equation}
        \begin{aligned}
            \frac{1}{\E\left\{\E(D_i/\sigma^2(D_i,X_i)|X_i)\left[1-\frac{\E(D_i/\sigma^2(D_i,X_i)|X_i)}{\E(1/\sigma^2(D_i,X_i)|X_i)}\right]\right\}}
        \end{aligned}
        \label{eq:PLM_SEB}
    \end{equation}
    which is smaller than $\E(\gamma^{*}(D_i,X_i)^2\sigma^2(D_i,X_i))$. 
    Intuitively, for a outcome function characterized by a partially linear model (and thus a constant treatment effect) but heteroskedasticity, an effective strategy to estimate the constant effect should assign greater weights to units with lower variances.
    But the weighting function $\gamma^*(d,x) = \frac{d-e(x)}{\E[(D-e(X))^2]}$ fails to achieve that because it does not account for variance differences among observations.
\end{example}

\subsection{Consistency of the standard error}
In this section, we consider the estimation of the variance of the minimax linear estimator with a fixed $\delta>0$, $\hat\psi_{\text{linear},\delta} = \sum_{i=1}^n k_\delta(Z_i)Y_i$. Let $\hat{f}$ be some preliminary estimate of the regression function $f \in \mathcal{F}$ and $\hat{\varepsilon}_i = Y_i - \hat{f}(Z_i)$ be the plug-in residuals. 
Define the variance estimator 
\begin{equation}
    \text{se}^2 = \sum_{i=1}^n k_{\delta}(Z_i)^2 \hat{\varepsilon}_i^2 + \frac{1}{n^2}\sum_{i=1}^n h(Z_i,\hat{f})^2 - \frac{1}{n}\hat{\psi}_{\text{linear},\delta}^2,
    \label{eq:se}
\end{equation}
and set 
\begin{equation}
    n \cdot \text{se}^2 = n \cdot \left[\sum_{i=1}^n k_{\delta}(Z_i)^2 \hat{\varepsilon}_i^2 + \frac{1}{n^2}\sum_{i=1}^n h(Z_i,\hat{f})^2 - \frac{1}{n}\hat{\psi}_{\text{linear},\delta}^2\right].
    \label{eq:se*n}
\end{equation}

\begin{theorem}
    Suppose that (i) $\frac{1}{n}\sum_{i=1}^n (\hat{\gamma}_\delta - \gamma^*)^2 \to_p 0$ where $\hat{\gamma}_\delta = nk_\delta$, 
(ii) $\hat{\psi}_{\text{linear},\delta} \to_p \psi(f)$, (iii) $\max_{1\leq i\leq n} |\hat{f}(Z_i)-f(Z_i)| \to_p 0$,
(iv) the function $f^*_\delta$ (that solves the modulus of continuity problem \eqref{eq:optmfstar} in Appendix B) satisfies the uniform continuity condition
\begin{equation}
    \sup_n \sup_{z,z': d_Z(z,z') \leq \eta} |f^*_\delta(z)-f^*_\delta(z')| \leq \bar{g}(\eta)
\end{equation}
where $\lim_{\eta \to 0} \bar{g}(\eta)=0$ and $d_Z$ is a metric on $\mathcal{Z}$, (v) for all $\eta>0$, $\min_{1\leq i\leq n} \sum_{j=1}^n I(d_Z(z_j,z_i)\leq \eta) \to \infty$,
(vi) there exists a constant $\tilde{C}$ such that $1/\tilde{C} \leq \sigma^2(Z_i) \leq \tilde{C}$ and $\E|\varepsilon_i|^{2+\eta} \leq \tilde{C}$, 
(vii) $\sum_{i=1}^n (h(Z_i,\hat{f}) - h(Z_i,f))^2/n \to_p 0$. Then it holds that 
\begin{equation}
    n \cdot \text{se}^2 \to_p \E(\gamma^*(Z_i)^2\sigma^2(Z_i))+\var (h(Z_i,f)).
\end{equation}
    \label{thm:se}
\end{theorem}

\begin{remark}
    If we are interested in ATE, i.e., $h(Z_i,f) = f(1,X_i) - f(0,X_i)$, a slightly stricter version of condition (iii) - $\max_{1\leq i\leq n, d\in\{0,1\}} |\hat{f}(d, X_i)-f(d, X_i)| \to_p 0$ - is sufficient for condition (vii). 
Specifically, we have
\begin{equation*}
    \begin{aligned}
        \frac{1}{n}\sum_{i=1}^n(h(Z_i,\hat{f}) - h(Z_i,f))^2 &\leq 2\cdot \frac{1}{n}\sum_{i=1}^n (\hat{f}(1, X_i) - f(1, X_i))^2 + (\hat{f}(0, X_i) - f(0, X_i))^2 \\
        & \leq 2\max_{1\leq i\leq n, d\in\{0,1\}} |\hat{f}(d, X_i)-f(d, X_i)|^2 \to_p 0.
    \end{aligned}
\end{equation*}
\end{remark}

\subsection{Root-$n$ consistency for heteroskedasticity-based weights}
\label{section:hetero}
In Section \ref{section:homo}, we showed that, even when the only information about the variance function of error terms is the upper bound, $\sigma^2(z) \leq \bar\sigma^2$, we can use homoskedasticity at level $\bar\sigma^2$ to compute minimax weights and obtain $\sqrt{n}$-consistency and asymptotic normality. 
In fact,  any positive constant $\bar\sigma^2$ works - once $n$ is large, the variance bound affects only the constraint that prevents the weights from exploding; it does not drive the first-order behavior. Intuitively, it is the worst-case bias constraint that pins down the Riesz representer (and hence the shape of the weights), while the variance bound merely caps their magnitude. 
In this section, we move to an idealized scenario where the full variance function $\sigma^2(z)$ is known. Although one rarely has this information in practice, analyzing this benchmark helps to sharpen our efficiency results and reveals what happens if we can fully adapt to heteroskedasticity.

Divide the regression equation \eqref{eq:yequation} by the conditional variance function $\sigma(Z_i)$ on both sides, yielding 
\begin{equation}
    Y_i^\diamond = f^\diamond(Z_i) + \varepsilon_i^\diamond
\end{equation}
where $ Y_i^\diamond:=Y_i/\sigma(Z_i)$, $f^\diamond(\cdot) := f(\cdot)/\sigma(\cdot)$, and $\varepsilon_i^\diamond:=\varepsilon_i/\sigma(Z_i)$.
Now $\var(\varepsilon_i^\diamond|Z_i) = 1$, so we are back in the homoskedastic setting of Section \ref{section:homo} after transformation. Define 
$$h^\diamond(Z, f^\diamond) := h(Z,f), \quad 
\psi^\diamond(f^\diamond) := \E(h^\diamond(Z, f^\diamond)),$$
which are linear in $f^\diamond$.
Because $\mathcal{F}$ is convex and centrosymmetric, the rescaled class
\begin{equation}
    \mathcal{F}^\diamond:= \left\{\frac{f(\cdot)}{\sigma(\cdot)}: f\in \mathcal{F}\right\},
\end{equation}
inherits the same properties. Hence, if $f \in \mathcal{F}$ then $f^\diamond \in \mathcal{F}^\diamond$. We, again, consider the optimal weights by minimizing $\beta$-quantile of excess length of one-sided intervals, denoted as $k^{\diamond}_{\delta}(Z_i)$.

Let $\gamma^{\diamond *}$ denote the Riesz representer of $\E[h^\diamond(Z, f^\diamond)]$ on the function class $\overline{\text{span}}\mathcal{F}^\diamond$. Because
\begin{equation}
    \E\left[h^\diamond(Z, f^\diamond)\right] = \E[h(Z,f)] = \E[\gamma^*(Z) f(Z)] = \E\left[\left(\gamma^{*}(Z)\sigma(Z)\right)f^\diamond(Z)\right],
\end{equation}
uniqueness of the Riesz representer on $\overline{\text{span}}\mathcal{F}^\diamond$ implies that $\gamma^{\diamond *}(Z)$ is exactly the projection of $\gamma^{*}(Z)\sigma(Z)$ onto $\overline{\text{span}}\mathcal{F}^\diamond$, i.e.,
\begin{equation}
    \gamma^{\diamond *}(Z) = \min_{g \in \overline{\text{span}}\mathcal{F}^\diamond} \E[(g(Z) - \gamma^{*}(Z)\sigma(Z))^2]
    \label{eq:RR_proj_hetero}
\end{equation}
Assuming $0<c_1\leq \sigma^2(\cdot) \leq c_2 < \infty$, the empirical process type conditions for $\mathcal{F}$, $\gamma^*$ and $h(\cdot,\mathcal{F})$ in Theorem \ref{thm:root-n} (or Corollary \ref{coro:asynormal})  continue to hold for $\mathcal{F}^\diamond$, $\gamma^{\diamond *}$, and $h^\diamond(\cdot,\mathcal{F}^\diamond)$ because multiplying by the bounded function $\sigma^{-1}(z)$ preserves Donsker-type and continuity properties.
Therefore, it is straightforward to show that the maximum bias of the minimax linear estimator $\hat{\psi}_{\text{linear},\delta}^{\diamond}$ using the heteroskedasticity information converges to 0 at a rate faster than $n^{-1/2}$ and thus the estimator achieves $\sqrt{n}$-consistency and asymptotic normality. Moreover, the additional condition for the semiparametric efficiency bound in Corollary \ref{coro:asynormal} is $\sigma^2(\cdot) \gamma^*(\cdot) \in \overline{\text{span}}\mathcal{F}$. Equivalently, we will expect semiparametric efficiency bound to achieve when $1 \cdot \gamma^{\diamond *}(\cdot) \in \overline{\text{span}}\mathcal{F}^\diamond$ holds. But this latter condition holds automatically since by construction $\gamma^{\diamond *} \in \overline{\text{span}}\mathcal{F}^\diamond$. This implies that the minimax linear estimator $\hat{\psi}_{\text{linear},\delta}^{\diamond}$ will always achieve efficiency if we know the true variance function. 

For notational convenience under heteroskedasticity case, we define one more notation $\gamma^\#(Z) := \gamma^{\diamond *}(Z)/\sigma(Z)$. Then the requirement that $\gamma^{\diamond *} \in a\mathcal{F}^\diamond$ is exactly equivalent to that $\gamma^\#(Z) \in a \mathcal{F}$. The Riesz property of $\gamma^{\diamond *}(Z)$ translates directly into that $\gamma^\#(Z)$ satisfies $\E[h(Z,f)] = \E[\gamma^\#(Z)f(Z)]$.

\begin{theorem}
    Assume that $f \in a\mathcal{F}$ and there exists a function $\gamma^\#$ satisfying (i) for any $f\in \mathcal{F}$, $\psi(f) = \E[\gamma^\#(Z) f(Z)]$ and (ii) $\gamma^\#(\cdot)\sigma^2(\cdot) \in a\mathcal{F}$ for some constant $a$. 
    Suppose that $0<c_1 \leq \sigma^2(z) \leq c_2 <\infty$ for any $z \in \mathcal{Z}$, under the conditions in Theorem \ref{thm:root-n}, for any given constant $\delta>0$, the linear estimator $$\hat{\psi}_{\text{linear},\delta}^{\diamond} = \sum_{i=1}^n k^{\diamond}_{\delta}(Z_i) Y_i$$ satisfies
	\begin{equation}
		\sqrt{n}\left(\hat{\psi}_{\text{linear},\delta}^{\diamond} - \psi(f)\right) \to_d \mathcal{N}(0, V^\#)
		\label{eq:asydist_hetero}
		\end{equation}
		where $V^\#=\E(\gamma^\#(Z_i)^2\sigma^2(Z_i)) + \var(h(Z_i,f))$. In addition, an estimator satisfying \eqref{eq:asydist_hetero} is asymptotically efficient on the model class $S$ if $S = \overline{\text{span}}\mathcal{F}$ and $\gamma^\#(\cdot)\sigma^2(\cdot) \in \overline{\text{span}}\mathcal{F}$.
		\label{thm:heteroknownV}
\end{theorem}

It is important to note that the Riesz representer $\gamma^*$ is unique within $ \overline{\text{span}} \mathcal{F}$. However, if $\overline{\text{span}}\mathcal{F}$ is a strict subspace of $L^2(P_Z)$, 
then its orthogonal complement $(\overline{\text{span}}\mathcal{F})^\bot$ is nontrivial. In that case, any function of the form
\begin{equation}
    \gamma^\# = \gamma^* + u, \quad u \in (\overline{\text{span}}\mathcal{F})^\bot,
\end{equation}
still satisfies
$\int \gamma^\#(z)f(z) dP(z) = \psi(f)$ for all $f \in \overline{\text{span}}\mathcal{F}$.
We call this function, $\gamma^\#$, that belongs to $L^2(P_Z)$ but may not belong to $\overline{\text{span}}\mathcal{F}$ a (possibly nonunique) \textit{representer} of $\psi$ on the function class $\overline{\text{span}}\mathcal{F}$. Therefore, Theorem \ref{thm:heteroknownV} has the following key takeaway: 
once we allow for the true variance function in forming the minimax weights, the resulting optimal weights need no longer track the Riesz representer $\gamma^*$ of $\psi(f)$ on $\mathcal{F}$. Instead, they are close to a representer $\gamma^\#$ which should satisfy two conditions: $\E(\gamma^\# f) = \psi(f)$ for all $f\in \mathcal{F}$ and $\sigma^2(\cdot) \gamma^{\#}(\cdot) \in a\mathcal{F}$ for some constant $a$.

\setcounter{example}{1}
\begin{example}
    (Partially Linear Model Coefficient, cont'd) When we rescale by the true variance, the Riesz representer of $\psi^\diamond(f^\diamond) = \beta$ for $f^\diamond \in \mathcal{F}^\diamond$ can shown to be 
        \begin{equation}
            \gamma^{\diamond *}(D_i,X_i) = \bar{C}\cdot\frac{1}{\sigma(D_i,X_i)}\left[D_i-\frac{\E(D_i/\sigma^2(D_i,X_i)|X_i)}{\E(1/\sigma^2(D_i,X_i)|X_i)}\right].
        \end{equation}
        where 
        \begin{equation}
        \bar{C}=\frac{1}{\E\left\{\E(D/\sigma^2(D,X)|X)\left[1-\E(D/\sigma^2(D,X)|X)/\E(1/\sigma^2(D,X)|X)\right]\right\}}.
    \end{equation}
        Indeed, for any $f^\diamond \in \mathcal{F}^\diamond$ with $f^\diamond (d,x) = \frac{d\beta}{\sigma(d,x)} + \frac{g(x)}{\sigma(d,x)}$, one checks by iterated conditioning that 
        \begin{equation*}
        \begin{aligned}
            \E(\gamma^{\diamond *}(D_i,X_i) f^\diamond(D_i,X_i)) &= \bar{C}\beta \cdot \E\left\{\frac{D_i}{\sigma^2(D_i,X_i)}\left[1-\frac{\E(D_i/\sigma^2(D_i,X_i)|X_i)}{\E(1/\sigma^2(D_i,X_i)|X_i)}\right]\right\}\\
            &+\bar{C}\cdot \E\left\{g(X_i)\left[\frac{0}{\E(1/\sigma^2(D_i,X_i)|X_i)}\right]\right\} \\
            &=\bar{C}\beta \cdot \frac{1}{\bar{C}}+0 = \beta,
        \end{aligned}
    \end{equation*}
    so it reproduces the target functional. If we then set
         $\gamma^{\#}(D_i,X_i):= \gamma^{\diamond *}(D_i,X_i)/\sigma(D_i,X_i)$, one verifies immediately that 
         $$\gamma^{\#}(D_i,X_i)= \bar{C}\cdot\frac{1}{\sigma^2(D_i,X_i)}\left[D_i-\frac{\E(D_i/\sigma^2(D_i,X_i)|X_i)}{\E(1/\sigma^2(D_i,X_i)|X_i)}\right].$$ is a representer but not the Riesz representer of $\psi(f)$ for $f \in \mathcal{F}$ because the function $\gamma^\#$ is not linear in $d$ due to $\sigma^2(d,x)$. Since it satisfies conditions (i) and (ii) in Theorem \ref{thm:heteroknownV}, the variance $V^\#$ of the minimax linear estimator $\hat{\psi}_{\text{linear},\delta}^{\diamond}$ should attain the efficiency bound. This indeed aligns with the bound \eqref{eq:PLM_SEB} known from the partially linear literature. Intuitively, by building $\sigma(\cdot)$ into the weights, $\gamma^\#$ down-weights observations with higher noise and thus attains the smallest possible variance.
\end{example}

Together, Sections \ref{section:homo} and \ref{section:hetero} show that if \(\sigma^2(Z_i)\gamma^*(Z_i) \in a\mathcal{F}\) (i.e., the variance function is relatively simple compared to the Riesz representer), then the minimax linear estimator, even with an incorrectly specified variance (assuming homoskedasticity), can still attain the semiparametric efficiency bound. 
    Otherwise, the variance function should be correctly specified to achieve the semiparametric efficiency bound.
    The minimax linear estimator with a correctly specified variance function automatically searches for the efficient weight function $\gamma^\#$ if it exists. Revisiting the example of partially linear model (constant treatment effects) under heteroskedasticity: the asymptotic variance of $\hat{\psi}_{\text{linear}, \delta}^\diamond$ by \eqref{eq:asydist_hetero} achieves the semiparametric efficiency bound
    while $\hat{\psi}_{\text{linear}, \delta}$ by \eqref{eq:asydist_homo} does not.

\section{Simulation Study}
\label{section:simulation}
We simulate from the following two-stage data-generating process 
\begin{equation}
    \begin{aligned}
        Y_i(1) &= f(1, X_i) + \varepsilon_{1,i}, \\
    Y_i(0) &= f(0,X_i) + \varepsilon_{0,i},\\
    D_i &\sim \text{Bernoulli}(f_D(X_i)), \\
    Y_i &= D_iY_i(1) + (1-D_i) Y_i(0).
    \end{aligned}
\end{equation}
The covariates $X_{i,j}$ are generated from a uniform distribution $U[0,1]$ across $i=1,...,n$ and $j = 1,...,p$ and the error terms $\varepsilon_{1,i}$, $\varepsilon_{0,i}$ are generated from a normal distribution with mean 0 and standard deviation $0.5$. We let the treatment probability be 
$$f_D(X_i) = \frac{1}{1+\exp\{X_i^\top \alpha\}},$$ 
where $\alpha = (-1,0,...,0)^\top$, and study three cases for the conditional regression function of $Y_i(1)$ and $Y_i(0)$:
\begin{itemize}
    \item Case 1: $f(d, x) = \sin(x^\top \beta_d)$, where $p = 1$, $\beta_0 = 1$, $\beta_1 = 2$.

    \item Case 2: $f(d, x) = \sin(x^\top \beta_d)$, where $p = 3$, $\beta_0 = (1,1,1)^\top$, $\beta_1 = (0.5,1.5,2)$.

    \item Case 3: $f(1, x) = \sin(1/(x+0.05))$ and $f(0, x) = \cos(1/(x+0.01))$, where $p = 1$.
\end{itemize}
Moreover, we keep the conditional regression function as in Case 1: $f(d, x) = \sin(x^\top \beta_d)$, where $p = 1$, $\beta_0 = 1$, $\beta_1 = 2$, and consider two different cases for treatment probability:
\begin{itemize}
    \item Case 4: $f_D(X_i) = 0.75 - 0.25\sqrt{1-X_{i,1}}$.
    \item Case 5: $f_D(X_i) = X_{i,1}$.
\end{itemize}
The parameter of interest is the average treatment effect on treated (ATT)\footnote{
For estimating ATT, we consider a feasible (minimax linear) estimator which assumes the treatment probability is unknown (see \ref{app:ATT}).
}:
\begin{equation}
    \text{ATT} := \E\left[f(1,X_i) - f(0,X_i)|D_i = 1\right] = 
    \E\left[\frac{D_i}{\E[D_i]}(f(1,X_i) - f(0,X_i))\right].
\label{eq:ATT_def}
\end{equation}
The corresponding Riesz representer for this parameter on $L^2(P_Z)$ space is 
\begin{equation}
    \gamma^*(D_i,X_i) = \frac{D_i}{\E[D_i]} - \frac{1-D_i}{\E[D_i]}\cdot \frac{\E[D_i|X_i]}{1-\E[D_i|X_i]}.
\end{equation}
In our Monte Carlo, we compare the oracle Riesz weights $\gamma^*(D_i,X_i)$ with two sets of weights:
\begin{itemize}
    \item the optimal weights $k_{\delta_n}$ by minimizing the worst-case RMSE (equivalently first solving the modulus of continuity problem for each $\delta$ and then choose $\delta_n$ by minimizing the square of maximum bias plus the variance),
    \item the optimal weights $k_{\delta^\star}$ by solving the modulus of continuity problem with a fixed $\delta^\star = 2$, which minimizes the 99\% quantile of worst-case excess length for one-sided 95\% CIs.
\end{itemize}
We evaluate the discrepancy by
\begin{equation}
    \text{Dis}_{\delta} = \frac{1}{n}\sum_{i=1}^n [nk_{\delta} - \gamma^*(D_i,X_i)]^2.
\end{equation}
We report the bias, worst-case bias, and RMSE of the minimax linear estimator with different Lipschitz constants and different criteria (i.e., fixed $\delta^\star$ and $\delta_n$).
\begin{equation}
    \begin{aligned}
        \text{Bias} &= \frac{1}{N_{\text{sim}}} \sum_{i=1}^{N_{\text{sim}}}\left(\hat{\psi}^{(i)} - \psi(f)\right) \\
        \text{MaxBias} &= \frac{1}{N_{\text{sim}}} \sum_{i=1}^{N_{\text{sim}}}\sup_{f \in \mathcal{F}} \left(\hat{\psi}^{(i)} - \frac{\sum_{j=1}^n D_j^{(i)}\left(f(1,Z_j^{(i)})-f(0,Z_j^{(i)})\right)}{\sum_{j=1}^n D_j^{(i)}}\right) \\
        \text{RMSE} &= \frac{1}{N_{\text{sim}}} \sum_{i=1}^{N_{\text{sim}}}\left(\hat{\psi}^{(i)} - \psi(f)\right)^2,
    \end{aligned}
\end{equation}
where we set the function class $\mathcal{F} = \{|f(d,x)-f(d,x')|\leq C||x-x'||_1, d=0,1\}$ and $N_{\text{sim}} = 500$. In addition, we also report the bias and RMSE of augmented minimax linear estimator where the augmentation part fits a local constant regression using the R package \textit{np} with the selection criterion of the bandwidths being least-squares cross-validation. 

\begin{table}[H]
    \centering
    \caption{Simulation Results under Case 1}
    \begin{tabular}{lccccccccc}
    \hline\hline
    \multicolumn{10}{l}{\textbf{Panel A: Minimax Linear, \(C = 2\)}} \\
    \midrule
    \multirow{2}{*}{$n$} 
        & \multirow{2}{*}{$\mathrm{Dis}_{\delta_n}$} 
        & \multirow{2}{*}{$\mathrm{Dis}_{\delta^\star}$} 
        & \multicolumn{3}{c}{$C=2,\ \delta_n$} 
        & \multicolumn{3}{c}{$C=2,\ \delta^\star$} \\ 
    \cmidrule(lr){4-6} \cmidrule(lr){7-9}
        &  &  & Bias & MaxBias & RMSE & Bias & MaxBias & RMSE &  \\ 
    \midrule
    100 & 0.6081 & 0.7023 & 0.0033 & 0.0425 & 0.1157 & 0.0027 & 0.0398 & 0.1168 \\ 
  250 & 0.3826 & 0.5255 & 0.0002 & 0.0204 & 0.0684 & -0.0003 & 0.0177 & 0.0698 \\ 
  500 & 0.2660 & 0.4128 & 0.0008 & 0.0119 & 0.0476 & 0.0009 & 0.0097 & 0.0484 \\ 
    \addlinespace[0.5em]
    \multicolumn{10}{l}{\textbf{Panel B: Minimax Linear, \(C = 1\)}} \\
    \midrule
    \multirow{2}{*}{$n$} 
        & \multirow{2}{*}{$\mathrm{Dis}_{\delta_n}$} 
        & \multirow{2}{*}{$\mathrm{Dis}_{\delta^\star}$} 
        & \multicolumn{3}{c}{$C=1,\ \delta_n$} 
        & \multicolumn{3}{c}{$C=1,\ \delta^\star$} \\ 
    \cmidrule(lr){4-6} \cmidrule(lr){7-9}
        &  &  & Bias & MaxBias & RMSE & Bias & MaxBias & RMSE &  \\ 
    \midrule
    100 & 0.3196 & 0.4531 & 0.0093 & 0.0290 & 0.1113 & 0.0053 & 0.0241 & 0.1135 \\ 
  250 & 0.2012 & 0.3397 & 0.0029 & 0.0142 & 0.0671 & 0.0007 & 0.0107 & 0.0681 \\ 
  500 & 0.1374 & 0.2638 & 0.0017 & 0.0083 & 0.0468 & 0.0008 & 0.0060 & 0.0476 \\ 
    \addlinespace[0.5em]
    \multicolumn{10}{l}{\textbf{Panel C: Minimax Linear, \(C = 3\)}} \\
    \midrule
    \multirow{2}{*}{$n$} 
        & \multirow{2}{*}{$\mathrm{Dis}_{\delta_n}$} 
        & \multirow{2}{*}{$\mathrm{Dis}_{\delta^\star}$} 
        & \multicolumn{3}{c}{$C=3,\ \delta_n$} 
        & \multicolumn{3}{c}{$C=3,\ \delta^\star$} \\ 
    \cmidrule(lr){4-6} \cmidrule(lr){7-9}
        &  &  & Bias & MaxBias & RMSE & Bias & MaxBias & RMSE &  \\ 
    \midrule
    100 & 0.8882 & 0.9000 & 0.0020 & 0.0548 & 0.1197 & 0.0022 & 0.0544 & 0.1195 \\ 
  250 & 0.5628 & 0.6735 & -0.0006 & 0.0258 & 0.0701 & -0.0008 & 0.0239 & 0.0711 \\ 
  500 & 0.3923 & 0.5347 & 0.0008 & 0.0150 & 0.0483 & 0.0010 & 0.0131 & 0.0492 \\ 
    \addlinespace[0.5em]
    \multicolumn{10}{l}{\textbf{Panel D: Augmented Minimax Linear}} \\
    \midrule
    \multirow{2}{*}{$n$} 
        & \multicolumn{2}{c}{\(\text{Aug }+\;C=2,\ \delta_n\)} 
        & \multicolumn{2}{c}{\(\text{Aug }+\;C=2,\ \delta^\star\)} 
        & \multicolumn{2}{c}{\(\text{Aug }+\;C=1,\ \delta_n\)} 
        & \multicolumn{2}{c}{\(\text{Aug }+\;C=3,\ \delta_n\)} 
        &  \\ 
    \cmidrule(lr){2-3} \cmidrule(lr){4-5} \cmidrule(lr){6-7} \cmidrule(lr){8-9}
        & Bias & RMSE & Bias & RMSE & Bias & RMSE & Bias & RMSE &  \\ 
    \midrule
    100 & 0.0011 & 0.1180 & 0.0010 & 0.1189 & 0.0028 & 0.1146 & 0.0009 & 0.1210 \\ 
  250 & -0.0008 & 0.0690 & -0.0010 & 0.0702 & -0.0002 & 0.0680 & -0.0011 & 0.0705 \\ 
  500 & 0.0004 & 0.0478 & 0.0007 & 0.0485 & 0.0001 & 0.0471 & 0.0006 & 0.0484 \\ 
    \hline\hline
    \end{tabular}
    \label{tab:case1}
\end{table}

\begin{table}[H]
    \centering
    \caption{Simulation Results under Case 2}
    \begin{tabular}{lccccccccc}
    \hline\hline
    \multicolumn{10}{l}{\textbf{Panel A: Minimax Linear, \(C = 2\)}} \\
    \midrule
    \multirow{2}{*}{$n$} 
        & \multirow{2}{*}{$\mathrm{Dis}_{\delta_n}$} 
        & \multirow{2}{*}{$\mathrm{Dis}_{\delta^\star}$} 
        & \multicolumn{3}{c}{$C=2,\ \delta_n$} 
        & \multicolumn{3}{c}{$C=2,\ \delta^\star$} \\ 
    \cmidrule(lr){4-6} \cmidrule(lr){7-9}
        &  &  & Bias & MaxBias & RMSE & Bias & MaxBias & RMSE &  \\ 
    \midrule
    100 & 2.0585 & 1.1892 & -0.0178 & 0.5406 & 0.1340 & -0.0156 & 0.5495 & 0.1245 \\ 
  250 & 2.0941 & 1.2139 & 0.0033 & 0.3871 & 0.0830 & 0.0013 & 0.3927 & 0.0770 \\ 
  500 & 2.0670 & 1.2258 & -0.0004 & 0.3033 & 0.0611 & -0.0007 & 0.3070 & 0.0580 \\ 
    \addlinespace[0.5em]
    \multicolumn{10}{l}{\textbf{Panel B: Minimax Linear, \(C = 1\)}} \\
    \midrule
    \multirow{2}{*}{$n$} 
        & \multirow{2}{*}{$\mathrm{Dis}_{\delta_n}$} 
        & \multirow{2}{*}{$\mathrm{Dis}_{\delta^\star}$} 
        & \multicolumn{3}{c}{$C=1,\ \delta_n$} 
        & \multicolumn{3}{c}{$C=1,\ \delta^\star$} \\ 
    \cmidrule(lr){4-6} \cmidrule(lr){7-9}
        &  &  & Bias & MaxBias & RMSE & Bias & MaxBias & RMSE &  \\ 
    \midrule
    100 & 1.6182 & 0.7869 & -0.0163 & 0.2716 & 0.1291 & -0.0142 & 0.2815 & 0.1205 \\ 
  250 & 1.6982 & 0.8007 & 0.0020 & 0.1942 & 0.0796 & 0.0002 & 0.2006 & 0.0742 \\ 
  500 & 1.7237 & 0.8149 & -0.0008 & 0.1520 & 0.0601 & -0.0005 & 0.1564 & 0.0557 \\ 
   \addlinespace[0.5em]
    \multicolumn{10}{l}{\textbf{Panel C: Minimax Linear, \(C = 3\)}} \\
    \midrule
    \multirow{2}{*}{$n$} 
        & \multirow{2}{*}{$\mathrm{Dis}_{\delta_n}$} 
        & \multirow{2}{*}{$\mathrm{Dis}_{\delta^\star}$} 
        & \multicolumn{3}{c}{$C=3,\ \delta_n$} 
        & \multicolumn{3}{c}{$C=3,\ \delta^\star$} \\ 
    \cmidrule(lr){4-6} \cmidrule(lr){7-9}
        &  &  & Bias & MaxBias & RMSE & Bias & MaxBias & RMSE &  \\ 
    \midrule
    100 & 2.1757 & 1.4142 & -0.0177 & 0.8106 & 0.1350 & -0.0161 & 0.8182 & 0.1271 \\ 
  250 & 2.1846 & 1.4462 & 0.0034 & 0.5805 & 0.0835 & 0.0017 & 0.5851 & 0.0786 \\ 
  500 & 2.1507 & 1.4509 & -0.0004 & 0.4548 & 0.0619 & -0.0006 & 0.4578 & 0.0589 \\ 
     \addlinespace[0.5em]
    \multicolumn{10}{l}{\textbf{Panel D: Augmented Minimax Linear}} \\
    \midrule
    \multirow{2}{*}{$n$} 
        & \multicolumn{2}{c}{\(\text{Aug }+\;C=2,\ \delta_n\)} 
        & \multicolumn{2}{c}{\(\text{Aug }+\;C=2,\ \delta^\star\)} 
        & \multicolumn{2}{c}{\(\text{Aug }+\;C=1,\ \delta_n\)} 
        & \multicolumn{2}{c}{\(\text{Aug }+\;C=3,\ \delta_n\)} 
        &  \\ 
    \cmidrule(lr){2-3} \cmidrule(lr){4-5} \cmidrule(lr){6-7} \cmidrule(lr){8-9}
        & Bias & RMSE & Bias & RMSE & Bias & RMSE & Bias & RMSE &  \\ 
    \midrule
    100 & -0.0157 & 0.1357 & -0.0141 & 0.1290 & -0.0147 & 0.1322 & -0.0156 & 0.1363 \\ 
  250 & 0.0034 & 0.0835 & 0.0018 & 0.0786 & 0.0022 & 0.0807 & 0.0035 & 0.0838 \\ 
  500 & -0.0003 & 0.0610 & -0.0009 & 0.0582 & -0.0008 & 0.0600 & -0.0003 & 0.0616 \\ 
    \hline\hline
    \end{tabular}
    \label{tab:case2}
\end{table}

\begin{table}[H]
    \centering
    \caption{Simulation Results under Case 3}
    \begin{tabular}{lccccccccc}
    \hline\hline
    \multicolumn{10}{l}{\textbf{Panel A: Minimax Linear, \(C = 2\)}} \\
    \midrule
    \multirow{2}{*}{$n$} 
        & \multirow{2}{*}{$\mathrm{Dis}_{\delta_n}$} 
        & \multirow{2}{*}{$\mathrm{Dis}_{\delta^\star}$} 
        & \multicolumn{3}{c}{$C=2,\ \delta_n$} 
        & \multicolumn{3}{c}{$C=2,\ \delta^\star$} \\ 
    \cmidrule(lr){4-6} \cmidrule(lr){7-9}
        &  &  & Bias & MaxBias & RMSE & Bias & MaxBias & RMSE &  \\ 
    \midrule
    100 & 0.6081 & 0.7023 & 0.0016 & 0.0425 & 0.1469 & 0.0011 & 0.0398 & 0.1480 \\ 
  250 & 0.3826 & 0.5255 & -0.0010 & 0.0204 & 0.0854 & -0.0018 & 0.0177 & 0.0867 \\ 
  500 & 0.2660 & 0.4128 & -0.0039 & 0.0119 & 0.0598 & -0.0040 & 0.0097 & 0.0607 \\ 
    \addlinespace[0.5em]
    \multicolumn{10}{l}{\textbf{Panel B: Minimax Linear, \(C = 1\)}} \\
    \midrule
    \multirow{2}{*}{$n$} 
        & \multirow{2}{*}{$\mathrm{Dis}_{\delta_n}$} 
        & \multirow{2}{*}{$\mathrm{Dis}_{\delta^\star}$} 
        & \multicolumn{3}{c}{$C=1,\ \delta_n$} 
        & \multicolumn{3}{c}{$C=1,\ \delta^\star$} \\ 
    \cmidrule(lr){4-6} \cmidrule(lr){7-9}
        &  &  & Bias & MaxBias & RMSE & Bias & MaxBias & RMSE &  \\ 
    \midrule
    100 & 0.3196 & 0.4531 & 0.0089 & 0.0290 & 0.1444 & 0.0044 & 0.0241 & 0.1460 \\ 
  250 & 0.2012 & 0.3397 & 0.0029 & 0.0142 & 0.0846 & -0.0004 & 0.0107 & 0.0853 \\ 
  500 & 0.1374 & 0.2638 & -0.0032 & 0.0083 & 0.0585 & -0.0040 & 0.0060 & 0.0597 \\ 
   \addlinespace[0.5em]
    \multicolumn{10}{l}{\textbf{Panel C: Minimax Linear, \(C = 3\)}} \\
    \midrule
    \multirow{2}{*}{$n$} 
        & \multirow{2}{*}{$\mathrm{Dis}_{\delta_n}$} 
        & \multirow{2}{*}{$\mathrm{Dis}_{\delta^\star}$} 
        & \multicolumn{3}{c}{$C=3,\ \delta_n$} 
        & \multicolumn{3}{c}{$C=3,\ \delta^\star$} \\ 
    \cmidrule(lr){4-6} \cmidrule(lr){7-9}
        &  &  & Bias & MaxBias & RMSE & Bias & MaxBias & RMSE &  \\ 
    \midrule
    100 & 0.8882 & 0.9000 & 0.0002 & 0.0548 & 0.1505 & 0.0006 & 0.0544 & 0.1504 \\ 
  250 & 0.5628 & 0.6735 & -0.0021 & 0.0258 & 0.0869 & -0.0025 & 0.0239 & 0.0874 \\ 
  500 & 0.3923 & 0.5347 & -0.0040 & 0.0150 & 0.0606 & -0.0037 & 0.0131 & 0.0616 \\ 
   \addlinespace[0.5em]
    \multicolumn{10}{l}{\textbf{Panel D: Augmented Minimax Linear}} \\
    \midrule
    \multirow{2}{*}{$n$} 
        & \multicolumn{2}{c}{\(\text{Aug }+\;C=2,\ \delta_n\)} 
        & \multicolumn{2}{c}{\(\text{Aug }+\;C=2,\ \delta^\star\)} 
        & \multicolumn{2}{c}{\(\text{Aug }+\;C=1,\ \delta_n\)} 
        & \multicolumn{2}{c}{\(\text{Aug }+\;C=3,\ \delta_n\)} 
        &  \\ 
    \cmidrule(lr){2-3} \cmidrule(lr){4-5} \cmidrule(lr){6-7} \cmidrule(lr){8-9}
        & Bias & RMSE & Bias & RMSE & Bias & RMSE & Bias & RMSE &  \\ 
    \midrule
    100 & -0.0026 & 0.1546 & -0.0023 & 0.1550 & -0.0019 & 0.1529 & -0.0022 & 0.1565 \\ 
  250 & -0.0027 & 0.0895 & -0.0027 & 0.0902 & -0.0024 & 0.0887 & -0.0028 & 0.0904 \\ 
  500 & -0.0035 & 0.0623 & -0.0033 & 0.0628 & -0.0037 & 0.0618 & -0.0033 & 0.0627 \\ 
    \hline\hline
    \end{tabular}
    \label{tab:case3}
\end{table}

\begin{table}[H]
    \centering
    \caption{Simulation Results under Case 4}
    \begin{tabular}{lccccccccc}
    \hline\hline
    \multicolumn{10}{l}{\textbf{Panel A: Minimax Linear, \(C = 2\)}} \\
    \midrule
    \multirow{2}{*}{$n$} 
        & \multirow{2}{*}{$\mathrm{Dis}_{\delta_n}$} 
        & \multirow{2}{*}{$\mathrm{Dis}_{\delta^\star}$} 
        & \multicolumn{3}{c}{$C=2,\ \delta_n$} 
        & \multicolumn{3}{c}{$C=2,\ \delta^\star$} \\ 
    \cmidrule(lr){4-6} \cmidrule(lr){7-9}
        &  &  & Bias & MaxBias & RMSE & Bias & MaxBias & RMSE &  \\ 
    \midrule
    100 & 0.5541 & 0.6579 & 0.0025 & 0.0397 & 0.1133 & 0.0021 & 0.0366 & 0.1145 \\ 
  250 & 0.3562 & 0.4977 & 0.0013 & 0.0193 & 0.0690 & 0.0009 & 0.0165 & 0.0700 \\ 
  500 & 0.2429 & 0.3862 & 0.0008 & 0.0113 & 0.0463 & 0.0008 & 0.0091 & 0.0473 \\ 
    \addlinespace[0.5em]
    \multicolumn{10}{l}{\textbf{Panel B: Minimax Linear, \(C = 1\)}} \\
    \midrule
    \multirow{2}{*}{$n$} 
        & \multirow{2}{*}{$\mathrm{Dis}_{\delta_n}$} 
        & \multirow{2}{*}{$\mathrm{Dis}_{\delta^\star}$} 
        & \multicolumn{3}{c}{$C=1,\ \delta_n$} 
        & \multicolumn{3}{c}{$C=1,\ \delta^\star$} \\ 
    \cmidrule(lr){4-6} \cmidrule(lr){7-9}
        &  &  & Bias & MaxBias & RMSE & Bias & MaxBias & RMSE &  \\ 
    \midrule
    100 & 0.2934 & 0.4251 & 0.0077 & 0.0273 & 0.1093 & 0.0041 & 0.0223 & 0.1115 \\ 
  250 & 0.1897 & 0.3234 & 0.0034 & 0.0136 & 0.0675 & 0.0016 & 0.0101 & 0.0686 \\ 
  500 & 0.1286 & 0.2465 & 0.0018 & 0.0079 & 0.0453 & 0.0009 & 0.0056 & 0.0462 \\ 
   \addlinespace[0.5em]
    \multicolumn{10}{l}{\textbf{Panel C: Minimax Linear, \(C = 3\)}} \\
    \midrule
    \multirow{2}{*}{$n$} 
        & \multirow{2}{*}{$\mathrm{Dis}_{\delta_n}$} 
        & \multirow{2}{*}{$\mathrm{Dis}_{\delta^\star}$} 
        & \multicolumn{3}{c}{$C=3,\ \delta_n$} 
        & \multicolumn{3}{c}{$C=3,\ \delta^\star$} \\ 
    \cmidrule(lr){4-6} \cmidrule(lr){7-9}
        &  &  & Bias & MaxBias & RMSE & Bias & MaxBias & RMSE &  \\ 
    \midrule
    100 & 0.8075 & 0.8430 & 0.0016 & 0.0508 & 0.1166 & 0.0018 & 0.0498 & 0.1169 \\ 
  250 & 0.5207 & 0.6372 & 0.0009 & 0.0243 & 0.0703 & 0.0007 & 0.0223 & 0.0711 \\ 
  500 & 0.3592 & 0.5012 & 0.0008 & 0.0141 & 0.0471 & 0.0010 & 0.0122 & 0.0480 \\ 
   \addlinespace[0.5em]
    \multicolumn{10}{l}{\textbf{Panel D: Augmented Minimax Linear}} \\
    \midrule
    \multirow{2}{*}{$n$} 
        & \multicolumn{2}{c}{\(\text{Aug }+\;C=2,\ \delta_n\)} 
        & \multicolumn{2}{c}{\(\text{Aug }+\;C=2,\ \delta^\star\)} 
        & \multicolumn{2}{c}{\(\text{Aug }+\;C=1,\ \delta_n\)} 
        & \multicolumn{2}{c}{\(\text{Aug }+\;C=3,\ \delta_n\)} 
        &  \\ 
    \cmidrule(lr){2-3} \cmidrule(lr){4-5} \cmidrule(lr){6-7} \cmidrule(lr){8-9}
        & Bias & RMSE & Bias & RMSE & Bias & RMSE & Bias & RMSE &  \\ 
    \midrule
   100 & 0.0008 & 0.1145 & 0.0009 & 0.1156 & 0.0022 & 0.1111 & 0.0008 & 0.1176 \\ 
  250 & 0.0007 & 0.0693 & 0.0006 & 0.0702 & 0.0010 & 0.0684 & 0.0006 & 0.0705 \\ 
  500 & 0.0005 & 0.0464 & 0.0007 & 0.0474 & 0.0005 & 0.0456 & 0.0006 & 0.0472 \\     \hline\hline
    \end{tabular}
    \label{tab:case4}
\end{table}

\begin{table}[H]
    \centering
    \caption{Simulation Results under Case 5}
    \begin{tabular}{lccccccccc}
    \hline\hline
    \multicolumn{10}{l}{\textbf{Panel A: Minimax Linear, \(C = 2\)}} \\
    \midrule
    \multirow{2}{*}{$n$} 
        & \multirow{2}{*}{$\mathrm{Dis}_{\delta_n}$} 
        & \multirow{2}{*}{$\mathrm{Dis}_{\delta^\star}$} 
        & \multicolumn{3}{c}{$C=2,\ \delta_n$} 
        & \multicolumn{3}{c}{$C=2,\ \delta^\star$} \\ 
    \cmidrule(lr){4-6} \cmidrule(lr){7-9}
        &  &  & Bias & MaxBias & RMSE & Bias & MaxBias & RMSE &  \\ 
    \midrule
    100 & 18.4114 & 18.0200 & 0.0269 & 0.0873 & 0.1636 & 0.0278 & 0.0918 & 0.1603 \\ 
  250 & 9.9514 & 9.6184 & 0.0080 & 0.0475 & 0.1014 & 0.0089 & 0.0468 & 0.1021 \\ 
  500 & 11.2525 & 10.8683 & 0.0049 & 0.0313 & 0.0760 & 0.0041 & 0.0292 & 0.0771 \\     \addlinespace[0.5em]
    \multicolumn{10}{l}{\textbf{Panel B: Minimax Linear, \(C = 1\)}} \\
    \midrule
    \multirow{2}{*}{$n$} 
        & \multirow{2}{*}{$\mathrm{Dis}_{\delta_n}$} 
        & \multirow{2}{*}{$\mathrm{Dis}_{\delta^\star}$} 
        & \multicolumn{3}{c}{$C=1,\ \delta_n$} 
        & \multicolumn{3}{c}{$C=1,\ \delta^\star$} \\ 
    \cmidrule(lr){4-6} \cmidrule(lr){7-9}
        &  &  & Bias & MaxBias & RMSE & Bias & MaxBias & RMSE &  \\ 
    \midrule
    100 & 18.2940 & 17.9671 & 0.0483 & 0.0655 & 0.1524 & 0.0464 & 0.0638 & 0.1532 \\ 
  250 & 10.2391 & 9.7982 & 0.0226 & 0.0385 & 0.0931 & 0.0200 & 0.0345 & 0.0946 \\ 
  500 & 11.8516 & 11.1093 & 0.0124 & 0.0247 & 0.0718 & 0.0108 & 0.0221 & 0.0712 \\ 
   \addlinespace[0.5em]
    \multicolumn{10}{l}{\textbf{Panel C: Minimax Linear, \(C = 3\)}} \\
    \midrule
    \multirow{2}{*}{$n$} 
        & \multirow{2}{*}{$\mathrm{Dis}_{\delta_n}$} 
        & \multirow{2}{*}{$\mathrm{Dis}_{\delta^\star}$} 
        & \multicolumn{3}{c}{$C=3,\ \delta_n$} 
        & \multicolumn{3}{c}{$C=3,\ \delta^\star$} \\ 
    \cmidrule(lr){4-6} \cmidrule(lr){7-9}
        &  &  & Bias & MaxBias & RMSE & Bias & MaxBias & RMSE &  \\ 
    \midrule
    100 & 18.9689 & 18.2913 & 0.0197 & 0.1104 & 0.1751 & 0.0215 & 0.1197 & 0.1685 \\ 
  250 & 10.1983 & 9.7982 & 0.0024 & 0.0575 & 0.1085 & 0.0048 & 0.0594 & 0.1069 \\ 
  500 & 11.2639 & 10.8378 & 0.0016 & 0.0369 & 0.0793 & 0.0020 & 0.0366 & 0.0790 \\ 
   \addlinespace[0.5em]
    \multicolumn{10}{l}{\textbf{Panel D: Augmented Minimax Linear}} \\
    \midrule
    \multirow{2}{*}{$n$} 
        & \multicolumn{2}{c}{\(\text{Aug }+\;C=2,\ \delta_n\)} 
        & \multicolumn{2}{c}{\(\text{Aug }+\;C=2,\ \delta^\star\)} 
        & \multicolumn{2}{c}{\(\text{Aug }+\;C=1,\ \delta_n\)} 
        & \multicolumn{2}{c}{\(\text{Aug }+\;C=3,\ \delta_n\)} 
        &  \\ 
    \cmidrule(lr){2-3} \cmidrule(lr){4-5} \cmidrule(lr){6-7} \cmidrule(lr){8-9}
        & Bias & RMSE & Bias & RMSE & Bias & RMSE & Bias & RMSE &  \\ 
    \midrule
   100 & 0.0118 & 0.1819 & 0.0117 & 0.1787 & 0.0163 & 0.1744 & 0.0101 & 0.1895 \\ 
  250 & -0.0023 & 0.1103 & -0.0012 & 0.1102 & 0.0017 & 0.1035 & -0.0047 & 0.1161 \\ 
  500 & -0.0023 & 0.0815 & -0.0024 & 0.0821 & -0.0009 & 0.0785 & -0.0033 & 0.0838 \\ 
 \hline\hline
    \end{tabular}
    \label{tab:case5}
\end{table}

Under Case 1 (see Table \ref{tab:case1}), the outcome regression function $f$ and the Riesz representer $\gamma^*$ both lie in the Lipschitz class $\mathcal{F}$, which in one dimension is Donsker; hence the conditions in Theorem \ref{thm:root-n} are satisfied. The columns 2-3 in Panel A of the Table \ref{tab:case1} report the discrepancy between the minimax-optimal weights (fixed $\delta^*$ and $\delta_n$) and the true Riesz representer.
As $n$ increases, the discrepancy get smaller and closer to $0$. The worst-case bias $\sup_{f\in\mathcal{F}} |\hat{\psi}_{\text{linear},\delta} - \psi(f)|$ gets closer to 0 and thus insensitive to the choice of the Lipschitz constant as the sample size $n$ grows. Moreover, Table \ref{tab:case1} also presents results for the augmented minimax linear estimator in Panel D. Both the pure and augmented estimators display similar bias and RMSE, indicating that augmentation yields no first‐order improvement when the pure minimax‐linear estimator already attains the semiparametric efficiency bound (Corollary \ref{coro:asynormal}). 

Under Case 2 (see Table \ref{tab:case2}),  we increase the covariate dimension to three, so that the Lipschitz class $\mathcal{F}$ is no longer Donsker.
Consequently, the discrepancy between optimal weights and Riesz representer remains large and does not decrease as $n$ grows. Although both the pure and augmented minimax linear estimators continue to exhibit good performance in terms of bias and RMSE, the worst-case bias of the pure estimator decays only slowly and varies markedly with the choice of Lipschitz constant. This pattern aligns with the findings in \citet{armstrong2021finite} that, in general cases, one must explicitly specify a Lipschitz bound to control worst-case bias when constructing honest confidence intervals. 

Under Case 3 (see Table \ref{tab:case3}), the outcome regression function $f$ lies in a one-dimensional Lipschitz class - albeit with a large Lipschitz constant. For example, $|f'(1,x)| \leq \frac{1}{(x+0.05)^2} \leq 400$ for $x\in [0,1]$. As $n$ grows, the distance between the optimal weights and Riesz representer become smaller, and both pure and augmented minimax linear estimators continue to perform well in terms of bias and RMSE.

Under Case 4 (see Table \ref{tab:case4}), we retain that the outcome regression function $f$ belongs to a one-dimension Lipschitz class but choose the Riesz representer to be $\gamma^*(0,x) \propto \frac{0.75-0.25\sqrt{1-x}}{0.25+0.25\sqrt{1-x}}$ whose first-order derivative becomes unbounded as $x$ approaches to 1. Hence $\gamma^*$ does not belong to any Lipschitz function class. Nevertheless, Table \ref{tab:case4} show that this boundary issue does not materially degrade finite-sample performance: the weight–Riesz discrepancy still decays with $n$ and bias, maximum bias, and RMSE remain comparable to Case 1. By contrast, under Case 5, we set $\gamma^*(0,x) \propto \frac{x}{1-x}$, which both fails any Lipschitz bound and diverges at $x = 1$ (reflecting the limited overlap issue). In this setting, the optimal weights cannot track the Riesz representer well according to the discrepancy results reported in Table \ref{tab:case5}.

\section{Empirical Application}
We apply the minimax linear estimation method to three classic policy questions, always targeting the ATT and reporting point estimates, na\"{i}ve and bias-aware confidence intervals. Throughout we assume that the conditional expectation function $f(d,x)$ belongs to a Lipschitz class with respect to a weighted norm:
\begin{equation}
    \mathcal{F}=\{|f(d,x)-f(d,x')|\leq C|x-x'|_{A},\ \forall \ d = 0,1\},
\end{equation} 
\begin{equation}
    |x|_A = \sum_{j=1}^p |A_{jj}x_j|,
\end{equation}
where $A=\text{diag}(A_{11},...,A_{dd})$ encodes priori bounds on the absolute value of the partial effects.
The diagonal entries $A_{jj}$ reflect a priori maximum change in the outcome from a one-unit change in each covariate. We choose $A$ such that its diagonal entries either modestly exceed the largest absolute OLS coefficients in the corresponding specifications or follow suggestions in the literature.

\subsection{Effect of minimum wage on youth employment}
We study the effect of an increase in minimum wage on teenage employment rates, using a subset of the data from \citet{callaway2021difference} (hereafter denoted CS), available in the R package \texttt{did}. 
The subset contains 500 county‐year observations from 2003–2007, but we focus only on two consecutive periods, 2006 and 2007. The treated sample is 131 counties first treated in 2007 - increased their minimum wage compared to the federal minimum wage while the control sample is 309 counties never treated before - the minimum wage was equal to the federal minimum wage. Define
\begin{equation}
    Y_i = (\text{lemp}_{i,2007} - \text{lemp}_{i,2006})
\end{equation}
where $\text{lemp}_{i,t}$ is the log of the county-level teen employment for county $i$ in year $t$. The treatment indicator is
\begin{equation}
    D_i = I\{\text{county } i \text{ first increases its minimum wage in year 2007}\}.
\end{equation} 
There are 131 counties that raised their minimum wage above the federal rate in 2007 (treated) and 309 counties that did not (controls)\footnote{We exclude 60 observations corresponding to counties that raised their minimum wage above the federal rate in 2004 or 2006.}.
Our single covariate, $X_i$, is the log of the population in year 2000. 
Table \ref{tab:emp_CS_sumstat} summarizes the outcome and covariate variables separately for treated and untreated counties.
\begin{table}[H]
    \centering
    \caption{Summary statistics for counties in the sample}
    \begin{tabular}{lcccc}
    \hline\hline
         & \multicolumn{2}{c}{Treated counties} & \multicolumn{2}{c}{Untreated counties} \\
         \cmidrule(lr){2-3} \cmidrule(lr){4-5} 
         & $Y$(change in lemp) & $X$(lpop) & $Y$(change in lemp) & $X$(lpop)\\ 
         \hline
       Mean  & -0.0038 & 3.4586 & 0.0222 & 3.1834 \\
       Median & -0.0056 & 3.3616 & 0.0164 & 3.1062 \\
       Std & 0.1590 & 1.2653 & 0.1631 & 1.2874 \\
       \hline \hline 
        \end{tabular}
        \label{tab:emp_CS_sumstat}
\end{table}
We set $A$ to be 0.01 which is plausible when $C = 1$. Because $X$ is univariate, this Lipschitz class $\mathcal{F}$ is Donsker. 
\begin{figure}[H]
    \centering
    \includegraphics[width=0.7\linewidth]{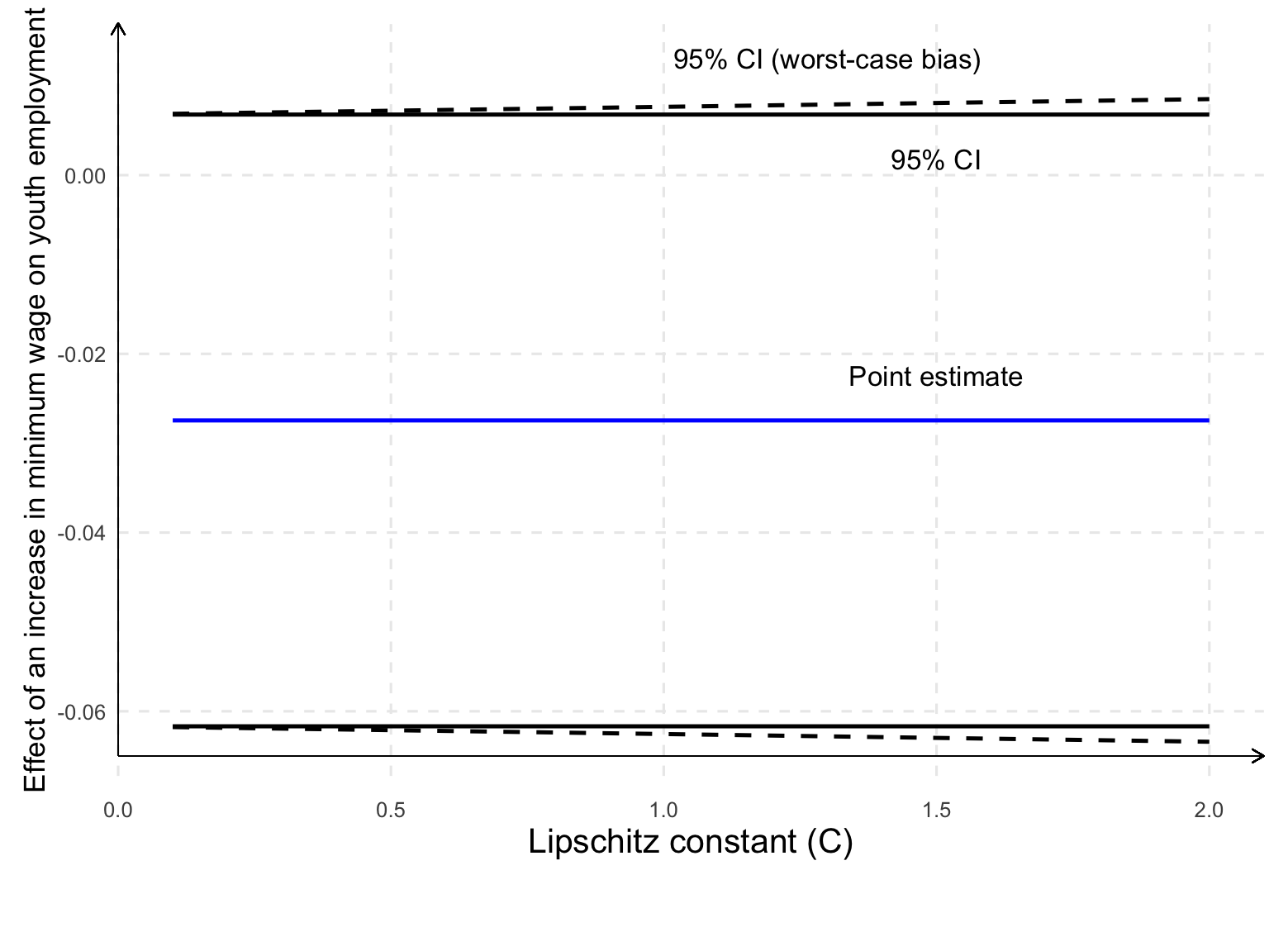}
    \caption{Point estimates and CIs for ATT in CS application as a function of the Lipschitz constant $C$.}
    \label{fig:CS}
\end{figure}
Figure \ref{fig:CS} displays, as $C$ varies over $[0.1,2]$, (i) the point estimate of ATT (blue line), (ii) the usual 95\% confidence interval\footnote{
The asymptotic variance splits into two parts—a conditional component and a marginal component (see \ref{app:ATT}). When the marginal part is estimated with the nearest-neighbor method of \citet{abadie2006large}, the estimate is negative. We therefore replace the nearest-neighbor step with a local-constant regression that fits $f(1,x)$ and $f(0,x)$ this yields a strictly positive estimate of the marginal variance. We then combine this positive marginal estimate with the conditional variance estimate to construct the standard error. Same for other applications.
} ignoring worst‐case bias (solid black lines), and (iii) the 95\% bias‐aware interval that adds the conditional worst‐case bias bound (dashed black lines). Varying the Lipschitz constant $C$ over its plausible range has little impact on our ATT estimate and on the width of the 95\% intervals. More importantly, the worst-case bias is negligible and is insensitive to $C$, as we expect.

\subsection{Effect of a job training program on earnings}
\begin{figure}[H]
    \centering
    \includegraphics[width=0.7\linewidth]{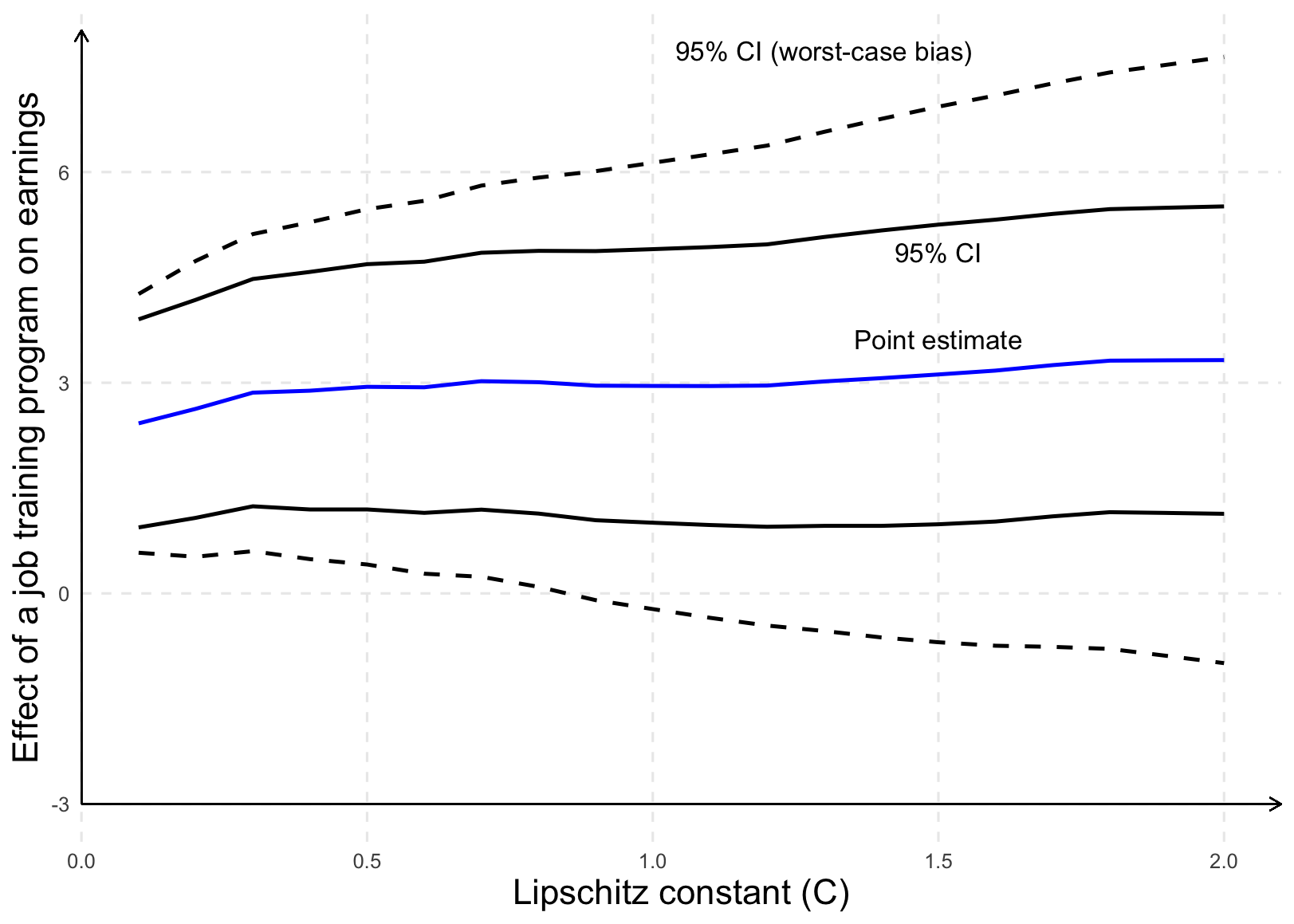}
    \caption{Point estimates and CIs for ATT in NSW application as a function of Lipschitz constant $C$}
    \label{fig:emp_NSW_set2}
\end{figure}

We consider the National Supported Work (NSW) application by using the same dataset as in \citet{dehejia1999causal}. The treated sample consists of 185 men in the NSW experimental sample who were randomly assigned to receive job training from December 1975 to January 1978, and the control sample is a nonexperimental sample of 2490 men taken from the PSID. We construct the outcome $Y_i$ by the difference between earnings in 1975 and earnings in 1978 in thousands of dollars. Let the covariate vector $X_i$ contain the variables: age, years of education, indicators for black and Hispanic, indicator for marriage, earnings in 1974, employment indicators for 1974. We assume that the unconfoundedness assumption holds conditional on $X_i$.
Table \ref{tab:emp_NSW_A2} reports the values of $A$\footnote{
The values of $A_{jj}$ in \cite{armstrong2021finite} are chosen based on the outcome - earnings in 1978. We keep their suggested values as the prior bound as they are more conservative for our outcome variable.
}:

\begin{table}[H]
    \centering 
    \caption{Diagonal elements of the weight matrix \(A\)}
    \begin{tabular}{cccccccc} 
    \toprule\toprule
      & Age & Edu   & Black & Hispanic & Married & Earnings for 1974 & Employed for 1974 \\ 
    \midrule
    $A^2$ & 0.15 & 0.60 & 2.50  & 2.50     & 2.50    & 0.50 & 0.10 \\
    \bottomrule\bottomrule
    \end{tabular}
    \label{tab:emp_NSW_A2}
\end{table}
With more than one continuous variable, the Lipschitz class is non-Donsker. Figure \ref{fig:emp_NSW_set2} shows that worst-case bias is sizable and highly sensitive to the constant $C$, illustrating the value of bias-aware inference under non-Donsker class.

\subsection{Effect of Title IIA of the National Job Training Partnership Act (JTPA) on 18-month earnings}
\begin{figure}[H]
    \centering
    \includegraphics[width=0.7\linewidth]{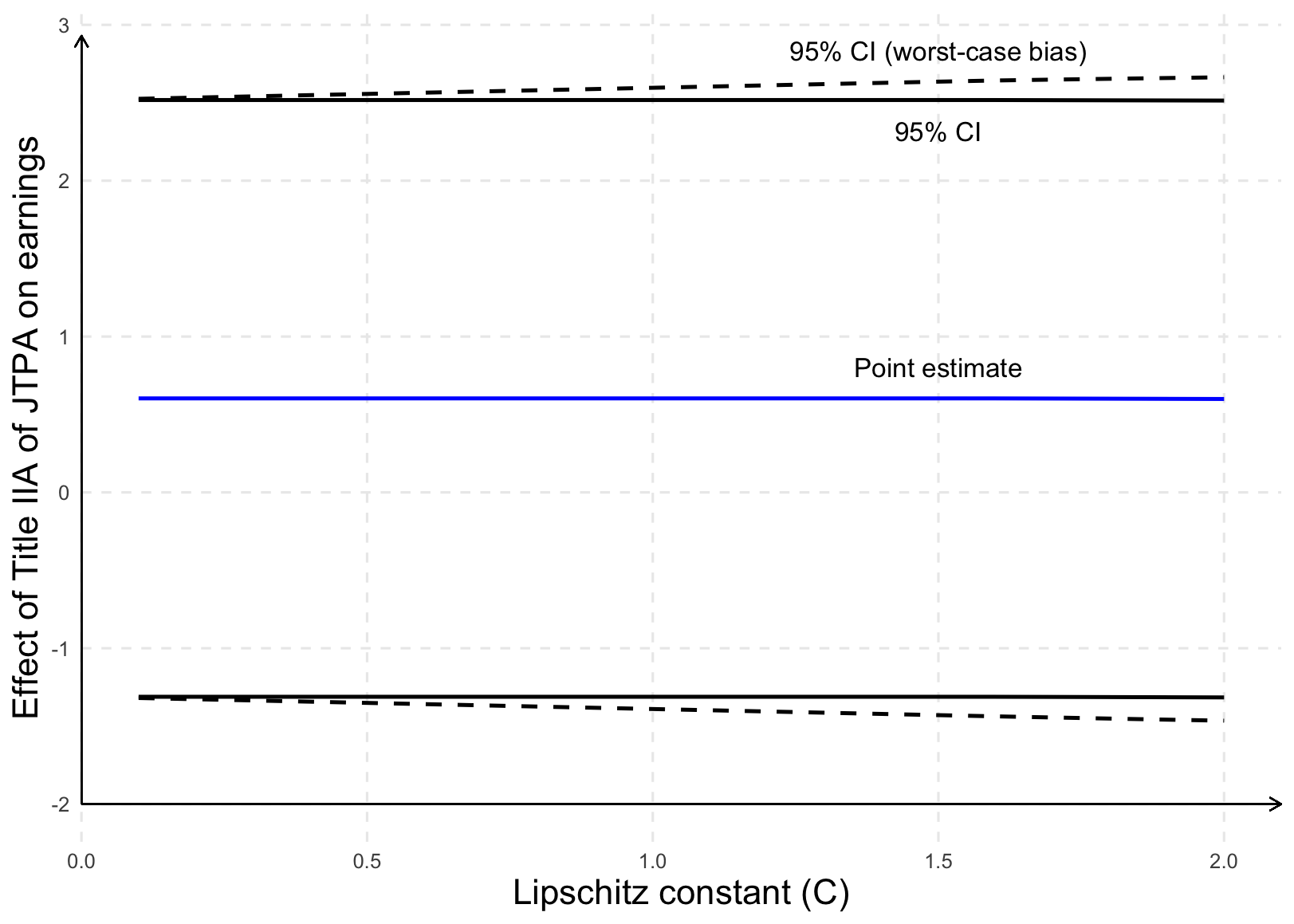}
    \caption{Point estimates and CIs for ATT in JTPA application using Specification 1 as a function of Lipschitz constant $C$}
    \label{fig:emp_JTPA_set1}
\end{figure}

\begin{figure}[H]
    \centering
    \includegraphics[width=0.7\linewidth]{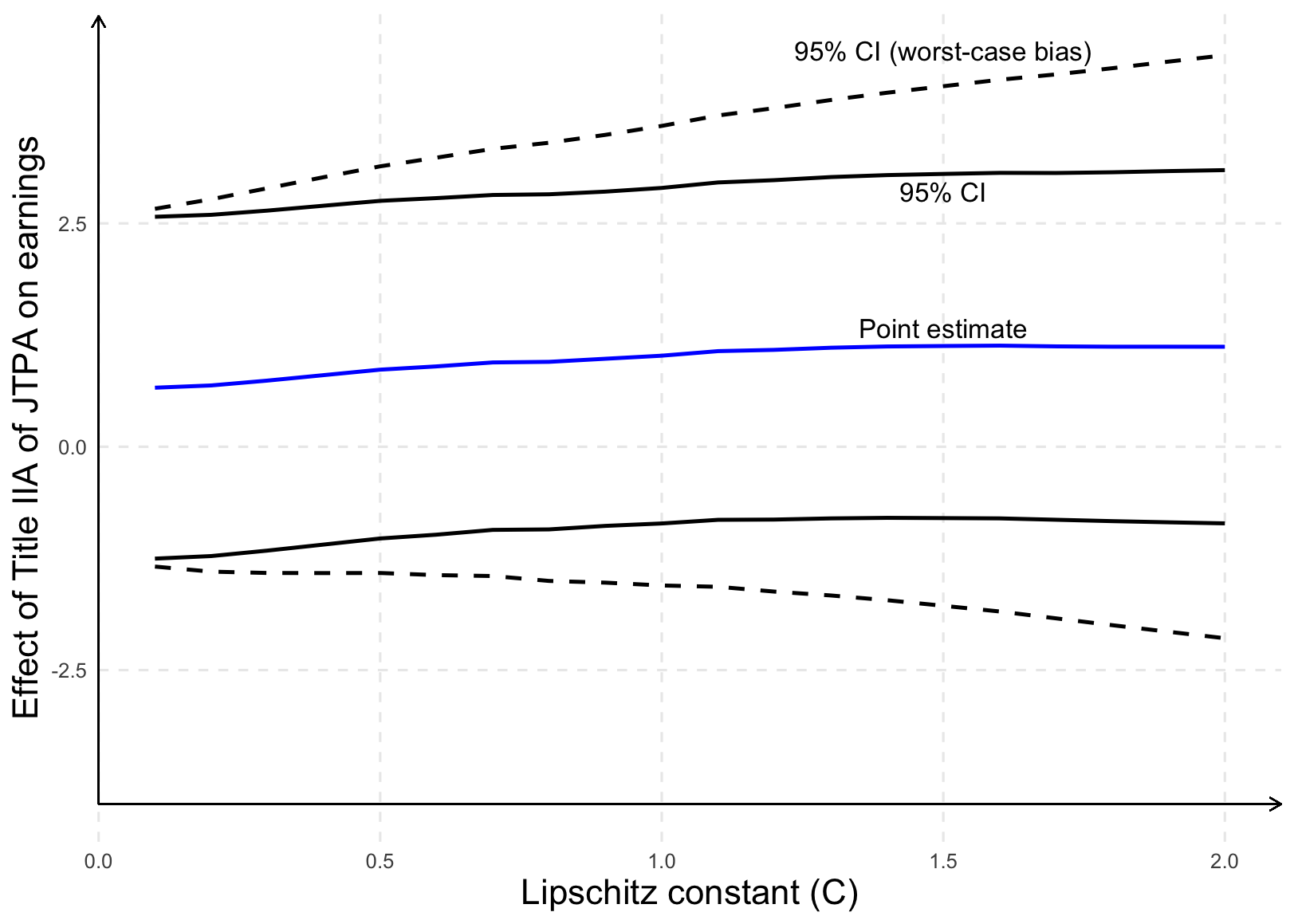}
    \caption{Point estimates and CIs for ATT in JTPA application using Specification 2 as a function of Lipschitz constant $C$}
    \label{fig:emp_JTPA_set2}
\end{figure}

We use the National JTPA experiment\footnote{We download the original dataset from \href{https://www.upjohn.org/data-tools/employment-research-data-center/national-jtpa-study}{https://www.upjohn.org/data-tools/employment-research-data-center/national-jtpa-study} and use two datasets \texttt{expbif.dta} and \texttt{earns.dta} for the following analysis.}, restricting to adults (age $\geq 22$) at the Decatur City, Illinois (site code “MD”). Our analytic sample \footnote{The original sample consists of 396 individuals in total and we drop 70 observations with missing data.} contains 326 individuals—219 (randomly) assigned to receive JTPA services and 107 assigned to the control group. The outcome variable $Y_i$ is total earnings over the 18 months after random assignment, expressed in thousands of dollars.
We consider two sets of control covariates as follows:
\begin{itemize}
    \item Specification 1: age 
    \item Specification 2: age, years of education, indicator for White, indicator for female, historical average annual earnings (in thousands of dollars)
\end{itemize}
Because assignment was randomized, unconfoundedness holds under either specification.
Tables \ref{tab:emp_JTPA_A} give the values of weight matrices $A^1$ and $A^2$ for two specifications:
\begin{table}[H]
    \centering 
    \caption{Diagonal elements of the weight matrix $A^1$ and \(A^2\)}
    \begin{tabular}{cccccc} 
    \toprule\toprule
      & Age & Edu   & White & Sex & HEarnings \\ 
    \midrule
    $A^1$ &  0.10 & - & - & - & -\\
    $A^2$ & 0.10 & 2.00 & 2.00 & 0.60 & 0.10\\
    \bottomrule\bottomrule
    \end{tabular}
    \label{tab:emp_JTPA_A}
\end{table}
Under Specification 1 (only one continuous covariate) the resulting function class is Donsker; under Specification 2 it is non-Donsker. Figures \ref{fig:emp_JTPA_set1} and \ref{fig:emp_JTPA_set2} plot the estimated ATT (blue) together with 95\% “naïve” intervals (solid black) and 95\% bias-aware intervals (dashed black) as a function of the Lipschitz constant $C$. In Specification 1, worst‐case bias remains negligible across all $C$. In Specification 2, bias is both substantial and highly sensitive to $C$, reflecting the non-negligibility of the worse-case bias.

\section{Concluding Remarks}
This paper develops the large‑sample theory for pure minimax linear weighting. Under standard Donsker‑type conditions, minimax weights converge to the Riesz representer and the estimator is root‑$n$, asymptotically normal, and semiparametrically efficient, making augmentation unnecessary; in this regime the worst‑case bias is second order and standard‑error–only CIs are valid. Outside this regime, the weights need not track the representer, worst‑case bias is non‑negligible and tuning‑sensitive, and bias‑aware CIs remain essential. Simulations and three applications visualize these two regimes and provide an operational rule for practice.

\clearpage

\appendix
\renewcommand{\thesection}{Appendix \Alph{section}}
\section{Proofs for Projection Problems}\label{app:projections}
\renewcommand{\theequation}{A.\arabic{equation}}
\setcounter{equation}{0}
\renewcommand{\thesubsection}{A.\arabic{subsection}}
\subsection{Proof for equation \eqref{eq:RR_ATE_PL_proj} in Example 2}
    Consider the optimization problem 
    \begin{equation}
        \min_{\gamma \in \mathcal{F}} \E\left[
        \left(\frac{D}{e(X)} - \frac{1-D}{e(X)} - \gamma(D,X)\right)^2 
        \right]
        \label{eqapp:RR_ATE_PL_proj}
    \end{equation}
    where $\mathcal{F} = \{f(d,x) = bd + g(x)|b\in \mathbb{R}, g \in L^2(P_X)\}$.
    Hilbert projection theorem guarantee us the existence and uniqueness of the solution to the optimization problem \eqref{eqapp:RR_ATE_PL_proj}. 
    It is equivalent to solve 
    \begin{equation*}
        \min_{b \in \mathbb{R}, g \in L^2(P_X)} \E\left[
            \left(\frac{D}{e(X)} - \frac{1-D}{1-e(X)} - (bD+g(X))\right)^2
        \right]
    \end{equation*}
    The FOC with respect to the function $g(x)$ is, for any direction $\tilde{g}(x) \in L^2(P_X)$,
    \begin{equation*}
        \E\left[\left(\frac{D}{e(X)} - \frac{1-D}{1-e(X)} - (bD+g(X))\right)\tilde{g}(X)\right] = 0,
    \end{equation*}
    and thus
    \begin{equation*}
        \E\left[\left(\frac{D}{e(X)} - \frac{1-D}{1-e(X)} - (bD+g(X))\right)|X\right] = 0 \text{ a.s. }
    \end{equation*}
    Because 
    \begin{equation*}
        \E\left[\frac{D}{e(X)} - \frac{1-D}{1-e(X)}\right] = 0,
    \end{equation*}
    it follows that 
    \begin{equation*}
        g(X) = -be(X) \text{ a.s. }
    \end{equation*}
    The FOC with respect to the scalar $b$ is, 
    \begin{equation*}
        \E\left[\frac{D^2}{e(X)}-\frac{D(1-D)}{1-e(X)}\right] - \E[D(bD+g(X))] = 0 \Rightarrow b\E[e(X)] = 1-\E[e(X)g(X)].
    \end{equation*}
    We substitute $g(X) = -be(X) \text{ a.s. }$ into the FOC for $b$ and get
    \begin{equation*}
        b\E[e(X)] = 1+b\E[e(X)^2] \Rightarrow b = \frac{1}{\E[e(X)(1-e(X))]} = \frac{1}{\E[(D-e(X))^2]}.
    \end{equation*}
    Therefore, the unique minimizer to \eqref{eqapp:RR_ATE_PL_proj} is 
    \begin{equation*}
        \frac{D-e(X)}{\E[(D-e(X))^2]}.
    \end{equation*}
    This shows that $\gamma_{\text{ATE, PL}}^*(d,x) = \frac{d-e(X)}{\E[(D-e(X))^2]}$ is exactly the $L^2$-projection of $\gamma_{\text{ATE, FULL}}^*(d,x) =\frac{D}{e(X)} - \frac{1-D}{1-e(X)}$ 
    onto the partially linear subspace $\mathcal{F}$.

\subsection{Proof for equation \eqref{eq:RR_proj_hetero}}
    The solution to the following optimization problem 
    \begin{equation}
         \min_{g \in \overline{\text{span}}\mathcal{F}^\diamond} \E[(g(Z) - \gamma^{*}(Z)\sigma(Z))^2]
    \end{equation}
    exists and is unique by the Hilbert projection theorem. We denote its minimizer as $\tilde\gamma$. The FOC with respect to $g$ is, for any function $\tilde{g} \in \overline{\text{span}}\mathcal{F}^\diamond$,
    \begin{equation}
        \E\left[(\tilde{\gamma}(Z) - \gamma^{*}(Z)\sigma(Z))\tilde{g}(Z)\right] = 0.
    \end{equation}
    Because $\E\left[h^\diamond(Z, f^\diamond)\right] = \E\left[\left(\gamma^{*}(Z)\sigma(Z)\right)f^\diamond(Z)\right]$ holds for any $f^\diamond(Z) \in \overline{\text{span}}\mathcal{F}^\diamond$, the FOC gives us 
    \begin{equation}
        \E\left[h^\diamond(Z, f^\diamond)\right] = \E\left[\left(\gamma^{*}(Z)\sigma(Z)\right)f^\diamond(Z)\right] = 
        \E\left[\tilde{\gamma}(Z)f^\diamond(Z)\right].
    \end{equation}
    By the Riesz representation theorem, the Riesz representer $\gamma^{\diamond*}$ for $\E\left[h^\diamond(Z, f^\diamond)\right]$ on the space $\overline{\text{span}}\mathcal{F}^\diamond$ should be unique. Therefore, 
    $\tilde\gamma = \gamma^{\diamond*}$

\section{Proofs for Asymptotic Results}\label{app:mainproofs}
\renewcommand{\theequation}{B.\arabic{equation}}
\setcounter{equation}{0}
\renewcommand{\thesubsection}{B.\arabic{subsection}}
\subsection{Proof for Theorem \ref{thm:consistency}}
The difference between the estimator and parameter of interest is 
    \begin{equation*}
        \begin{aligned}
            \hat{\psi}_{\text{linear}} - \psi(f) &= \frac{1}{n}\sum_{i=1}^n \hat{\gamma}_i Y_i - \frac{1}{n}\sum_{i=1}^n \gamma^*(Z_i) f(Z_i) + \left[\frac{1}{n}\sum_{i=1}^n\gamma^*(Z_i) f(Z_i) - \E(h(Z_i,f))\right] \\
            &= \frac{1}{n}\sum_{i=1}^n (\hat{\gamma}_i - \gamma^*(Z_i)) f(Z_i) + \left[\frac{1}{n}\sum_{i=1}^n\gamma^*(Z_i) f(Z_i) - \E(h(Z_i,f))\right] + \frac{1}{n}\sum_{i=1}^n \hat{\gamma}_i \varepsilon_i.
        \end{aligned}
    \end{equation*}
    For the first term, we have
    \begin{equation*}
        \frac{1}{n}\sum_{i=1}^n (\hat{\gamma}_i - \gamma^*(Z_i)) f(Z_i) \leq \sqrt{\frac{1}{n}\sum_{i=1}^n (\hat{\gamma}_i - \gamma^*(Z_i))^2 \cdot \frac{1}{n}\sum_{i=1}^n f(Z_i)^2} =o_p(1)\cdot O_p(1) = o_p(1)
    \end{equation*}
    since $f \in L^2(P_Z)$. For the second term, we have
    \begin{equation*}
        \frac{1}{n}\sum_{i=1}^n \gamma^*(Z_i) f(Z_i) - \E(h(Z_i,f)) = \frac{1}{n}\sum_{i=1}^n\gamma^*(Z_i) f(Z_i) - \E(\gamma^*f) \to_p 0
    \end{equation*}
    by Riesz representation theorem and WLLN. For the third term, we have 
    \begin{equation*}
        \frac{1}{n}\sum_{i=1}^n \hat{\gamma}_i \varepsilon_i = \frac{1}{n}\sum_{i=1}^n \gamma^*(Z_i) \varepsilon_i + \frac{1}{n}\sum_{i=1}^n (\hat{\gamma}_i - \gamma^*(Z_i)) \varepsilon_i \to_p 0.
    \end{equation*}
    Because $\varepsilon_i$ is mean independent of the regressors and $\gamma^* \in L^2(P_Z)$, it follows that $\E(\gamma^*(Z_i)\varepsilon_i) = 0$ and $\var(\gamma^*(Z_i)\varepsilon_i) \leq \E[(\gamma^*(Z_i))^2]\bar{\sigma}^2 < \infty$. 
    By WLLN, we have $\frac{1}{n}\sum_{i=1}^n \gamma^*(Z_i) \varepsilon_i \to 0$. The term $\frac{1}{n}\sum_{i=1}^n (\hat{\gamma}_i - \gamma^*(Z_i)) \varepsilon_i$ converges to 0 in probability since $\E(\varepsilon_i^2|Z_i) \leq \bar{\sigma}^2 <\infty$. 
    Thus we have 
    \begin{equation*}
        \hat{\psi}_{\text{linear}} - \psi(f) \to_p 0.
    \end{equation*}

\subsection{Proof for Theorem \ref{thm:root-n}}
We begin by proving a key lemma used in the overall proof procedure, following Section A.5.1 (p.16) and Section B (p.19-20) in the supplementary materials of \cite{hirshberg2021augmented}.
To characterize the size of a class $\mathcal{G}$, we will use its Rademacher complexity, $R_n(\mathcal{G})=\E\sup_{g\in\mathcal{G}}|n^{-1}\sum_{i=1}^n \epsilon_i g(Z_i)|$ where $\epsilon_i = \pm 1$ each with probability $1/2$ independently and independently of the sequence $Z_1,...,Z_n$,
as well as the uniform bound $M_\infty (\mathcal{G}) = \sup_{g\in \mathcal{G}}||g||_\infty
$.
\begin{lemma}
        Under the conditions in Theorem \ref{thm:root-n}, 
        there exists a sequence $r=o(n^{-1/4})$ and a constant $\eta_Q\in (0,1)$ such that, for any $\epsilon>0$, there exists $\eta_M>0$, condition (B.1) holds with probability at least $1-\epsilon-o(1)$,
        \begin{equation}
            \begin{aligned}
                \E_n f^2 \geq \eta_Q \E f^2 \quad &\text{when } \E f^2 \geq r^2, \\
                |M(f)| \leq \eta_M \E f^2 \quad &\text{when } \E f^2 \geq r^2, \\
                |M(f)| \leq \eta_M r^2 \quad &\text{when } \E f^2 \leq r^2,
            \end{aligned}
            \label{eq:condF}
        \end{equation}
        where $M(f):=(\E_n-\E)[\gamma^*f-h(\cdot,f)]$.
        \label{lem:Donskerclass}
    \end{lemma}

    \begin{proof}
        Step 1. By Corollary 3.3 in \cite{mendelson2020extending}, for any $\eta_Q \in (0,1)$, the event 
        \[\E_n f^2 \geq \eta_Q \E f^2 \text{ for all } f \in \mathcal{F} \text{ when } \E f^2 \geq r^2 \]
        holds
    with probability $(1-2\exp\{-c_2nr^2/M_\infty^2(\mathcal{F})\})$ if $R_n(\mathcal{F} \cap c_0rB) \leq c_1 r^2/M_\infty(\mathcal{F})$, where $c_0$, $c_1$, $c_2$ are constants that only depend on $\eta_Q$.
    By Lemma 3.4 and 3.2 in \citet{bartlett2005}, there is a unique positive $r_Q$ satisfying the fixed point condition $R_n(\mathcal{F} \cap c_0rB) \leq c_1 r^2/M_\infty(\mathcal{F})$ with equality and it is satisfied for all $r \geq r_Q$. Note that $M_\infty(\mathcal{F}):=\sup_{f\in \mathcal{F}}||f||_\infty$ is bounded by the uniform boundedness condition of $\mathcal{F}$ and $B$ is defined as the unit ball in $L^2(P)$.
    Because $\mathcal{F}$ is Donsker, the corresponding Rademacher processes are asymptotically equicontinuous in the sense that
    $\sqrt{n} R_n(\mathcal{F} \cap s_n B) \to 0$ when $s_n \to 0$.
    Applying Lemma 9 in \cite{hirshberg2021augmented} with $\tau_n(r)=R_n(\mathcal{F}\cap c_0 r B)$ yields a sequence $\bar{r}_n = o(n^{-1/4})$ for which 
    $R_n(\mathcal{F} \cap c_0 \bar{r}_n B) \leq c_1 \bar{r}_n^2/M_\infty(\mathcal{F})$ for sufficiently large $n$.
    By the definition of $r_Q$, it holds that $r_Q \leq \bar{r}_n$ and thus $r_Q = o(n^{-1/4})$. 

    Step 2. By Markov's inequality, with probability $1-\epsilon$, we have 
    \begin{equation*}
        \begin{aligned}
            \sup_{h \in h_{\gamma}(\cdot, \mathcal{F} \cap rB)} |(\E_n-\E)h| &\leq \frac{1}{\epsilon} \E\left\{ \sup_{h \in h_{\gamma}(\cdot, \mathcal{F} \cap rB)}|(\E_n-\E)h|\right\}, \\    
            & \leq \frac{2}{\epsilon} R_n(h_\gamma(\cdot,  \mathcal{F} \cap rB))
        \end{aligned}
    \end{equation*}
    where $h_{\gamma}(z, f):= \gamma^*(z) f(z) - h(z, f)$ and the second inequality is based on Lemma 2.3.1 (symmetrization) in \cite{vaart2023}. 
    By Lemma 3.4 and 3.2 in \citet{bartlett2005}, there is a unique positive $r_M$ satisfying the fixed point condition $2R_n(h_\gamma(\cdot,  \mathcal{F} \cap rB))/\epsilon \leq \eta_M r^2$ with equality and it is satisfied for all $r \geq r_M$, where $\eta_M>0$.
    In other words, the probability of $$\sup_{h \in h_{\gamma}(\cdot, \mathcal{F} \cap rB)} |(\E_n-\E)h| \leq \eta_M r_M^2,$$ where $r_M = \inf\{r>0: 2R_n(h_\gamma(\cdot,  \mathcal{F} \cap rB))/\epsilon \leq \eta_M r^2\}$, is $1-\epsilon$.
    For each linear map $\chi \in \{f \mapsto h(\cdot,f), f \mapsto \gamma^* f\}$ define 
    $\omega_\chi(r):= \sup_{f\in \mathcal{F}\cap r B} ||\chi(f)||_{L^2(P_Z)}$.
    Note that $h(\cdot, \mathcal{F}\cap rB) \subseteq h(\cdot, \mathcal{F}) \cap \omega_h(r)B$ and $\gamma^* (\mathcal{F} \cap rB) \subseteq \gamma^*\mathcal{F} \cap \omega_{\gamma^*}(r)B$. 
    It follows that $R_n(h(\cdot, \mathcal{F} \cap rB)) \leq R_n(
    h(\cdot, \mathcal{F}) \cap \omega_h(r)B
    )$ and $R_n(\gamma^* (\mathcal{F} \cap rB)) \leq R_n(\gamma^*\mathcal{F} \cap \omega_{\gamma^*}(r)B)$.
    Because $\mathcal{F}$, $h(\cdot, \mathcal{F})$ and $\gamma^*\mathcal{F}$ are Donsker, \citet{hirshberg2021augmented} showed at the beginning of the proof of their Theorem 1 (p.20 of their supplementary material) that 
    $\sqrt{n}R_n(\chi(\mathcal{F})\cap s_n B) \to 0$ as $s_n \to 0$ and $\lim_{r\to 0} \omega_\chi(r) = 0$ for $\chi \in \{\gamma^*f, h(\cdot, f)\}$.
    Thus, for any sequence $r \to 0$, $\sqrt{n}R_n(h(\cdot, \mathcal{F}\cap r B))\to 0$ and likewise for $\gamma^*(\mathcal{F}\cap rB)$. As $R_n(h_\gamma(\cdot, \mathcal{G})) \leq R_n(h(\cdot,\mathcal{G})) + R_n(\gamma^*\mathcal{G})$ for any $\mathcal{G}$, we have $\sqrt{n} R_n(h_\gamma(\cdot, \mathcal{F}\cap rB)) \to 0$.
    Now apply Lemma 9 in \cite{hirshberg2021augmented} with $\tau_n(r)=R_n(h_\gamma(\cdot,\mathcal{F}\cap r B))$ to obtain $r_M = o(n^{-1/4})$, using the same logic in Step 1.

    Step 3. Lemma 5 in \cite{hirshberg2021augmented} is a scaling device, showing that if $|M(f)| \leq \eta_M r^2$ for $\E f^2 \leq r^2$, then $|M(f)| \leq \eta_M \E f^2$ for $\E f^2 \geq r^2$.

    Step 4. Define a $\tilde{r}_n$ satisfying $\lim_{n\to \infty} \tilde{r}_n = 0$ and $n\tilde{r}_n^2 \to \infty$. From Steps 1-2, 
    $r_Q = o(n^{-1/4})$ and $r_M = o(n^{-1/4})$. Let $r:=\max\{r_Q, r_M, \tilde{r}_n\}$ and thus $r=o(n^{-1/4})$ and $nr^2 \to \infty$. The probability of the intersection of the three events in Steps 1-3 is at least $1-\epsilon -o(1)$.
    \end{proof}
    
We now begin the proof of Theorem \ref{thm:root-n}.
When the function class $\mathcal{F}$ is convex and centrosymmetric, the optimal weights $k(Z_i)$ that minimize \eqref{eq:onesidedCI_excess} for a given $\beta$, 
can be characterized by solving the modulus of continuity problem as shown in \cite{donoho1994statistical}. 
Specifically, given $\delta >0$, let $f^*_\delta$ solve
\begin{equation}
    \max_{f\in \mathcal{F}} \frac{2}{n} \sum_{i=1}^n h(Z_i,f) \quad s.t. \sum_{i=1}^n f(Z_i)^2 \leq \frac{\delta^2}{4}.
    \label{eq:optmfstar}
\end{equation}
The corresponding linear weights are given by 
\begin{equation}
    k_\delta(Z_i) = 2\omega_n'(\delta)f^*_\delta(Z_i)/\delta,
    \label{eq:optmweights}
\end{equation}
where 
\begin{equation}
    \omega_n(\delta) := \frac{2}{n}\sum_{i=1}^n h(Z_i,f^*_\delta),
\end{equation}
and the maximum bias of the linear estimator 
\begin{equation}
    \hat{\psi}_{\text{linear},\delta} = \sum_{i=1}^n k_{\delta}(Z_i) Y_i,
\end{equation} 
is given by
\begin{equation}
     \overline{\text{bias}}_{\mathcal{F},h} = (\omega_n(\delta)-\delta\omega_n'(\delta))/2.
     \label{eq:homo_worstbias}
\end{equation}
    Define $\tilde{f}^*_\delta$ as the solution to the problem 
\begin{equation}
    \max_{f\in \mathcal{F}} \frac{2}{n} \sum_{i=1}^n \gamma^*(Z_i)f(Z_i) \quad s.t. \sum_{i=1}^n f(Z_i)^2 \leq \frac{\delta^2}{4},
    \label{eq:optmftilde}
\end{equation}
and $\tilde{\omega}_n(\delta) := 2\sum_{i=1}^n \gamma^*(Z_i)\tilde{f}^*_\delta(Z_i)/n$. By the definitions, we have 
\begin{equation}
    \begin{aligned}
        \frac{1}{2}[\omega_n(\delta) - \tilde{\omega}_n(\delta)] &= \E_n(h(Z_i,f^*_\delta)) - \E_n(\gamma^*\tilde{f}^*_\delta)    \\
        &=\E_n(h(Z_i,f^*_\delta)) - \E_n(h(Z_i,\tilde f^*_\delta)) - \E_n(\gamma^*\tilde{f}^*_\delta - h(Z_i,\tilde f^*_\delta)) \\
        &=\E_n(h(Z_i,\Delta^*_\delta)) - M(\tilde f^*_\delta)
    \end{aligned}
    \label{eq:optmvalff}
\end{equation}
where $\Delta^*_\delta:=f^*_\delta-\tilde f^*_\delta$ and $M(f):=(\E_n-\E)[\gamma^*f - h(Z_i,f)]$. Because of the optimality of $\tilde f^*_\delta$, we have 
\begin{equation}
    0\leq \frac{1}{2}\tilde{\omega}_n(\delta) - \E_n \left(\gamma^*(Z_i)f^*_\delta(Z_i)\right) = \underbrace{-\E_n(h(Z_i,\Delta^*_\delta))}_{\leq 0 \text{ by the optimality of } f^*_\delta} - M(\Delta^*_\delta).
\label{eq:optmptff}
\end{equation}
Given the conditions of Theorem \ref{thm:consistency}, condition \eqref{eq:condF} is satisfied for $r = o(n^{-1/4})$ with probability approaching 1 by Lemma \ref{lem:Donskerclass}.
Note that $\Delta^*_\delta$ satisfies that $\Delta^*_\delta \in 2\mathcal{F}$ and $\E_n(\Delta^*_\delta)^2 \leq \delta^2/2n$.
We aim to first establish a upper bound for $\E_n(h(Z_i,g))$ for any function $g$ satisfying the following conditions: 
\begin{equation}
    g \in 2\mathcal{F}, \quad \E_n g^2 \leq \delta^2/2n, \quad \E_n(h(Z_i,g)) \geq 0, \quad \E_n(h(Z_i,g)) + M(g) \leq 0
    \label{eq:condFordelta/g}
\end{equation}
if $\mathcal{F}$ satisfies the condition \eqref{eq:condF}.

If $\E g^2 \geq 4r^2$, then $\E_n g^2 \geq \eta_Q \E g^2 \geq 4\eta_Q r^2$. Since $\E_n g^2 \leq \delta^2/2n$, we must have $\E g^2  \leq \delta^2/2\eta_Q \cdot 1/n$. 
Combining this upper bound on $\E g^2$ with the second line of \eqref{eq:condF} implies that $|M(g)| \leq (\delta^2\eta_M/4\eta_Q) \cdot 1/n$. To make sure the last two inequalities in the condition \eqref{eq:condFordelta/g} hold, 
it has to hold that $|\E_n (h(Z_i, g))| \leq |M(g)|\leq (\delta^2\eta_M/4\eta_Q) \cdot 1/n$.

If $\E g^2 \leq 4r^2$, then $|M(g)| \leq 2\eta_M r^2$ and thus it has to be $|\E_n (h(Z_i, g))| \leq |M(g)|\leq 2\eta_M r^2$ for the condition \eqref{eq:condFordelta/g} to hold.

Because $\mathcal{F}$ satisfies the condition \eqref{eq:condF} for $r = o(n^{-1/4})$ with probability approaching 1, we have $\sqrt{n}\E_n(h(Z_i,g)) \to_p 0$ for any function $g$ satisfying the condition \eqref{eq:condFordelta/g}. 
It follows that $\sqrt{n}\E_n(h(Z_i,\Delta^*_\delta)) \to_p 0$ by the fact that $\Delta^*_\delta$ satisfies the condition \eqref{eq:condFordelta/g}.
Following the similar logic, we could show that $\sqrt{n} M(\tilde f^*_\delta) \to_p 0$ by the facts that $\tilde f^*_\delta \in \mathcal{F}$ and $\E_n \tilde f^{*2}_\delta \leq \delta^2/4n$.
Equation \eqref{eq:optmvalff} gives us that 
\begin{equation}
    \sqrt{n}[\omega_n(\delta) - \tilde{\omega}_n(\delta)] \to_p 0.
    \label{eq:omega_rootnconverg}
\end{equation}
Equation \eqref{eq:omega_rootnconverg} holds for any value within a small neighborhood of $\delta$. Since $\gamma^*(Z_i) \in a\mathcal{F}$, the solution to the optimization problem \eqref{eq:optmftilde} is
\begin{equation}
    \tilde f^*_\delta(Z_i) = \frac{\delta}{2}\frac{\gamma^*(Z_i)}{\sqrt{\sum_{i=1}^n \gamma^*(Z_i)^2}},
    \label{eq:tildefstar}
\end{equation}
which gives 
\begin{equation}
    \tilde{\omega}_n(\delta) = \delta \sqrt{\E_n[\gamma^*(Z_i)^2]}/\sqrt{n},\quad \tilde{\omega}_n'(\delta) = \sqrt{\E_n[\gamma^*(Z_i)^2]}/\sqrt{n}.
    \label{eq:tildeomega}
\end{equation}
It follows that $\sqrt{n}\tilde{\omega}_n(\delta) \to_p \delta \sqrt{\E(\gamma^*(Z_i)^2)}$ since $\gamma^*(Z_i) \in L^2(P_Z)$.
Therefore, $\sqrt{n}\omega_n(\delta_0) \to_p \delta_0 \sqrt{\E(\gamma^*(Z_i)^2)}$ for all value $\delta_0$ in a small neighborhood of $\delta$.

The modulus function $\omega_n(\delta)$ is nondecreasing, nonnegative (by the definition) and concave on $[0,\infty)$. Because if $f^*_{\delta_1}$ attain the modulus at $\delta_1$ and similarly for $\delta_2$, then,
for $\lambda \in [0,1]$, $f^*_\lambda = \lambda f^*_{\delta_1} + (1-\lambda) f^*_{\delta_2}$ satisfies $\sum_{i=1}^n f^*_\lambda(Z_i)^2 \leq \frac{1}{4}[\lambda \delta_1+(1-\lambda)\delta_2]^2$
so that $\omega_n(\lambda \delta_1+(1-\lambda)\delta_2) \geq \lambda \omega_n(\delta_1)+ (1-\lambda)\omega_n(\delta_2)$.
This implies that the superdifferential of $\omega_n(\delta)$ at $\delta$, denoted as $\partial \omega_n(\delta)$, is non-empty for any $\delta>0$. Thus we let $\omega_n'(\delta)$
denote an element in the superdifferential set.
Lemma G.3. in the supplementary materials of \citet{armstrong2018optimal} implies $\sqrt{n}\omega_n'(\delta) \to_p \sqrt{\E(\gamma^*(Z_i)^2)}$, because $\delta \sqrt{\E(\gamma^*(Z_i)^2)}$ is nonnegative and concave on $\delta \in [0,\infty)$.
Recall that the maximum bias is given by $(\omega_n(\delta)-\delta\omega_n'(\delta))/2$, which can be rewritten by
\[\frac{1}{2}(\omega_n(\delta)-\delta\omega_n'(\delta)) = \frac{1}{2} \left[\delta \sqrt{\E[\gamma^*(Z_i)^2]}-\delta \cdot \sqrt{\E[\gamma^*(Z_i)^2]}\right] + o_p(1/\sqrt{n}) = o_p(1/\sqrt{n}).\]
Therefore, the maximum bias of the minimax linear estimator converges to 0 at a rate faster than $\sqrt{n}$.

Define $\rho_n:= \frac{4n}{\delta^2} \E_n[f_\delta^* \tilde{f}_\delta^*]$. Then by a simple decomposition,
\begin{equation*}
    \E_n\left[(f_\delta^* -\tilde{f}_\delta^*)^2\right] = \frac{\delta^2}{2n}(1-\rho_n)
\end{equation*}
Using equations \eqref{eq:tildefstar} and \eqref{eq:tildeomega}, we have 
\begin{equation*}
    \frac{1}{2}\tilde{\omega}_n(\delta)\rho_n = \frac{\delta}{2n}\sqrt{\sum_{i=1}^n \gamma^*(Z_i)^2} \cdot \frac{2}{\delta} \cdot \frac{\sum_{i=1}^n f^*_\delta(Z_i)\gamma^*(Z_i)}{\sqrt{\sum_{i=1}^n \gamma^*(Z_i)^2}} = \E_n[\gamma^*f^*_\delta].
\end{equation*}
It follows that 
\begin{equation*}
    \begin{aligned}
        1-\rho_n &= \frac{\frac{1}{2}\tilde{\omega}_n(\delta) - \E_n[\gamma^*f^*_\delta]}{\frac{1}{2}\tilde{\omega}_n(\delta)} = 
    \frac{-\E_n[h(Z,\Delta_\delta^*)]-M(\Delta_\delta^*)}{\frac{1}{2}\tilde{\omega}_n(\delta)} \\
    & \leq \frac{\sqrt{n}\left(|\E_n[h(Z,\Delta_\delta^*)]|+|M(\Delta_\delta^*)|\right)}{\sqrt{n}\tilde{\omega}_n(\delta)/2} \to_p 0.
    \end{aligned}
\end{equation*}
Therefore, $\E_n\left[(f_\delta^* -\tilde{f}_\delta^*)^2\right] = o_p(1/n)$.
Define $\hat{\gamma}^\delta_i = nk_\delta(Z_i) = \frac{2n\omega'_n(\delta)}{\delta}f^*_\delta(Z_i)$. Note that the Riesz representer can be rewritten as 
\(\gamma^*(Z_i) = \frac{2n\tilde{\omega}'_n(\delta)}{\delta}\tilde{f}^*_\delta(Z_i)\) by the expressions in \eqref{eq:tildefstar} and \eqref{eq:tildeomega}. We have 
\begin{equation*}
    \begin{aligned}
        \hat{\gamma}^\delta_i - \gamma^*(Z_i) &= \frac{2n}{\delta}\left[
        \omega'_n(\delta)f^*_\delta(Z_i) - \tilde{\omega}'_n(\delta)\tilde{f}^*_\delta(Z_i) \right] \\
        &= \frac{2n}{\delta}\left[
        (\omega'_n(\delta)-\tilde{\omega}'_n(\delta))f^*_\delta(Z_i) - \tilde{\omega}'_n(\delta)
        (\tilde{f}^*_\delta(Z_i)-f^*_\delta(Z_i)) \right].
    \end{aligned}
\end{equation*}
We square this difference and average over $i$ to get 
\begin{equation*}
    \begin{aligned}
        &\quad \frac{1}{n}\sum_{i=1}^n \left[\hat{\gamma}^\delta_i - \gamma^*(Z_i)\right]^2\\
        &= \frac{4n^2}{\delta^2} \cdot \frac{1}{n}\sum_{i=1}^n \left[
        (\omega'_n(\delta)-\tilde{\omega}'_n(\delta))f^*_\delta(Z_i) - \tilde{\omega}'_n(\delta)
        (\tilde{f}^*_\delta(Z_i)-f^*_\delta(Z_i)) \right]^2 \\
        &\leq \frac{8n^2}{\delta^2} \cdot \left\{
        (\omega'_n(\delta)-\tilde{\omega}'_n(\delta))^2 \cdot \frac{1}{n}\sum_{i=1}^n [f^*_\delta(Z_i)]^2 + [\tilde{\omega}'_n(\delta)]^2 \cdot \frac{1}{n}\sum_{i=1}^n
        [f^*_\delta(Z_i)-\tilde{f}^*_\delta(Z_i)]^2 \right\}.
    \end{aligned}
\end{equation*}
This converges in probability to zero because $\tilde{\omega}'_n(\delta) = O_p(1/\sqrt{n})$, $\omega'_n(\delta)-\tilde{\omega}'_n(\delta) = o_p(1/\sqrt{n})$ and constraints of the optimization problems \eqref{eq:optmfstar} and \eqref{eq:optmftilde}.

\subsection{Proof for Corollary \ref{coro:asynormal}}
By Theorem \ref{thm:root-n}, the maximum bias of $\hat{\psi}_{\text{linear},\delta}$ coverages to 0 at a rate faster than $\sqrt{n}$.
Therefore, the estimation error can be decomposed as:
\begin{equation}
    \begin{aligned}
        \hat{\psi}_{\text{linear},\delta} - \psi(f) &= \sum_{i=1}^n k_{\delta}(Z_i)f(Z_i) + \sum_{i=1}^n k_{\delta}(Z_i)\varepsilon_i - \E[h(Z_i,f)] \\
        &= \sum_{i=1}^n k_{\delta}(Z_i)\varepsilon_i + \left(\frac{1}{n}\sum_{i=1}^n h(Z_i,f)-\E[h(Z_i,f)]\right) \\
        &+ \left(\sum_{i=1}^n k_{\delta}(Z_i)f(Z_i) - \frac{1}{n}\sum_{i=1}^n h(Z_i,f)\right) \\
        &= \sum_{i=1}^n k_{\delta}(Z_i)\varepsilon_i + \left(\frac{1}{n}\sum_{i=1}^n h(Z_i,f)-\E[h(Z_i,f)]\right) + o_p(1/\sqrt{n}).
    \end{aligned}
    \label{eq:esterrordecomp_homo}
\end{equation}
Note that the optimal weights $k_{\delta}(Z_i)$, or equivalently, $\hat{\gamma}_{\delta}(Z_i)=nk_{\delta}(Z_i)$ satisfy \eqref{eq:fixeddetla_weightapprox}. 
Conditional on $Z_1,...,Z_n$, with the probability of $1-\epsilon$,
it holds that
\begin{equation}
    \begin{aligned}
        \left|\sum_{i=1}^n k_{\delta}(Z_i)\varepsilon_i - \frac{1}{n}\sum_{i=1}^n\gamma^*(Z_i)\varepsilon_i \right|&=\left| \frac{1}{n}\sum_{i=1}^n\left[\hat{\gamma}_{\delta}(Z_i)-\gamma^*(Z_i)\right]\varepsilon_i \right|\\
        & \leq \epsilon^{-1/2}n^{-1/2} \sqrt{\frac{1}{n}\sum_{i=1}^n \left[\hat{\gamma}_{\delta}(Z_i)-\gamma^*(Z_i)\right]^2 \bar{\sigma}^2}
    \end{aligned}
    \label{eq:noiseterm_converg}
\end{equation}
where the inequality follows from the Chebyshev's inequality.  Therefore, 
\begin{equation}
    \sqrt{n}\left|\sum_{i=1}^n k_{\delta}(Z_i)\varepsilon_i - \frac{1}{n}\sum_{i=1}^n\gamma^*(Z_i)\varepsilon_i \right| \to 0
\end{equation}
holds unconditionally with probability approaching 1, by \eqref{eq:HWwights_RP} and $\bar{\sigma}^2<\infty$.
Equation \eqref{eq:esterrordecomp_homo} can be further decomposed into 
\begin{equation}
    \begin{aligned}
        \hat{\psi}_{\text{linear},\delta} - \psi(f) 
        = \frac{1}{n}\sum_{i=1}^n \gamma^*(Z_i)\varepsilon_i + \left(\frac{1}{n}\sum_{i=1}^n h(Z_i,f)-\E[h(Z_i,f)]\right) + o_p(1/\sqrt{n}).
    \end{aligned}
\end{equation}
Because $\gamma^*(Z_i)\varepsilon_i + h(Z_i,f)$ is $i.i.d.$ across $i=1,...,n$, it holds that, by central limit theorem, 
\begin{equation}
    \sqrt{n}\left(\hat{\psi}_{\text{linear},\delta} - \psi(f)\right) \to_d \mathcal{N}(0, V)
\label{eq:appendix_asydist_homo}
\end{equation}
where $V=\E(\gamma^*(Z_i)^2\sigma^2(Z_i)) + \var(h(Z_i,f))$, which is equal to the semiparametric efficiency bound  if $\gamma^*(Z_i) \sigma^2(Z_i) \in \overline{\text{span}}\mathcal{F}$ (\cite{hirshberg2021augmented}).
Thus the linear estimator $\hat{\psi}_{\text{linear},\delta}$ is $\sqrt{n}$-consistent and asymptotically efficient.

\subsection{Proof for Theorem \ref{thm:se}}

    (1) We want to show that $n \cdot \sum_{i=1}^n k_{\delta}(Z_i)^2 \hat{\varepsilon}_i^2 \to_p \E(\gamma^*(Z_i)^2\sigma^2(Z_i))$.

It suffices to show that $\sum_{i=1}^n k_{\delta}(Z_i)^2 \hat{\varepsilon}_i^2/\sum_{i=1}^n k_{\delta}(Z_i)^2 \sigma(Z_i)^2 \to_p 1$ 
and $\frac{1}{n} \cdot \sum_{i=1}^n (\hat{\gamma}_{\delta}(Z_i)^2 - \gamma^*(Z_i)^2) \sigma(Z_i)^2 \to_p 0$.
The latter one can be easily shown because 
\begin{equation*}
    \begin{aligned}
        \left|\frac{1}{n} \cdot \sum_{i=1}^n (\hat{\gamma}_{\delta}(Z_i)^2 - \gamma^*(Z_i)^2) \sigma(Z_i)^2\right| &\leq \left[\frac{1}{n}\sum_{i=1}^n(\hat{\gamma}_{\delta}(Z_i) - \gamma^*(Z_i))^2\right]^{1/2}\cdot  \\
        &\quad \left[\frac{1}{n}\sum_{i=1}^n(\hat{\gamma}_{\delta}(Z_i) + \gamma^*(Z_i))^2\right]^{1/2}\bar{\sigma}^2,
    \end{aligned}
\end{equation*}
which converges to 0 in probability by the facts that $\frac{1}{n}\sum_{i=1}^n(\hat{\gamma}_{\delta}(Z_i) - \gamma^*(Z_i))^2 \to_p 0$ and $\frac{1}{n}\sum_{i=1}^n(\hat{\gamma}_{\delta}(Z_i) + \gamma^*(Z_i))^2 \leq 2 \cdot \left[ \frac{1}{n}\sum_{i=1}^n(\hat{\gamma}_{\delta}(Z_i) - \gamma^*(Z_i))^2 + \frac{1}{n}\sum_{i=1}^n(2\gamma^*(Z_i))^2\right]$ is bounded in probability.

Assume that there exists some $c>0$ such that $\max_{1\leq i \leq n} f^*_\delta(Z_i)^2 > c^2$ infinitely often. Let $\eta$ be small enough so that $\bar{g}(\eta)\leq c/2$. Then, for $n$ such that this holds and $k_n$ achieving this maximum,
\begin{equation*}
    \sum_{i=1}^n f^*_\delta(Z_i)^2 \geq \sum_{i=1}^n \frac{c^2}{4}I(d_Z(Z_i, Z_{k_n})\leq \eta) \to \infty.
\end{equation*}
This is a contradiction since $\sum_{i=1}^n f^*_\delta(Z_i)^2 \leq \delta^2/4$ and thus $\max_{1\leq i \leq n} f^*_\delta(Z_i)^2 \to 0$. Note that 
\begin{equation*}
    \frac{\sum_{i=1}^n k_{\delta}(Z_i)^2 \hat{\varepsilon}_i^2}{\sum_{i=1}^n k_{\delta}(Z_i)^2 \sigma^2(Z_i)} -1 = \frac{\sum_{i=1}^n f_\delta^*(Z_i)^2 \left[(f(Z_i)-\hat{f}(Z_i) + \varepsilon_i)^2 - \sigma^2(Z_i)\right]}{\sum_{i=1}^n f^*_{\delta}(Z_i)^2 \sigma^2(Z_i)}.
\end{equation*}
Because $\sum_{i=1}^n f^*_\delta(Z_i)^2 = \delta^2/4$ and the conditional variance function is lower bounded, it suffices to show the numerator, which can be rewritten as,
\begin{equation*}
    \sum_{i=1}^n f_\delta^*(Z_i)^2(\varepsilon_i^2 - \sigma^2(Z_i)) + \sum_{i=1}^n f_\delta^*(Z_i)^2(f(Z_i)-\hat{f}(Z_i))^2 + \sum_{i=1}^n f_\delta^*(Z_i)^2\varepsilon_i(f(Z_i)-\hat{f}(Z_i))
\end{equation*}
converges to 0 in probability. Specifically, the second term is bounded by $\max_{1\leq i\leq n} |\hat{f}(Z_i)-f(Z_i)|^2 \cdot \delta^2/4$, which converges to 0 in probability by condition (iii). 
Similarly, the last term is bounded by $\max_{1\leq i\leq n} |\hat{f}(Z_i)-f(Z_i)| \cdot (\sum_{i=1}^n f_\delta^*(Z_i)^2 |\varepsilon_i|)$, and the conditional expectation of the latter term is bounded.
Therefore, the last term converges to 0 in probability. Theorem 2 in \cite{von1965inequalities} shows that the absolute $(1+\eta)$ conditional moment of the first term could be bounded by a constant times 
\begin{equation*}
    \sum_{i=1}^n (f_\delta^*(Z_i)^2)^{1+\eta} \E[|\varepsilon_i^2 - \sigma^2(Z_i)|^{1+\eta}] \leq \left(\max_{1\leq i\leq n} f_\delta^*(Z_i)^2\right)^{\eta} \left(\sum_{i=1}^n f_\delta^*(Z_i)^2\right) \max_{1\leq i\leq n} \E[|\varepsilon_i^2 - \sigma^2(Z_i)|^{1+\eta}]
\end{equation*}
where converges to 0 since $\max_{1\leq i\leq n} f_\delta^*(Z_i)^2 \to 0$ as shown earlier. Thus the first term converges to 0 in probability.

(2) We want to show that $\frac{1}{n}\sum_{i=1}^n h(Z_i,\hat{f})^2 - \hat{\psi}_{\text{linear},\delta}^2 \to_p \var (h(Z_i,f)) = \E(h(Z_i,f)^2) - \psi(f)^2$.
We have 
\begin{equation*}
    \begin{aligned}
        &\left|\frac{1}{n}\sum_{i=1}^n h(Z_i,\hat{f})^2 - \hat{\psi}_{\text{linear},\delta}^2 - \E(h(Z_i,f)^2) + \psi(f)^2\right|\\
        &\leq \underbrace{\left|\frac{1}{n}\sum_{i=1}^n h(Z_i,\hat{f})^2 - h(Z_i,f)^2\right|}_{(a)} + \underbrace{\left|\frac{1}{n}\sum_{i=1}^n h(Z_i,f)^2-\E(h(Z_i,f)^2)\right| + \left| \hat{\psi}_{\text{linear},\delta}^2 - \psi(f)^2\right|}_{(b)}.
    \end{aligned}
\end{equation*}
Specifically, the second term $(b) = o_p(1)$ and the first term $(a)$ can be bounded by 
\begin{equation*}
    (a) \leq \left[\frac{1}{n}\sum_{i=1}^n(h(Z_i,\hat{f}) - h(Z_i,f))^2\right]^{1/2} \left[\frac{1}{n}\sum_{i=1}^n(h(Z_i,\hat{f}) + h(Z_i,f))^2\right]^{1/2}.
\end{equation*}
The conclusion follows since $\frac{1}{n}\sum_{i=1}^n(h(Z_i,\hat{f}) - h(Z_i,f))^2 \to_p 0$ and $\frac{1}{n}\sum_{i=1}^n(h(Z_i,\hat{f}) + h(Z_i,f))^2$ is bounded in probability.

\subsection{Proof for Theorem \ref{thm:heteroknownV}}
 Theorem \ref{thm:root-n} can be directly applied because those conditions also hold for $\mathcal{F}^\diamond$, $h^\diamond(\cdot, \mathcal{F}^\diamond)$ and $\gamma^{\diamond *}$. 
    In other words, the worst-case bias converges to 0 at a rate faster than root-$n$ and it holds that
    \begin{equation}
        \frac{1}{n}\sum_{i=1}^n \left[\hat{\gamma}^\diamond_{\delta}(Z_i) - \gamma^{\diamond *} \right]^2 = o_p(1),
    \end{equation}
    where $\hat{\gamma}^\diamond_{\delta}(Z_i) = k^\diamond_{\delta}(Z_i)/n$.
    The estimation error can be decomposed as: 
    \begin{equation}
        \begin{aligned}
            \hat{\psi}^\diamond_{\text{linear}, \delta} - \psi(f) &= \sum_{i=1}^n k^\diamond_{\delta}(Z_i)(f^\diamond(Z_i) + \varepsilon_i^\diamond)-\E[h(Z_i,f)] \\
            &= \sum_{i=1}^n k^\diamond_{\delta}(Z_i) \varepsilon_i^\diamond + \left(\frac{1}{n}\sum_{i=1}^n h(Z_i,f) - \E[h(Z_i,f)]\right) + o_p(1/\sqrt{n})
        \end{aligned}
        \label{eq:esterrordecomp_hetero_hetero}
    \end{equation}
     By the condition of $0<c_1\leq \sigma^2(\cdot) \leq c_2 < \infty$, it follows that 
    \begin{equation}
        \frac{1}{n}\sum_{i=1}^n \frac{\left[\hat{\gamma}^\diamond_{\delta}(Z_i) - \gamma^{\diamond *} \right]^2}{\sigma^2(Z_i)} \leq \frac{1}{c_1} \cdot \frac{1}{n}\sum_{i=1}^n \left[\hat{\gamma}^\diamond_{\delta}(Z_i) - \gamma^{\diamond *} \right]^2 = o_p(1).
    \end{equation}
    The equation \eqref{eq:esterrordecomp_hetero_hetero} can be further decomposed into 
    \begin{equation}
        \begin{aligned}
            \hat{\psi}^\diamond_{\text{linear}, \delta} - \psi(f)
            = \sum_{i=1}^n \gamma^{\#}(Z_i) \varepsilon_i + \left(\frac{1}{n}\sum_{i=1}^n h(Z_i,f) - \E[h(Z_i,f)]\right) + o_p(1/\sqrt{n}),
        \end{aligned}
    \end{equation}
    where $\gamma^{\#}(\cdot) = \gamma^{\diamond *}(\cdot)/\sigma(\cdot)$, by the Chebyshev's inequality.
    Finally we could show
    \begin{equation}
        \sqrt{n}\left(\hat{\psi}^\diamond_{\text{linear}, \delta} - \psi(f)\right) \to_d \mathcal{N}(0,V)
        \label{eq:asydist_hetero_hetero}
    \end{equation}
    where $V = \E(\gamma^\#(Z_i)^2\sigma^2(Z_i))+\var(h(Z_i,f))$.
    The efficiency of the minimax linear estimator, which is asymptotically linear and represented by the formula:
    \[
    \frac{1}{n}\sum_{i=1}^n\left(\gamma^\# \varepsilon_i + \frac{1}{n}\sum_{i=1}^n h(Z_i,f) - \E[h(Z_i,f)]\right),
    \]
    can be demonstrated by following the methodology outlined in Appendix B.1 of \citet{hirshberg2021augmented}. This proof requires substituting \(\gamma^*\) with \(\gamma^\#\), under the assumption that there exists a representer \(\gamma^\#\) such that \(\gamma^\#(\cdot)\sigma^2(\cdot) \in \overline{\text{span}}\mathcal{F}\) holds true.

\section{Uniform Inference}
\renewcommand{\theequation}{C.\arabic{equation}}
\renewcommand{\thetheorem}{C.\arabic{theorem}}
\renewcommand{\thelemma}{C.\arabic{lemma}}
\setcounter{equation}{0}
\setcounter{theorem}{0}
\setcounter{lemma}{0}
Fix $P_Z$ for the marginal distribution of $Z$.
For each $f \in \mathcal{F}$ and $Q \in \mathcal{Q}$, let $P_{f,Q}$ denote the joint law of $W = (Y,Z)$ generated by 
\begin{equation*}
    Y = f(Z) + \varepsilon, \quad \varepsilon|Z \sim Q,\quad \E_{Q}[\varepsilon|Z] = 0, \quad \var_{Q}[\varepsilon|Z]=\sigma^2(Z).
\end{equation*}
Let $\{W_i\}_{i=1}^n$ be i.i.d. draws from $P_{f,Q}$. A sequence of confidence sets $\mathcal{C}_n$ is said to have asymptotic coverage at least $1-\alpha$ uniformly over $\mathcal{F}$ and $\mathcal{Q}$ if 
\begin{equation}
    \liminf_{n\to \infty} \inf_{f\in \mathcal{F},\ Q\in \mathcal{Q}} P_{f,Q}\left[\psi(f) \in \mathcal{C}_n\right] \geq 1-\alpha.
\end{equation}
This section states sufficient conditions under which the usual standard-error-based CI centered at $\hat{\psi}_{\text{linear},\delta}$ attains uniform coverage $1-\alpha$ over $\mathcal{F}$ and $\mathcal{Q}$ asymptotically. We use the notation $\overset{d}{\underset{\mathcal{F},\mathcal{Q}}{\to}}$ to mean convergence in distribution uniformly over $f \in \mathcal{F}$ and $Q\in \mathcal{Q}$, and similarly for $\overset{p}{\underset{\mathcal{F},\mathcal{Q}}{\to}}$.
\begin{theorem}
    Assume the conditions of Theorem \ref{thm:root-n} and in addition: (i) $\sigma^2(Z) \leq \bar{\sigma}^2$ for all $Z \in \mathcal{Z}$ and all $Q \in \mathcal{Q}$; (ii) the uniform central limit theorem holds:
    \begin{equation}
        \sqrt{n}\cdot \frac{\sum_{i=1}^n \rho_{f,Q}(W_i)/n -\E_{f,Q}[\rho_{f,Q}(W_i)]}{\sqrt{V_{f,Q}}} \overset{d}{\underset{\mathcal{F},\mathcal{Q}}{\to}} \mathcal{N}(0,1)
        \label{eq:unif_CLT}
    \end{equation}
    where $V_{f,Q}:=\var_{f,Q}[\rho_{f,Q}]$ and $\rho_{f,Q}$ is defined in \eqref{eq:rhor1r2}; (iii) there is an estimate $\hat{se}_{f,Q}$ such that $\frac{\hat{se}_{f,Q}}{\sqrt{V_{f,Q}/n}}\overset{p}{\underset{\mathcal{F},\mathcal{Q}}{\to}} 1$. Then the CI centered at $\hat{\psi}_{\text{linear},\delta}$
    \begin{equation*}
        \mathcal{C}_n = [\hat{\psi}_{\text{linear},\delta}-\hat{se}_{f,Q} z_{1-\alpha/2},\  \hat{\psi}_{\text{linear},\delta}+\hat{se}_{f,Q} z_{1-\alpha/2}],
    \end{equation*}
    achieves uniform asymptotic coverage:
    \begin{equation}
    \liminf_{n\to \infty} \inf_{f\in \mathcal{F}, Q\in \mathcal{Q}} 
    P_{f,Q} \left[ 
    \psi(f) \in \mathcal{C}_n
    \right] \geq 1-\alpha.
    \label{eq:unif_CI}
\end{equation}
\end{theorem}
\begin{proof}
    By the definition of $\hat{\psi}_{\text{linear},\delta}$ and the error decomposition in \eqref{eq:esterrordecomp_homo},
\begin{equation}
    \hat{\psi}_{\text{linear},\delta} - \psi(f) =\frac{1}{n} \sum_{i=1}^n \rho_{f,Q}(W_i) + r_{1,n} + r_{2,n},
\end{equation}
where
\begin{equation}
    \begin{aligned}
        \rho_{f,Q}(W_i) &= \gamma^*(Z_i)\varepsilon_i + h(Z_i,f)-\E[h(Z_i,f)], \\
        r_{1,n} &= \sum_{i=1}^n k_{\delta}(Z_i)f(Z_i) - \frac{1}{n}\sum_{i=1}^n h(Z_i,f), \\
        r_{2,n} &= 
        \sum_{i=1}^n k_{\delta}(Z_i)\varepsilon_i - \frac{1}{n}\sum_{i=1}^n \gamma^*(Z_i)\varepsilon_i.
    \end{aligned}
    \label{eq:rhor1r2}
\end{equation}
By Theorem \ref{thm:root-n}, 
\begin{equation}
    \sqrt{n}r_1 = o_p(1) \quad \text{uniformly over $f\in \mathcal{F}$.}
    \label{eq:unif_r1}
\end{equation}
Conditional on $Z_1,...,Z_n$, under $f_n \in \mathcal{F}$ and $Q_n \in \mathcal{Q}$, with the probability of $1-\epsilon$, \eqref{eq:noiseterm_converg} holds by the Chebyshev's inequality.  It follows that
\begin{equation}
    \sqrt{n}\left|\sum_{i=1}^n k_{\delta}(Z_i)\varepsilon_i - \frac{1}{n}\sum_{i=1}^n\gamma^*(Z_i)\varepsilon_i \right| \overset{p}{\to} 0
\end{equation}
holds unconditionally for any $f_n \in \mathcal{F}$ and $Q_n \in \mathcal{Q}$ by the uniform bounded variance assumption (i) and 
$\frac{1}{n}\sum_{i=1}^n \left[\hat{\gamma}_{\delta}(Z_i)-\gamma^*(Z_i)\right]^2 = o_p(1)$.
Therefore, 
\begin{equation}
    \sqrt{n}r_2 = o_p(1) \quad \text{uniformly over $f\in \mathcal{F}$ and $Q\in \mathcal{Q}$.}
    \label{eq:unif_r2}
\end{equation}
Equation \eqref{eq:unif_CI} follows from \eqref{eq:unif_r1}, \eqref{eq:unif_r2} and conditions (ii) and (iii).
\end{proof}
Condition (ii) will follow by applying Lemma \ref{lem:suffcond_UCLT} to show convergence under arbitrary sequences $f_n \in \mathcal{Q}$ and $Q_n \in \mathcal{Q}$ as long as \eqref{eq:LindcondforUCLT} holds for every such sequence $(f_n, Q_n)$.
\begin{lemma}
\label{lem:suffcond_UCLT}
    Let $W_{1,n},...,W_{n,n}$ be a triangular array of $i.i.d.$ random variables with a distribution $P_{f_n, Q_n}$ where $f_n \in \mathcal{F}$ and $Q_n \in \mathcal{Q}$. Suppose there exists $K>0$ and $\eta>0$ such that for all $n$ and all $i\leq n$,
    \begin{equation}
        \E_{f_n, Q_n}\left[
        \frac{|\phi_{f_n, Q_n}(W_{i,n}) - \E_{f_n, Q_n}[\phi_{f_n, Q_n}(W_{i,n})]|^{2+\eta}}{V_{f_n, Q_n}^{1+\eta/2}}
        \right] \leq K.
        \label{eq:LindcondforUCLT}
    \end{equation}
    Then under $(f_n, Q_n)$,
    \begin{equation}
    \sqrt{n}\cdot \frac{\sum_{i=1}^n \phi_{f_n, Q_n}(W_{i,n})/n -\E_{f_n, Q_n}[\phi_{f_n, Q_n}(W_{i,n})]}{\sqrt{V_{f_n, Q_n}}} \overset{d}{\to} \mathcal{N}(0,1).
\end{equation}
\end{lemma}
\begin{proof}
    The conclusion is a direct application of Lemma 11.4.1 in \citet{lehmann2005testing}.
\end{proof}

\section{Estimating the Average Treatment Effect on Treated}\label{app:ATT}
\renewcommand{\theequation}{D.\arabic{equation}}
\setcounter{equation}{0}
Let $Z = (D,X)$ where $D \in \{0,1\}$ is a binary treatment indicator and $X$ is a univariate observed covariate. In the potential outcomes framework, each unit has a pair of potential responses $\{Y(1), Y(0)\}$ and we observe only $Y = D Y(1) + (1-D)Y(0)$. Define the conditional mean $f(d,x) = \E[Y|D=d,X=x]$. We assume that potential outcomes $(Y(1), Y(0))$ are mean independent of the treatment indicator $D$ conditional on the control covariate $X$ (i.e., unconfoundedness assumption). Under this assumption, it holds that 
\begin{equation}
    f(d,x) = \E[Y|D=d,X=x] = \E[Y(d)|X=x].
\end{equation}
We are interested in estimating the ATT:
\begin{equation}
    \tau = \frac{\psi(f)}{p} = \frac{\E[D(f(1,X)-f(0,X))]}{\E[D]},
\end{equation}
where $\psi(f) = \E[h(Z, f)]$ is a linear functional on $f$ with 
$h(z, f) = d(f(1,x)-f(0,x))$, and $p=\E[D]$ is the marginal treatment probability.
The function 
\begin{equation}
    \gamma^*(D, X) = D + (1-D)\frac{e(X)}{1-e(X)},
\end{equation}
satisfies $\psi(f) = \E[\gamma^*(D,X)f(D,X)]$ for any square integrable function $f$.

Let $\mathcal{F}$ be a Lipschitz function class with constant $C$ and one covariate:
\begin{equation}
    \mathcal{F} = \{f: |f(d,x) - f(d,x')| \leq C|x-x'|, \text{ for } d=0,1\}
\end{equation}
Let $\hat\psi_\delta$ be the minimax linear estimator  with weights $k_\delta$ solving the optimization problem 
\begin{equation}
    \min_{\{k(Z_i)\}_{i=1}^n} \left\{\sup_{f\in \mathcal{F}}\left[
    \sum_{i=1}^n k(Z_i)f(Z_i) - \frac{1}{n} \sum_{i=1}^n D_i(f(1,X_i)-f(0,X_i))
    \right] + \delta \sqrt{\sum_{i=1}^n k(Z_i)^2}\right\}
\end{equation}
and let $\hat{p} = n_1/n$ with $n_1= \sum_{i=1}^n D_i$, which is a (minimax) linear estimator for $p$.
It is straightforward to obtain 
\begin{equation}
    \hat{p} - p = \frac{1}{n}\sum_{i=1}^n D_i - p.
    \label{eq:IF_p}
\end{equation}
Corollary \ref{coro:asynormal} tells us that, if $f \in a\mathcal{F}$ and $\gamma^* \in a \mathcal{F}$, then the minimax linear estimator $\hat\psi_\delta$ is asymptotically linear:
\begin{equation}
    \begin{aligned}
        &\quad \hat{\psi}_\delta - \psi \\
    &= \frac{1}{n}\sum_{i=1}^n 
    \left[D_i(Y_i-f(1,X_i)) - \frac{(1-D_i)e(X_i)}{1-e(X_i)}(Y_i-f(0,X_i)) + 
    D_i(f(1,X_i)-f(0,X_i))\right] - \psi \\
    &+o_p(1/\sqrt{n}).
    \end{aligned}
    \label{eq:IF_psi}
\end{equation}
Equations \eqref{eq:IF_p} and \eqref{eq:IF_psi} give that 
\begin{equation}
    \sqrt{n}\begin{pmatrix}
        \hat{\psi}_\delta - \psi \\
        \hat{p} - p
    \end{pmatrix} \to_d 
    \mathcal{N}\left(
\boldsymbol{0},
\Sigma
    \right)
\end{equation}
with
\begin{equation*}
    \Sigma = \begin{pmatrix}
        \sigma_\psi^2 & \sigma_{\psi,p} \\
        \sigma_{\psi,p} & \sigma^2_p
    \end{pmatrix},
\end{equation*}
where 
\begin{align*}
    \sigma_\psi^2 &= \E\left[e(X)\sigma^2(1,X) + \frac{e(X)^2\sigma^2(0,X)}{(1-e(X))}+\beta(X)^2 e(X) - \psi^2\right], \\
    \sigma^2_p &= p(1-p),\\
    \sigma_{\psi,p} &= \E\left[\beta(X)e(X)(1-p)\right], \\
    \beta(X) &= f(1,X) - f(0,X), \ \sigma^2(d,x) = \E[(Y(d)-f(d,X))^2|X=x].
\end{align*}
Because $\tau = g(\psi, p) = \frac{\psi}{p}$ where $p >0$, by the Delta method, we have 
\begin{equation}
    \sqrt{n}(g(\hat\psi, \hat p) - g(\psi, p)) \to_d \mathcal{N}(0, V),
\end{equation}
where $V = \nabla  g^\top \Sigma \nabla  g$ and $\nabla  g^\top = (1/p, -\psi/p^2)$.
Therefore, 
\begin{align*}
    V &= \frac{1}{p^2} \sigma^2_\psi + \frac{\psi^2}{p^4} \sigma_p^2 - \frac{2\psi}{p^3} \sigma_{\psi,p} \\
    &= \frac{1}{p^2} \tau^2_\psi + \frac{\tau^2}{p^2} \sigma_p^2 - \frac{2\tau}{p^2} \sigma_{\psi,p}.
\end{align*}
by the fact that $\tau = \psi/p$.
Because
\begin{align*}
    I_1 &:= \frac{1}{p^2} \sigma^2_\psi = 
    \E\left[\frac{e(X)\sigma^2(1,X)}{p^2} + \frac{e(X)^2\sigma^2(0,X)}{p^2(1-e(X))}+\frac{\beta(X)^2 e(X)}{p^2} - \tau^2\right], \\
    I_2 &:= \frac{\tau^2}{p^2} \sigma_p^2 = 
    \frac{\tau^2 (1-p)}{p},\\
    I_3 &:= \frac{\tau}{p^2} \sigma_{\psi,p} = \E\left[
        \frac{\tau \beta(X)e(X)(1-p)}{p^2} 
        \right] = 
        \frac{\tau^2(1-p)}{p},
\end{align*}
we can further simplify $V$ by 
\begin{equation}
\begin{aligned}
    V &= I_1 + I_2 - 2I_3 \\
    &=\E\left[\frac{e(X)\sigma^2(1,X)}{p^2} + \frac{e(X)^2\sigma^2(0,X)}{p^2(1-e(X))}\right]
    + 
    \frac{1}{p^2}\E\left[
        \beta(X)^2 e(X) - p^2 \tau^2 - \tau^2 p(1-p) 
    \right] \\
    &= \E\left[\frac{e(X)\sigma^2(1,X)}{p^2} + \frac{e(X)^2\sigma^2(0,X)}{p^2(1-e(X))}\right]
    + 
    \frac{1}{p^2}\E\left[
        \beta(X)^2 e(X) - \tau^2 p
    \right].
    \label{eq:var_estimator}
    \end{aligned}
\end{equation}
Let $V_{SEB}$ denotes the semiparametric efficiency bound (SEB) for ATT (e.g. \citealp{hahn1998role}):
\begin{equation}
        V_{SEB} = \E\left[\frac{e(X)\sigma_1^2(X)}{p^2} + \frac{e(X)^2\sigma_0^2(X)}{p^2(1-e(X))}+\frac{(\beta(X) - \tau)^2 e(X)}{p^2}\right].\,
    \label{eq:SEBforATT_Hahn}
\end{equation}
Then we compare $V$ in \eqref{eq:var_estimator} with $V_{SEB}$ in \eqref{eq:SEBforATT_Hahn} by 
\begin{align*}
    V- V_{SEB}
    &= \frac{1}{p^2}\E\left[
        \beta(X)^2 e(X) - \tau^2 p  - 
        \left(
            \beta(X)^2 e(X) + \tau^2 e(X) - 2\tau \beta(X)e(X)
        \right)
    \right] \\
    &=\frac{1}{p^2}\E\left[
        - \tau^2 p  - \tau^2 p +  2\tau^2 p
    \right] = 0
\end{align*}
Therefore, the estimator $\frac{\hat{\psi}_\delta}{\hat{p}}$ is $\sqrt{n}$-consistent and asymptotically normally distributed, with asymptotic variance equal to the SEB.

It is worthy to note that the estimator $\frac{\hat{\psi}_\delta}{\hat{p}}$ is equivalent to a minimax linear estimator $\tilde{\tau}_\delta$ with weights $\tilde{k}_\delta$ solving the following problem:
\begin{equation}
    \min_{\{k(Z_i)\}_{i=1}^n} \left\{\sup_{f\in \mathcal{F}}\left[
    \sum_{i=1}^n k(Z_i)f(Z_i) - \frac{1}{n_1} \sum_{i=1}^n D_i(f(1,X_i)-f(0,X_i))
    \right] + \delta \sqrt{\sum_{i=1}^n k(Z_i)^2}\right\}.
\end{equation}
Therefore, in the simulation section, we consider the performance of this (feasible) minimax linear estimator $\tilde{\tau}_\delta$ with its weights $\tilde{k}_\delta$ under different scenarios.

To estimate the asymptotic variance of the minimax linear estimator, we decompose it into two components. The first term (conditional variance) is 
$$\E\left[\frac{e(X)\sigma_1^2(X)}{p^2} + \frac{e(X)^2\sigma_0^2(X)}{p^2(1-e(X))}\right].$$
Under the conditions of Theorem \ref{thm:se} and since $n_1/n \to_p p$ with $\epsilon\leq p \leq 1-\epsilon, \epsilon>0$, this component can be consistently estimated by 
$$n \cdot \sum_{i=1}^n \tilde{k}_\delta^2\hat\varepsilon_i^2.$$
The second term (marginal variance) is 
$$\frac{1}{p^2}\E\left[(\beta(X) - \tau)^2 e(X))\right],$$ 
or equivalently $$\frac{1}{p^2}\E\left[
        \beta(X)^2 e(X) - \tau^2 p
    \right].$$ 
Under conditions of Theorem \ref{thm:se} and $n_1/n \to_p p$, a consistent estimator is 
$$\frac{n}{n_1^2}\sum_{i=1}^n D_i (\hat{f}(1,X_i)-\hat{f}(0,X_i))^2 - \frac{n}{n_1}\tilde{\tau}_\delta^2.$$ Alternatively, one may use the nearest‐neighbor estimator described in Theorem 7 of \citet{abadie2006large} to estimate the second component.

\section{Estimating Nonlinear Functional of Multiple Regression Functions}\label{app:nonlinear}
\renewcommand{\theequation}{E.\arabic{equation}}
\setcounter{equation}{0}
We observe a random sample $\{(Z_{1,i},Y_{1,i},...,Z_{K,i},Y_{K,i})\}_{i=1}^n$.
Define $\bar{Z}_i := (Z_{1,i},...,Z_{K,i})$.
For each $k = 1,...,K$, let the regression function
\begin{equation}
    f_k(z) := \E[Y_k|Z_k=z], \quad z \in \mathcal{Z}_k,
\end{equation}
belong to a function space $\mathcal{F}_k$. For each $k = 1,...,K$, the map $f_k \mapsto h_k(z,f_k)$ is linear for each fixed $z\in \mathcal{Z}_K$ and we define
\begin{equation}
    \psi_k = \E[h_k(Z_k, f_k)].
\end{equation}
Let $g: \mathbb{R}^K \to \mathbb{R}$ be continuously differentiable. The parameter of interest is
\begin{equation}
    \varphi = g(\psi_1,...,\psi_K).
\end{equation}
Under certain conditions (for example, the conditions in Corollary \ref{coro:asynormal}), for each $k=1,...,K$, the minimax linear estimator $\hat{\psi}_k$ has an asymptotically linear representation 
\begin{equation}
    \hat{\psi}_k - \psi_k = \frac{1}{n}\sum_{i=1}^n \zeta_k(Y_{k,i}, Z_{k,i}) + o_p(1/\sqrt{n}),
\end{equation}
where 
\begin{equation}
    \zeta_k(Y_{k,i}, Z_{k,i}) = 
    \gamma^*_k(Z_{k,i})(Y_{k,i} - f_k(Z_{k,i})) + h(Z_{k,i}, f_k(Z_{k,i})) - \psi_k.
\end{equation}
It follows that 
\begin{equation}
    \sqrt{n}\left(\begin{pmatrix}
        \hat\psi_1\\
        \vdots \\
        \hat\psi_K
    \end{pmatrix} - 
    \begin{pmatrix}
        \psi_1\\
        \vdots \\
        \psi_K
    \end{pmatrix}
    \right) \to_d \mathcal{N}(\boldsymbol{0}, \Sigma),
\end{equation}
where $\Sigma = \var(\boldsymbol{\zeta}(\bar Y_i, \bar Z_i))$ and $\boldsymbol{\zeta}(\bar Y_i, \bar Z_i) = (\zeta_1(Y_{1,i}, Z_{1,i}),...,\zeta_K(Y_{K,i}, Z_{K,i}))^\top$.
Define $\hat{\varphi} = g(\hat{\psi}_1,...,\hat{\psi}_K)$.
Because $g$ is differentiable, the Delta method gives 
\begin{equation}
    \sqrt{n}(\hat{\varphi} - \varphi) \to_d \mathcal{N}(0, V_\varphi),
\end{equation}
where $V_\varphi = \nabla g(\psi_1,...,\psi_K)^\top \Sigma \nabla g(\psi_1,...,\psi_K) $.
\renewcommand{\theexample}{E.\arabic{example}}
\setcounter{example}{0}
\begin{example} (ATT)
The ATT example in \ref{app:ATT} shows the construction when 
\(K=2\) with \(Z_{1,i}=(D_i,X_i)\), \(Y_{1,i}=Y_i\) for the outcome
functional and \(Z_{2,i}=1\), \(Y_{2,i}=D_i\) for the treatment–probability functional.
\end{example}

\begin{example}
    (LATE) We observe $\{Y_i, W_i, D_i, X_i)\}_{i=1}^n$,
    where $W_i \in \{0,1\}$ is an instrument (e.g. eligibility letter), $D_i \in \{0,1\}$ is the treatment actually taken and $X_i$ are observed covariates. Potential outcomes are $(D_i(0), D_i(1), Y_i(0), Y_i(1))$.  Define the principal stratum $S=(D(1),D(0)) \in \{(1,1), (1,0), (0,0), (0,1)\}$ and label the four cases by 
always-taker (AT), complier (CP), never-taker (NT), defier (DF). We assume 
\begin{itemize}
    \item (E.i) (Unconfoundedness) $(D_i(0), D_i(1), Y_i(0), Y_i(1)) \indep W_i |X_i$;
    \item (E.ii) (Monotonicity) $P(D_{i}(0)>D_i(1)) = 0$, i.e., the measure of defiers is zero;
    \item (E.iii) (Existence of compliers) $P(D_i(1)>D_i(0)) >0$;
    \item (E.iv) (Positivity) $P(W_i = 1|X_i=x) \in (\epsilon, 1-\epsilon)$ for any $x \in \mathcal{X}$ and $\epsilon>0$. 
\end{itemize}
Let $Z_{1,i} = (W_i, X_i)$, $Y_{1,i} = Y_i$, $Z_{2,i} = (W_i, X_i)$, $Y_{2,i} = D_i$, and 
\begin{align*}
    f_1(w,x) &:= \E[Y|W=w, X=x], \ \varepsilon_{1,i} = Y_i - f_1(W_i,X_i)\\
    f_2(w,x) &:= \E[D|W=w, X=x], \ \varepsilon_{2,i} = D_i - f_2(W_i,X_i) \\
    e(x) &:= P[W=1|X]
\end{align*}
Our target parameter is 
\begin{equation}
    \varphi = \frac{\psi_1}{\psi_2},\quad \psi_k = \E[f_k(1,X) - f_k(0,X)], \ k=1,2,
\end{equation}
which is the LATE for compliers under Assumptions (D.i) - (D.iv). Both linear functional have the same Riesz representer (if it belongs to $\mathcal{F}_1$ and $\mathcal{F}_2$ up to some constant):
\begin{equation}
    \gamma^*_1(w,x) = \gamma^*_2(w,x) = \frac{w}{e(x)} - \frac{1-w}{1-e(x)}.
\end{equation}
Because $\psi_2>0$, by the Delta method, we have 
\begin{equation}
    \sqrt{n}\left(
    \frac{\psi_1}{\psi_2}-\frac{\hat\psi_1}{\hat\psi_2}\right) \to_d \mathcal{N}(0, V_{LATE}),
\end{equation}
where 
\begin{align*}
     V_{LATE}&= \frac{1}{\psi_2^2}\E\left[
            \frac{\sigma_1^2(1,X)-2\varphi \sigma_{12}(1,X)+\varphi^2 \sigma_2^2(1,X)}{e(X)}
        \right] \\
        & +
        \frac{1}{\psi_2^2}\E\left[
            \frac{\sigma_1^2(0,X)-2\varphi \sigma_{12}(0,X)+\varphi^2 \sigma_2^2(0,X)}{1-e(X)}
        \right] \\
        &+ \frac{1}{\psi_2^2}\E\left[
            \left(f_1(1,X)-f_1(0,X) - \varphi (f_2(1,X)-f_2(0,X)) \right)^2
        \right],\\
        \sigma_1^2(w,x) &= \var[Y|W=w,X=x], \\ 
    \sigma_2^2(w,x) &= \var[D|W=w,X=x], \\  
    \sigma_{12}(w,x) &= \cov(Y,D|W=w,X=x).
\end{align*}
This variance achieves the SEB for estimating LATE (e.g. \citealp{frolich2007nonparametric}).

\end{example}
In addition, one may consider a class of nonlinear functional parameters. Fix a map
\begin{equation*}
    h: \mathcal{Z}_1 \times \cdots \times \mathcal{Z}_K \times \mathcal{F}_1 \times \cdots \times \mathcal{F}_K \to \mathbb{R},
\end{equation*}
and assume that $h(\bar{z}, \cdot)$ is (Fr\'{e}chet) differentiable in each of the functional arguments $f_1$,...,$f_K$. The parameter of interest is the nonlinear functional 
\begin{equation}
    \vartheta = \E\left[h(\bar{Z}, f_1,..., f_K)\right]
\end{equation}
This framework includes, for example, mean-squared differences such as $\E[(f_1-f_2)^2]$. For this target, the non-linearity enters inside the expectation, so a pure (minimax) linear estimator may fail to attain desirable asymptotic properties. 
A natural remedy is an augmented correction obtained from the first-order Taylor expansion of $h$ around the true regression functions. Asymptotic properties of this augmented minimax linear estimator are left for future work. Section 5 in \citet{chernozhukov2022automatic} shows a closely related construction for Auto-DML estimators.

\section{Donsker Class Examples}\label{app:donsker}
\renewcommand{\theequation}{F.\arabic{equation}}
\setcounter{equation}{0}
\renewcommand{\theexample}{F.\arabic{example}}
\setcounter{example}{0}
Let $(\mathcal{Z}, \mathcal{A}, P)$ be a probability space and let $\mathcal{F} \subset L_2(P)$ be a set of square-integrable functions. The empirical process indexed by $\mathcal{F}$ is defined by 
    \begin{equation}
        \mathbb{G}_n(f) := \sqrt{n}(\E_n - \E)(f), \quad f\in \mathcal{F}
    \end{equation}
where $\E_n$ is the empirical measure based on an $i.i.d.$ sample $Z_1,...,Z_n \sim P$. 
\begin{definition}
    \textnormal{The function class $\mathcal{F}$ is $P$-Donsker if}
    \begin{equation}
        \mathbb{G}_n \to_d \mathbb{G}
    \end{equation}
    where the limit $\mathbb{G}$ is a tight Borel measurable element in the space $\ell^\infty(\mathcal{F})$.
\end{definition}

\begin{example}\textnormal{(H\"{o}lder classes)
Define the differential operator for any vector $k = (k_1,...,k_p) \in \mathbb{N}_0^p$ of $p$ nonnegative integers}
\begin{equation}
    D^k = \frac{\partial^{k_\cdot}}{\partial z_1^{k_1} \cdots \partial z_p^{k_p}}
\end{equation}
\textnormal{where $k_\cdot = \sum_{i=1}^p k_i$.
Define the H\"{o}lder norm for a function $f$ as} $$||f||_\alpha = \max_{0\leq k_\cdot \leq \lfloor \alpha \rfloor } \sup_{z \in \mathcal{Z}} |D^k f(z)| + \max_{k_\cdot = \lfloor \alpha \rfloor} \sup_{z_1 \not = z_2 \in \mathcal{Z}} \frac{|D^k f(z_1) - D^k f(z_2)|}{||z_1 - z_2||^{\alpha - \lfloor \alpha \rfloor}}$$ \textnormal{where $\alpha \in (0,\infty)$ and $\lfloor \alpha \rfloor$ is the greatest integer strictly smaller than $\alpha$.
The H\"{o}lder space $C^\alpha(\mathcal{Z})$ of order $\alpha$ is the space of all continuous functions $f:\mathcal{Z} \to \mathbb{R}$ for which the norm $||f||_\alpha$ is well defined and finite.}

\textnormal{Let $H^\alpha(\mathcal{Z})_M = \{f\in H^\alpha(\mathcal{Z}):||f||_\alpha \leq M\}$. Corollary 2.7.2 in \citet{vaart2023} shows that $H^\alpha(\mathcal{Z})_1$ with a bounded and convex set $\mathcal{Z} \subset \mathbb{R}^p$ is $P$-Donsker when $\alpha>p/2$.\footnote{This can be further relaxed to $\mathcal{Z} = \mathbb{R}^p$ by Corollary 2.7.3 in \citet{vaart2023}.}
}
\end{example}

\begin{example}\textnormal{(Elliptical classes)
    Let $\{\xi_j\}$ be a sequence of measurable functions satisfying $P(\xi_i \xi_j)=0$ for each $i \not= j$ and $\sum_{j=1}^\infty P(\xi_j(Z)^2) <\infty$. Define the ellipsoid $\mathcal{F}=\{\sum_{j=1}^\infty \beta_j \xi_j: \sum_{j=1} \beta_j^2 \leq 1, \text{series converges pointwise}\}$. $\mathcal{F}$ is $P$-Donsker by Theorem 2.13.2 in \citet{vaart2023}.}
\end{example}

\begin{example} \textnormal{(RKHS) Let $\mathcal{K}: \mathcal{Z}\times \mathcal{Z} \to \mathbb{R}$ be a measurable positive definite kernel with a corresponding reproducing kernel Hilbert space (RKHS) $\mathcal{H}_\mathcal{K}$. Assume $\mathcal{Z}$ is a separable metric space (such as $\mathbb{R}^p$), the kernel $\mathcal{K}$ is continuous as a real function of one variable with the other kept fixed and is bounded on the diagonal:}
\begin{equation}
    \sup_{z\in \mathcal{Z}} \mathcal{K}(z,z) < \infty.
\end{equation}
\textnormal{Let $\mathcal{F} = \{f\in \mathcal{H}_\mathcal{K}: ||f||_{\mathcal{H}_\mathcal{K}} \leq 1\}$.
By Theorem 1 in \citet{carcamo2024uniform}, $\mathcal{F}$ is $P$-Donsker.}
\end{example}

\begin{example}
    \textnormal{(Shallow neural nets) Define a function $f:[0,1]^p \to [0,1]$ be of Barron class with constant $C>0$ if there exists a measurable function $F: \mathbb{R}^p \to \mathbb{C}$ and some $c \in [-C,C]$ satisfying }
    \begin{equation}
        f(z) = c+\int_{\mathbb{R}^p }(e^{iz\cdot \eta}-1) \cdot F(\eta) d\eta \text{ for all } z \in [0,1]^p
        \label{eq:def_Barron_1}
    \end{equation}
    and 
    \begin{equation}
        \int_{\mathbb{R}^p } |\eta|\cdot |F(\eta)| d\eta \leq C.
        \label{eq:def_Barron_2}
    \end{equation}
    \textnormal{This condition goes back to \citet{barron1993universal}, who proposes it as a function-space characterization tailored to feedforward networks with one layer of sigmoidal nonlinearities (i.e., a single hidden layer). It requires that $f$ admits a Fourier representation with a finite first absolute moment of its Fourier transform. Barron class functions admit dimension-free approximation by shallow nets: Theorem 1 in \citet{barron1993universal} shows that there exist networks with one hidden layer with $m$ nodes whose integrated squared approximation error is $O(1/m)$, independent of the input dimension $p$.}
    
    \textnormal{We write $\mathcal{F}$ for the class of all functions defined by \eqref{eq:def_Barron_1} and \eqref{eq:def_Barron_2} with $C=1$. By Proposition 4.4 in \citet{petersen2021optimal}, $\mathcal{F}$ is $P$-Donsker.}
\end{example}

\section{Figures in Empirical Applications}\label{app:emp}
\renewcommand{\thefigure}{G.\arabic{figure}}
\setcounter{figure}{0}
\begin{figure}[H]
    \centering
    \includegraphics[width=0.8\linewidth]{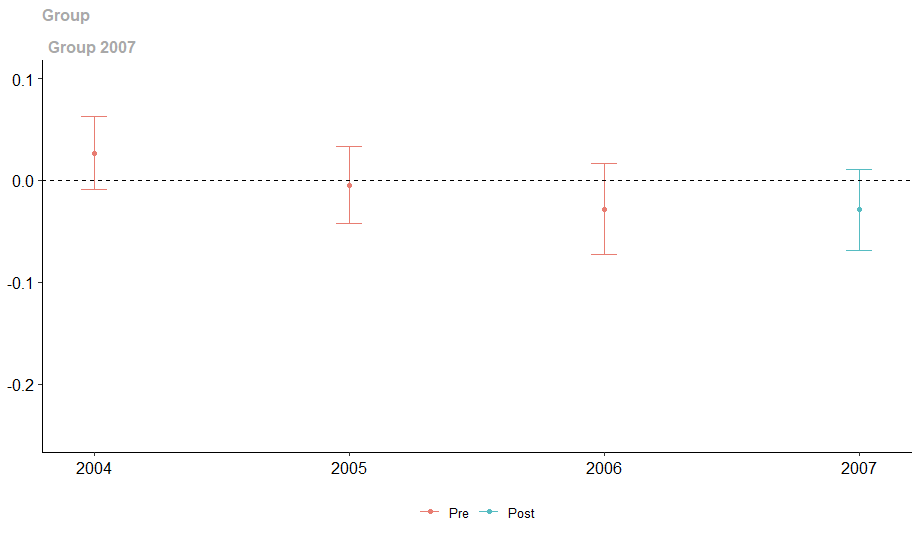}
    \caption{CS estimates for ATT in different time periods}
    \label{fig:CS_paralleltrend}
\end{figure}

\clearpage
\bibliographystyle{chicago}
\bibliography{ref}

\begin{thebibliography}{}

\bibitem[\protect\citeauthoryear{Abadie and Imbens}{Abadie and Imbens}{2006}]{abadie2006large}
Abadie, A. and G.~W. Imbens (2006).
\newblock Large sample properties of matching estimators for average treatment effects.
\newblock {\em Econometrica\/}~{\em 74\/}(1), 235--267.

\bibitem[\protect\citeauthoryear{Armstrong and Koles{\'a}r}{Armstrong and Koles{\'a}r}{2018}]{armstrong2018optimal}
Armstrong, T.~B. and M.~Koles{\'a}r (2018).
\newblock Optimal inference in a class of regression models.
\newblock {\em Econometrica\/}~{\em 86\/}(2), 655--683.

\bibitem[\protect\citeauthoryear{Armstrong and Koles{\'a}r}{Armstrong and Koles{\'a}r}{2021}]{armstrong2021finite}
Armstrong, T.~B. and M.~Koles{\'a}r (2021).
\newblock Finite-sample optimal estimation and inference on average treatment effects under unconfoundedness.
\newblock {\em Econometrica\/}~{\em 89\/}(3), 1141--1177.

\bibitem[\protect\citeauthoryear{Athey, Imbens, and Wager}{Athey et~al.}{2018}]{athey2018approximate}
Athey, S., G.~W. Imbens, and S.~Wager (2018).
\newblock Approximate residual balancing.
\newblock {\em Journal of the Royal Statistical Society. Series B (Statistical Methodology)\/}~{\em 80\/}(4), 597--623.

\bibitem[\protect\citeauthoryear{Barron}{Barron}{1993}]{barron1993universal}
Barron, A.~R. (1993).
\newblock Universal approximation bounds for superpositions of a sigmoidal function.
\newblock {\em IEEE Transactions on Information theory\/}~{\em 39\/}(3), 930--945.

\bibitem[\protect\citeauthoryear{Bartlett, Bousquet, and Mendelson}{Bartlett et~al.}{2005}]{bartlett2005}
Bartlett, P.~L., O.~Bousquet, and S.~Mendelson (2005).
\newblock Local rademacher complexities.
\newblock {\em The Annals of Statistics\/}~{\em 33\/}(4), 1497--1537.

\bibitem[\protect\citeauthoryear{Ben-Michael, Feller, Hirshberg, and Zubizarreta}{Ben-Michael et~al.}{2021}]{ben2021balancing}
Ben-Michael, E., A.~Feller, D.~A. Hirshberg, and J.~R. Zubizarreta (2021).
\newblock The balancing act in causal inference.
\newblock {\em arXiv preprint arXiv:2110.14831\/}.

\bibitem[\protect\citeauthoryear{Bruns-Smith, Dukes, Feller, and Ogburn}{Bruns-Smith et~al.}{2025}]{bruns2025augmented}
Bruns-Smith, D., O.~Dukes, A.~Feller, and E.~L. Ogburn (2025).
\newblock Augmented balancing weights as linear regression.
\newblock {\em Journal of the Royal Statistical Society Series B: Statistical Methodology\/}, qkaf019.

\bibitem[\protect\citeauthoryear{Callaway and Sant’Anna}{Callaway and Sant’Anna}{2021}]{callaway2021difference}
Callaway, B. and P.~H. Sant’Anna (2021).
\newblock Difference-in-differences with multiple time periods.
\newblock {\em Journal of Econometrics\/}~{\em 225\/}(2), 200--230.

\bibitem[\protect\citeauthoryear{C{\'a}rcamo, Cuevas, and Rodr{\'\i}guez}{C{\'a}rcamo et~al.}{2024}]{carcamo2024uniform}
C{\'a}rcamo, J., A.~Cuevas, and L.-A. Rodr{\'\i}guez (2024).
\newblock A uniform kernel trick for high and infinite-dimensional two-sample problems.
\newblock {\em Journal of Multivariate Analysis\/}~{\em 202}, 105317.

\bibitem[\protect\citeauthoryear{Chamberlain}{Chamberlain}{1992}]{chamberlain1992efficiency}
Chamberlain, G. (1992).
\newblock Efficiency bounds for semiparametric regression.
\newblock {\em Econometrica: Journal of the Econometric Society\/}, 567--596.

\bibitem[\protect\citeauthoryear{Chan, Yam, and Zhang}{Chan et~al.}{2016}]{chan2016globally}
Chan, K. C.~G., S.~C.~P. Yam, and Z.~Zhang (2016).
\newblock Globally efficient non-parametric inference of average treatment effects by empirical balancing calibration weighting.
\newblock {\em Journal of the Royal Statistical Society Series B: Statistical Methodology\/}~{\em 78\/}(3), 673--700.

\bibitem[\protect\citeauthoryear{Chernozhukov, Chetverikov, Demirer, Duflo, Hansen, Newey, and Robins}{Chernozhukov et~al.}{2018}]{chernozhukov2018double}
Chernozhukov, V., D.~Chetverikov, M.~Demirer, E.~Duflo, C.~Hansen, W.~Newey, and J.~Robins (2018).
\newblock Double/debiased machine learning for treatment and structural parameters.
\newblock {\em The Econometrics Journal\/}~{\em 21\/}(1), C1--C68.

\bibitem[\protect\citeauthoryear{Chernozhukov, Newey, Singh, and Syrgkanis}{Chernozhukov et~al.}{2022}]{chernozhukov2022dynamic}
Chernozhukov, V., W.~Newey, R.~Singh, and V.~Syrgkanis (2022).
\newblock Automatic debiased machine learning for dynamic treatment effects and general nested functionals.
\newblock {\em arXiv preprint arXiv:2203.13887\/}.

\bibitem[\protect\citeauthoryear{Chernozhukov, Newey, Quintas-Martinez, and Syrgkanis}{Chernozhukov et~al.}{2024}]{chernozhukov2024automatic}
Chernozhukov, V., W.~K. Newey, V.~Quintas-Martinez, and V.~Syrgkanis (2024).
\newblock Automatic debiased machine learning via riesz regression.
\newblock {\em arXiv preprint arXiv:2104.14737\/}.

\bibitem[\protect\citeauthoryear{Chernozhukov, Newey, and Singh}{Chernozhukov et~al.}{2022}]{chernozhukov2022automatic}
Chernozhukov, V., W.~K. Newey, and R.~Singh (2022).
\newblock Automatic debiased machine learning of causal and structural effects.
\newblock {\em Econometrica\/}~{\em 90\/}(3), 967--1027.

\bibitem[\protect\citeauthoryear{Dehejia and Wahba}{Dehejia and Wahba}{1999}]{dehejia1999causal}
Dehejia, R.~H. and S.~Wahba (1999).
\newblock Causal effects in nonexperimental studies: Reevaluating the evaluation of training programs.
\newblock {\em Journal of the American statistical Association\/}~{\em 94\/}(448), 1053--1062.

\bibitem[\protect\citeauthoryear{Donoho}{Donoho}{1994}]{donoho1994statistical}
Donoho, D.~L. (1994).
\newblock Statistical estimation and optimal recovery.
\newblock {\em The Annals of Statistics\/}~{\em 22\/}(1), 238--270.

\bibitem[\protect\citeauthoryear{Fr{\"o}lich}{Fr{\"o}lich}{2007}]{frolich2007nonparametric}
Fr{\"o}lich, M. (2007).
\newblock Nonparametric iv estimation of local average treatment effects with covariates.
\newblock {\em Journal of Econometrics\/}~{\em 139\/}(1), 35--75.

\bibitem[\protect\citeauthoryear{Hahn}{Hahn}{1998}]{hahn1998role}
Hahn, J. (1998).
\newblock On the role of the propensity score in efficient semiparametric estimation of average treatment effects.
\newblock {\em Econometrica\/}, 315--331.

\bibitem[\protect\citeauthoryear{Heckman}{Heckman}{1988}]{heckman1988minimax}
Heckman, N.~E. (1988).
\newblock Minimax estimates in a semiparametric model.
\newblock {\em Journal of the American Statistical Association\/}~{\em 83\/}(404), 1090--1096.

\bibitem[\protect\citeauthoryear{Hirano, Imbens, and Ridder}{Hirano et~al.}{2003}]{hirano2003efficient}
Hirano, K., G.~W. Imbens, and G.~Ridder (2003).
\newblock Efficient estimation of average treatment effects using the estimated propensity score.
\newblock {\em Econometrica\/}~{\em 71\/}(4), 1161--1189.

\bibitem[\protect\citeauthoryear{Hirshberg, Maleki, and Zubizarreta}{Hirshberg et~al.}{2019}]{hirshberg2019minimax}
Hirshberg, D.~A., A.~Maleki, and J.~R. Zubizarreta (2019).
\newblock Minimax linear estimation of the retargeted mean.
\newblock {\em arXiv preprint arXiv:1901.10296\/}.

\bibitem[\protect\citeauthoryear{Hirshberg and Wager}{Hirshberg and Wager}{2021}]{hirshberg2021augmented}
Hirshberg, D.~A. and S.~Wager (2021).
\newblock Augmented minimax linear estimation.
\newblock {\em The Annals of Statistics\/}~{\em 49\/}(6), 3206--3227.

\bibitem[\protect\citeauthoryear{Javanmard and Montanari}{Javanmard and Montanari}{2014}]{javanmard2014confidence}
Javanmard, A. and A.~Montanari (2014).
\newblock Confidence intervals and hypothesis testing for high-dimensional regression.
\newblock {\em The Journal of Machine Learning Research\/}~{\em 15\/}(1), 2869--2909.

\bibitem[\protect\citeauthoryear{Kallus}{Kallus}{2020}]{kallus2020generalized}
Kallus, N. (2020).
\newblock Generalized optimal matching methods for causal inference.
\newblock {\em Journal of Machine Learning Research\/}~{\em 21\/}(62), 1--54.

\bibitem[\protect\citeauthoryear{Lehmann and Romano}{Lehmann and Romano}{2005}]{lehmann2005testing}
Lehmann, E.~L. and J.~P. Romano (2005).
\newblock {\em Testing statistical hypotheses}.
\newblock Springer.

\bibitem[\protect\citeauthoryear{Low}{Low}{1995}]{low1995bias}
Low, M.~G. (1995).
\newblock Bias-variance tradeoffs in functional estimation problems.
\newblock {\em The Annals of Statistics\/}~{\em 23\/}(3), 824--835.

\bibitem[\protect\citeauthoryear{Ma, Chiou, and Wang}{Ma et~al.}{2006}]{ma2006efficient}
Ma, Y., J.-M. Chiou, and N.~Wang (2006).
\newblock Efficient semiparametric estimator for heteroscedastic partially linear models.
\newblock {\em Biometrika\/}~{\em 93\/}(1), 75--84.

\bibitem[\protect\citeauthoryear{Mendelson}{Mendelson}{2020}]{mendelson2020extending}
Mendelson, S. (2020).
\newblock Extending the scope of the small-ball method.
\newblock {\em arXiv preprint arXiv:1709.00843\/}.

\bibitem[\protect\citeauthoryear{Petersen and Voigtlaender}{Petersen and Voigtlaender}{2021}]{petersen2021optimal}
Petersen, P. and F.~Voigtlaender (2021).
\newblock Optimal learning of high-dimensional classification problems using deep neural networks.
\newblock {\em arXiv preprint arXiv:2112.12555\/}.

\bibitem[\protect\citeauthoryear{Rosenbaum and Rubin}{Rosenbaum and Rubin}{1983}]{rosenbaum1983central}
Rosenbaum, P.~R. and D.~B. Rubin (1983).
\newblock The central role of the propensity score in observational studies for causal effects.
\newblock {\em Biometrika\/}~{\em 70\/}(1), 41--55.

\bibitem[\protect\citeauthoryear{van~der Vaart and Wellner}{van~der Vaart and Wellner}{2023}]{vaart2023}
van~der Vaart, A.~W. and J.~A. Wellner (2023).
\newblock {\em Weak Convergence and Empirical Processes: With Applications to Statistics\/} (2 ed.).
\newblock Springer Cham.

\bibitem[\protect\citeauthoryear{Von~Bahr and Esseen}{Von~Bahr and Esseen}{1965}]{von1965inequalities}
Von~Bahr, B. and C.-G. Esseen (1965).
\newblock Inequalities for the $r$th absolute moment of a sum of random variables, $1\leq r\leq 2$.
\newblock {\em The Annals of Mathematical Statistics\/}, 299--303.

\bibitem[\protect\citeauthoryear{Wang and Zubizarreta}{Wang and Zubizarreta}{2020}]{wang2020minimal}
Wang, Y. and J.~R. Zubizarreta (2020).
\newblock Minimal dispersion approximately balancing weights: asymptotic properties and practical considerations.
\newblock {\em Biometrika\/}~{\em 107\/}(1), 93--105.

\bibitem[\protect\citeauthoryear{Wong and Chan}{Wong and Chan}{2018}]{wong2018kernel}
Wong, R.~K. and K.~C.~G. Chan (2018).
\newblock Kernel-based covariate functional balancing for observational studies.
\newblock {\em Biometrika\/}~{\em 105\/}(1), 199--213.

\bibitem[\protect\citeauthoryear{Zhang and Zhang}{Zhang and Zhang}{2014}]{zhang2014confidence}
Zhang, C.-H. and S.~S. Zhang (2014).
\newblock Confidence intervals for low dimensional parameters in high dimensional linear models.
\newblock {\em Journal of the Royal Statistical Society: Series B: Statistical Methodology\/}, 217--242.

\bibitem[\protect\citeauthoryear{Zubizarreta}{Zubizarreta}{2015}]{zubizarreta2015stable}
Zubizarreta, J.~R. (2015).
\newblock Stable weights that balance covariates for estimation with incomplete outcome data.
\newblock {\em Journal of the American Statistical Association\/}~{\em 110\/}(511), 910--922.

\end{thebibliography}

\end{document}